# Skybridge: 3-D Integrated Circuit Technology Alternative to CMOS

Mostafizur Rahman, Santosh Khasanvis, Jiajun Shi, Mingyu Li, and Csaba Andras Moritz

**Continuous scaling of CMOS has been the major catalyst in miniaturization of integrated circuits (ICs) and crucial for global socio-economic progress. However, scaling to sub-20nm technologies is proving to be challenging as MOSFETs are reaching their fundamental limits[1] and interconnection bottleneck[2] is dominating IC operational power and performance. Migrating to 3-D, as a way to advance scaling, has eluded us due to inherent customization and manufacturing requirements in CMOS that are incompatible with 3-D organization. Partial attempts with die-die[3] and layer-layer[4] stacking have their own limitations[5]. We propose a 3-D IC fabric technology, Skybridge™, which offers paradigm shift in technology scaling as well as design. We co-architect Skybridge's core aspects, from device to circuit style, connectivity, thermal management, and manufacturing pathway in a 3-D fabric-centric manner, building on a uniform 3-D template. Our extensive bottom-up simulations, accounting for detailed material system structures, manufacturing process, device, and circuit parasitics, carried through for several designs including a designed microprocessor, reveal a 30-60x density, 3.5x performance/watt benefits, and 10X reduction in interconnect lengths vs. scaled 16-nm CMOS. Fabric-level heat extraction features are shown to successfully manage IC thermal profiles in 3-D. Skybridge can provide continuous scaling of integrated circuits beyond CMOS in the 21st century.**

As CMOS scaling options are exhausted by fundamental limitations, device and circuit integration in the third-dimension could provide a possible pathway without extensively relying on ultra-scaled transistors. So far, however, the migration of CMOS to 3-D has been unattainable. The CMOS fabric architecture uses complementary MOSFETs in an inverted logic, where both pull-up and pull-down transistors share the same input. The C-MOSFETs have opposite doping profiles and each MOSFET contains multiple doping regions. In order to achieve correct circuit operation, these MOSFETs have to be carefully sized and doped precisely in a 3-D stack. In terms of connectivity, a 3-D implementation of CMOS circuits would imply that each input signal has to be vertically routed twice for C-MOSFETs. Mapping such connectivity in 3-D even for a 4 fan-in logic gate, where pull-down transistors are stacked and pull-up transistors are isolated, or vice versa, would result in connectivity bottlenecks; for a large circuit these issues would become unmanageable. In terms of manufacturing, CMOS in 3-D would imply extreme lithography to create various vertical shapes for 3-D C-MOSFETs with each MOSFET doped precisely in isolated 3-D regions, which is impractical. In addition to these, there is no heat extraction capability inherent to CMOS to prevent thermal hotspot development. Admittedly, since the inception of vertical devices in 2000[6] there has been no success in the realization of 3-D CMOS despite a significant industrial push.

In contrast to CMOS, that has evolved focusing on the device especially and requires a largely component-centric assembly, the Skybridge fabric shifts to a fabric-centric mindset and provides an integrated solution for all technology aspects. First, it starts with a regular array of uniform vertical nanowires that forms the Skybridge *template* (Fig. 1A). Second, its doping requirement is uniform, without regions, and done once at the wafer level. Finally, the various features of the fabric are realized through functionalizing this template with material deposition techniques. All inserted material structure features, regarding device, circuit style, connectivity, thermal management, are co-architected for 3-D requirements, compatibility, manufacturability, and overall efficiency, even if tradeoffs needed to be made on individual aspects.

Vertical Junctionless transistors that do not require doping variations are implemented on these nanowires, and are accommodated in new Skybridge circuit styles supporting both logic and volatile memory in 3-D and using only single-type and uniformly sized transistors. Further, nanowires are linked with structures referred to as Skybridge Bridges for connectivity. Noise management is similarly supported in the fabric. Various heat extraction features are accommodated in the same template with fabric-level features architected to be used in circuits preventing hotspot development. In contrast to CMOS's system-level heat management, Skybridge's fabric-level heat-extraction features are an integral part and a new dimension of 3-D circuit design.

Lithographic precision is required only for patterning vertical nanowires. The definition of all active components is based on a multi-layer material deposition, which is lower cost, and can be controlled to few Angstrom's precision. Several groups have already demonstrated individual manufacturing steps that are required for this type of fabric assembly. Our group has shown the Junctionless transistor[7] experimentally and through detailed simulations (Supplementary Section 2.1 and 8). We have also done a comprehensive fabric technology evaluation by fully designing and verifying several circuits including a microprocessor. This evaluation uses a detailed bottom-up approach, from the material system layer to architecture, with 3-D TCAD device modeling, process simulation, extracted behavioral models, and interconnect parasitics specific to circuit layouts captured in HSPICE simulations (Supplementary Section 3).

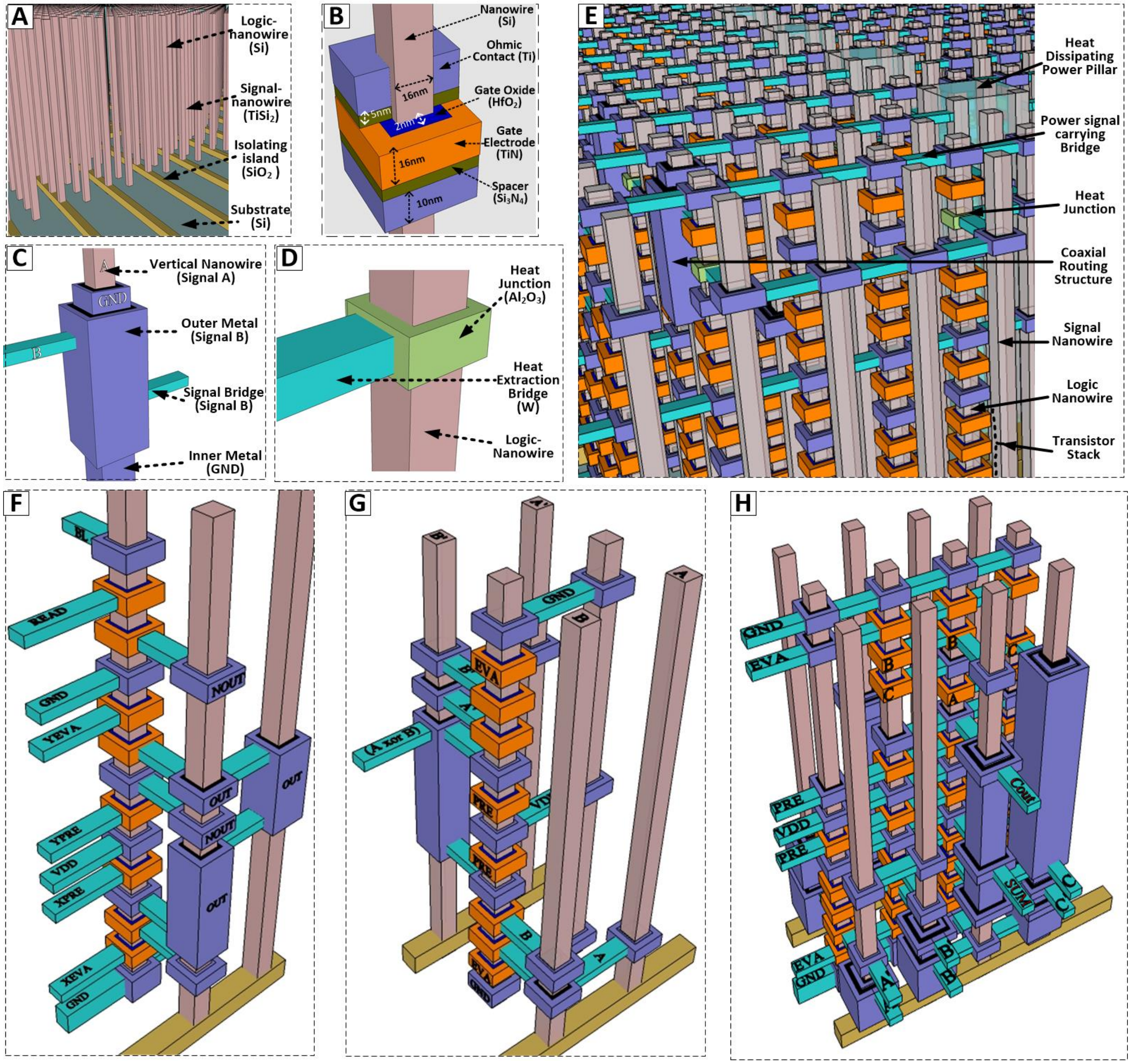

**Figure 1 | Fabric core components and material choices, fabric view and 3-D circuit implementation examples**. A) Arrays of regular crystal vertical Si nanowires, B) Vertical Gate-All-Around Junctionless nanowire transistor, C) Routing features: Bridges and Coaxial structures, D) Heat Extraction features: Heat Junction and Bridge (additional feature Heat Dissipating Power Pillar is shown in E). E) Skybridge fabric implementation utilizing core components (Abstract view). F-H shows 3-D circuit examples in Skybridge. F) Skybridge's volatile memory cell, G) XOR gate with AND-of-NAND dynamic logic, H) 3-D full adder implementation using combination of NAND-NAND and AND-of-NAND logics.

Following the 3-D templated design and assembly principles, the core Skybridge components include vertical semiconducting single crystalline nanowires (Fig. 1A), vertical Gate-All-Around Junctionless transistors (Fig. 1B), Bridges and Coaxial Routing structures (Fig. 1C), Heat Extraction Junctions (Fig. 1D) and Heat Dissipating Power Pillars (Fig. 1E). The 3-D integrated Junctionless device has a very simple structure with no abrupt doping junctions for Drain/Channel/Source regions and device behavior that is primarily modulated by the workfunction difference between the gate and the channel (detailed simulation results can be found in Supplementary Section 2.1). The simplified device structure implies that sequential material deposition techniques can be used to construct these devices in the vertical direction, and *a priori* wafer level doping is sufficient for fabric implementation. These devices are used in a 3-D Compound Dynamic circuit style with only single type uniformly sized transistors (Supplementary Section 2.2) and intrinsic noise mitigation. The tuning knobs for Skybridge circuit implementations are cascading choices and compound gates, dual rail vs. single rail implementations, and fan-in. Elementary logic circuits are single-stage NAND, XOR, and AND-of-NAND. Each vertical nanowire contains logic gates with a stack of transistors connected to Input/Output/Global signals using connecting Bridges. The Bridges carry signals and span the required distance by hopping nanowires, facilitated by Coaxial Routing structures, without perturbing normal operation or introducing coupling noise (Supplementary Section 1 and 2.2.2). Fig. 1E shows an abstract view of Skybridge fabric utilizing core components. The density benefit of such vertical integration is obvious from an XOR and full adder logic implementations shown in Figs. 1G and 1H; they are completed on just one and four transistor carrying nanowires respectively.

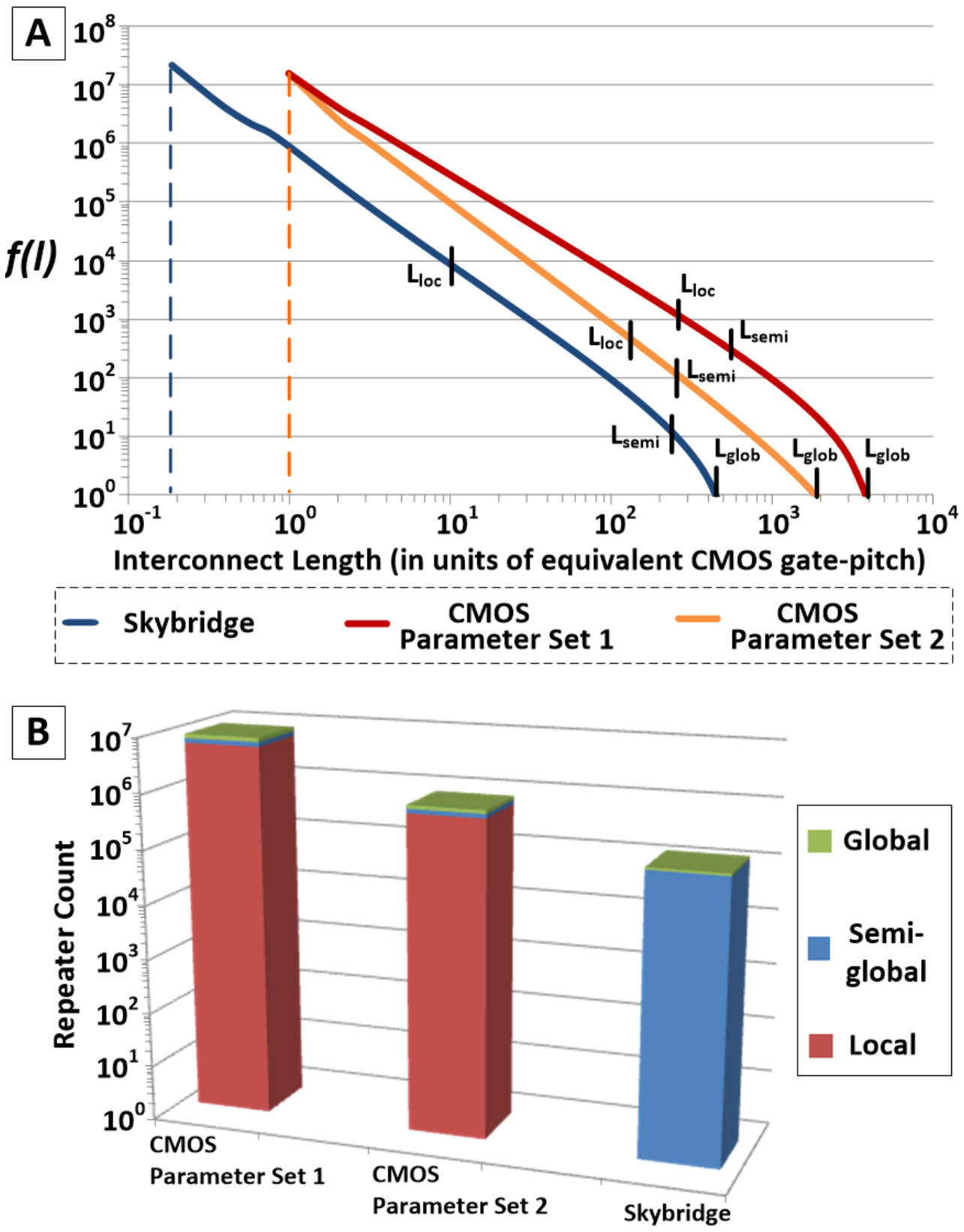

**Figure 2 | Comparison of interconnect distribution and estimated repeater count in Skybridge and CMOS, for an integrated circuit consisting of 10 million gates.** A) Interconnect distribution estimating the number of interconnects of a given length (in gate-pitches). Skybridge reduces the length of interconnects significantly, by almost 10x for the longest interconnect. B) Estimated count of repeaters based on the interconnect distribution in (A). Parameters for Skybridge: $k$=5.39, $p$=0.577 (Rent's parameters), average fan-out = 2.018. For CMOS, Parameter Set 1: $k$=4, $p$=0.66, average fan-out = 3; and Parameter Set 2: $k$=3.416, $p$=0.473, average fan-out = 1.7.

Additionally to logic, a high-performance volatile memory is a key requirement in integrated circuits. CMOS SRAMs use complementary-doped devices in a complex layout where device sizing needs to be highly customized for operational stability; such circuit is not suitable for 3-D. The Skybridge volatile RAM design follows the Skybridge circuit style and is mapped on to a single nanowire (Fig. 1F) without customizations. There are no write/read stability concerns as in SRAMs, and significant benefits are achieved in all aspects when scaled (Supplementary Section 2.3 and 6.2.2).

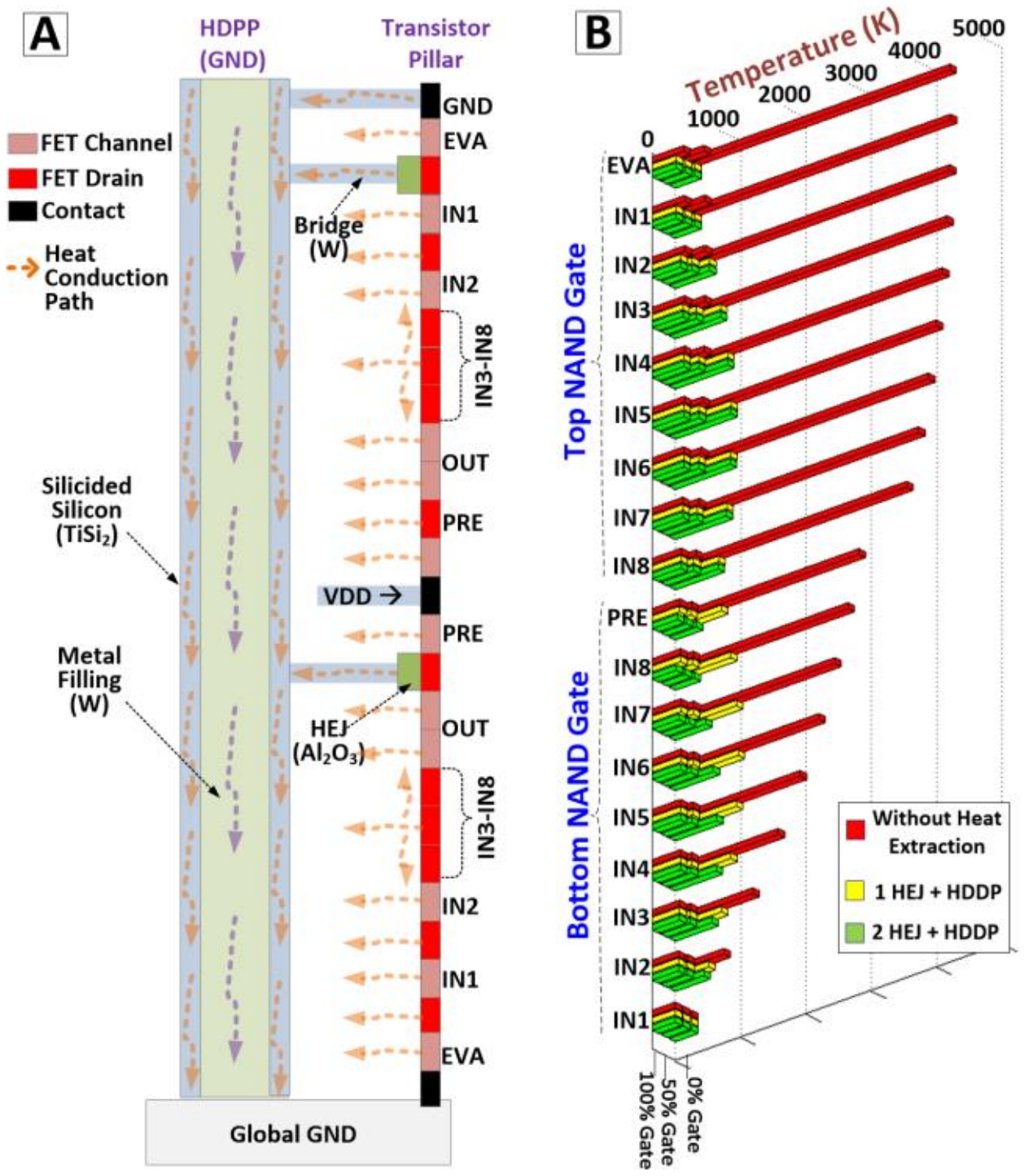

**Figure 3 | Heat management in 3-D.** A) A logic-nanowire implementing compound dynamic circuit using two logic gates; configuration for heat extraction with Skybridge includes Heat Junction (HEJ), Heat Bridges, and Heat Dissipating Power Pillars (HDPP). B) Thermal profile of each transistor for various scenarios; the thermal profiles include effects of heat conduction through transistor Gates, when the Gates are 0%, 50%, and 100% thermally conductive.

In nanoscale CMOS connectivity bottlenecks escalate and global interconnect delays are dominant. To mitigate this, long interconnects are broken down into shorter segments where each segment is driven by large area repeaters. While this mitigates long interconnect delays, it has introduced new challenges with scaling in terms of significant increase in power consumption and repeater count, also resulting in area overhead. 3-D connectivity requirements for Skybridge circuits are met through intrinsic routing features in addition to routing with top metal layers. We have quantified the 3-D connectivity implications for a 10 million logic gate design using predictive interconnect distribution models (Supplementary Section 4). Our results show (Fig. 2A) that Skybridge interconnect lengths are up to 10x shorter than in CMOS for the longest wire, and Local and Semi-Global interconnects are dominant. The repeater requirements are up-to 100x less vs. CMOS (Fig. 2B). This has huge implications for overall power consumption and density improvements of Skybridge-based large-scale circuit architectures such as superscalar processors and multi-cores.

Managing the 3-D thermal profile is supported at the fabric-level in Skybridge. Architected structures are used in a synergistic manner to mitigate thermal challenges. Heat Extraction Junctions are specialized junctions that are

**Table 1 | Benchmarking: 16nm Skybridge vs. 16nm CMOS**

| | Throughput (operations/sec) | | Power (μW) | | Area ($\mu m^2$) | |
|---|---|---|---|---|---|---|
| | CMOS | SB | CMOS | SB | CMOS | SB |
| **WISP-4 Microprocessor** | 4.39e09 | 5.1e09 | 886 | 301 | 289 | 9.52 |
| **RAM (cell)** | 50e09 | 49.5e09 | 1.4*<br>8.2† | 0.33*<br>0.16† | 0.065 | 0.014 |
| **4-Bit CLA** | 9.9e9 | 10.4e9 | 235 | 19.4 | 18.7 | 0.76 |
| **8-Bit CLA** | 4.5e9 | 5.7e9 | 287 | 23.5 | 64.7 | 1.34 |
| **16-BitCLA** | 2.4e9 | 3.7e9 | 297 | 27.8 | 130.2 | 2.15 |

**SB**, Skybridge; **CLA**, Carry Look Ahead Adder. *Active Power, †Leakage Power.

employed to extract heat from heated regions in a logic-implementing nanowire without affecting its operation; the extracted heat is then carried by Heat Extraction Bridges and is dissipated through Heat Dissipation Power Pillars to the bulk silicon substrate.

To quantify the effectiveness of fabric-level heat extraction, we have performed detailed thermal modeling considering effects of nanoscale dimensions for the worst-case static heat scenario. Fig. 3A shows a logic-nanowire implementing a Skybridge style compound dynamic circuit using two logic gates with each having maximum fan-in. Fig. 3B shows the thermal profile of all the transistors. Up-to 90% reduction in temperature from 4307K to 400K was achieved primarily by using heat extraction features in the topmost region of the nanowire; by using two Heat Junctions and Bridges the average temperature was reduced by 85% and found to be below threshold temperature for modern microprocessors[8]. The temperature profile can be further reduced, as per requirements, by utilizing more heat extraction features (Supplementary Section 5).

A fully functional 4-bit microprocessor, several higher bit-width circuits, and Skybridge volatile memory circuits were implemented and compared with equivalent 16-nm CMOS implementations. Our benchmarking shows Skybridge's RAM design being 4.6x denser, and consuming 4.2x and 51.2x less active and leakage power, at similar performance (Table 1), when compared with scaled 16nm high performance 6T-SRAM (Supplementary Section 6.2.2). The Skybridge processor implementation is 30x denser, consumes 2.94x less power and operates at 16% higher frequency (Supplementary Section 7). In addition, 4, 8, and 16 bit Carry Look Ahead (CLA) adder designs were created and evaluated. As shown in Table I, the 16-bit CLA achieves 60.5x density, 10.6x power, and 54% performance (or ~16.5x performance/Watts) benefits (Supplementary Section 6.2.1). For even larger multi-million to billion transistor designs these benefits would improve further due to the projected 10x shorter interconnect lengths and 100x reduced repeater counts. We estimate that Skybridge could enable more than two orders of magnitude density and close to two orders of magnitude performance/Watt improvements. Further vertical scaling and fabric refinement can also be expected.

In summary, we have introduced and extensively validated a novel 3-D IC fabric technology. Its advantages vs. CMOS are tremendous in all respects. Skybridge can drive several fields of study and continue to advance ICs in the 21st century. Its benefits would be touching all facets of human life.

## METHODS SUMMARY

**Device simulation.** V-GAA Junctionless device was simulated using 3-D Sentaurus Process and Device simulators[9,10]. Process simulator created the device structure emulating actual process flow; process parameters such as ion implantation dosage, anneal duration and temperature, deposition parameters etc. were similar as our experimental process parameters for Junctionless device demonstration[7]. This device structure was then used in Sentaurus Device simulations to extract device characteristics accounting for nanoscale confinement, surface and coulomb scattering, and mobility degradation effects.

**3-D circuit simulation.** TCAD device simulation data was used to generate an HSPICE compatible device model using Datafit Software for circuit simulations. Using the device model and the 3-D circuit specific interconnect parasitic in HSPICE, core logic and memory circuit validation, fan-in sensitivity and noise mitigation analysis were performed. The same methodology was extended for high bit-width arithmetic and microprocessor evaluation; all circuit designs were according to Skybridge design rules and layout guidelines (Supplementary Section 9) that are based on manufacturing assumptions (Supplementary Section 8) at 16nm. For logic benchmarking, equivalent CMOS circuits were designed using state-of-the-art CAD tools, and were scaled to 16nm node using standard scaling parameters[11,12]. 6T-SRAM was evaluated with HSPICE using extracted scaling rules[13].

**Interconnect estimation.** Predictive models based on Rent's rule [14,15] were developed and used for analytical estimation of interconnect lengths, distributions, and repeater requirements. Rent's parameters for Skybridge were extracted from designed circuits; CMOS parameters were determined from designed circuits and from literature[14]. Estimations were done for a 10 million logic-gate design.

**Thermal simulation.** Circuit-level thermal evaluation was done using detailed modeling and simulation. The thermal model was built at the transistor-level granularity[16], and effects of nanoscale dimensions[17] were taken into consideration. An HSPICE equivalent thermal resistance network was built for the circuit, and HSPICE simulations were done for the worst-case static heat dissipation scenario.

**Acknowledgements** This work was supported by the Center for Hierarchical Manufacturing (CHM, NSF DMI-0531171), and NSF (CCF-0508382) grants.

# Supplementary Information

# SkyBridge: 3-D Integrated Circuit Technology Alternative to CMOS

## Table of Contents

## SUPPLEMENTARY DOCUMENT ORGANIZATION

This supplementary document provides extensive technical details and evaluation of the Skybridge fabric. We discuss core fabric components, material choices and structures, device, logic and memory circuit styles, fabric evaluation methodologies, connectivity implications, thermal management, 3-D circuit design rules and guidelines, high bit-width arithmetic circuits, microprocessor design, benchmarking results and associated manufacturing pathway. Since this is a multi-disciplinary document, we tried to ensure that each section is self-contained as much as possible. For readability, every section has a very brief summary and overview sub-sections preceding the main discussion.

## 1. CORE SKYBRIDGE COMPONENTS

Section Summary

This section details key components of the Skybridge fabric.

Overview

Skybridge fabric design follows a fabric-centric mindset assembling structures on a 3-D uniform template of single crystal vertical nanowires, keeping 3-D requirements, compatibility, and overall efficiency as its central goal. All active components and fabric features are formed on these nanowires through material depositions. In this fabric, 3-D device, circuit, connectivity, and thermal management issues are solved by carefully architecting towards 3-D organization. From architectural perspective, this is in stark contrast to the CMOS component-centric mindset, where transistors are the primary design components and the main technology scaling factor, wherein circuits, interconnection network, power and system level heat-management schemes are engineered to accommodate these transistors.

Beyond the Skybridge template based on the uniform single-doped vertical silicon nanowires, the key components functionalized include vertical Gate-All-Around (V-GAA) Junctionless transistors, Bridges, Coaxial routing structures, Heat Extraction Junctions (HEJs) and large area Heat Dissipating Power Pillars (HDPPs). V-GAA Junctionless transistors are stacked on the vertical nanowires and are interconnected for realizing 3-D circuits. Local interconnection is primarily through unique routing features: Bridges and Coaxial routing structures. The heat management features HEJs and HDPPs are used in conjunction with Bridges to extract and dissipate heat from heated regions in the logic implementing nanowires.

Fabric Components

### *1.1 Vertical Silicon Nanowires*

Regular Arrays of single crystal vertical silicon nanowires are fundamental building blocks of Skybridge fabric. All logic and memory functionalities are achieved in these nanowires. These nanowires are classified such that some of them are used as (i) *logic nanowires* to accommodate logic gates with each gate consisting of a stack of vertical transistors, and (ii) *signal nanowires* to carry Input/Output/Global signals themselves and facilitate routing of other signals for logic gates. All the nanowires are heavily doped; this is necessary for the V-GAA Junctionless transistors employed and for metal silicidation. The nanowires that are used for Input/Output/Global signal routing are silicided to reduce their electrical resistance.

Fig. S1.1A shows arrays of regular vertical silicon nanowires that are patterned from highly doped silicon substrate with discrete $SiO_2$ islands (see Section 8.1 and 8.2 for wafer preparation and Nanowire Patterning). The $SiO_2$ islands are used to isolate signal-carrying nanowires from contacting the bulk silicon substrate.

### *1.2 Vertical Gate-All-Around Junctionless Nanowire Transistors*

Active devices in this fabric are *n-type vertical Gate-All-Around (V-GAA) Junctionless nanowire transistors*. Junctionless transistors are well-suited for Skybridge's 3-D implementation, since they eliminate the requirement of precision doping in 3-D. Junctionless transistors have uniform doping across drain, channel and source regions; their behavior is modulated by the workfunction difference between the gate and the heavily doped channel. In addition, there is no requirement for raised S/D structure for Contact formation: contacting the low workfunction metal with heavily n-doped source and drain regions can form a good Ohmic contact. In Section 2.1, we present more details of V-GAA device characteristics through 3-D TCAD process and device simulations. Previously, we have also experimentally validated the Junctionless device concept [7].

In Skybridge, structural simplicity of Junctionless transistors is exploited to easily form devices in vertical direction. As shown in Fig. S1.1B, V-GAA Junctionless transistors are formed by just depositing materials; in the beginning drain contact metal (Ti) layer is deposited, and is followed by spacer ($Si_3N_4$), gate oxide ($HfO_2$), gate electrode (TiN), spacer ($Si_3N_4$) and source metal (Ti) layer deposition. Since depositing materials forms the devices, there is no requirement for lithographic or doping precision. A wafer/IC level *a priori* doping is sufficient for devices and contacts (see section 8 for the envisioned manufacturing pathway).

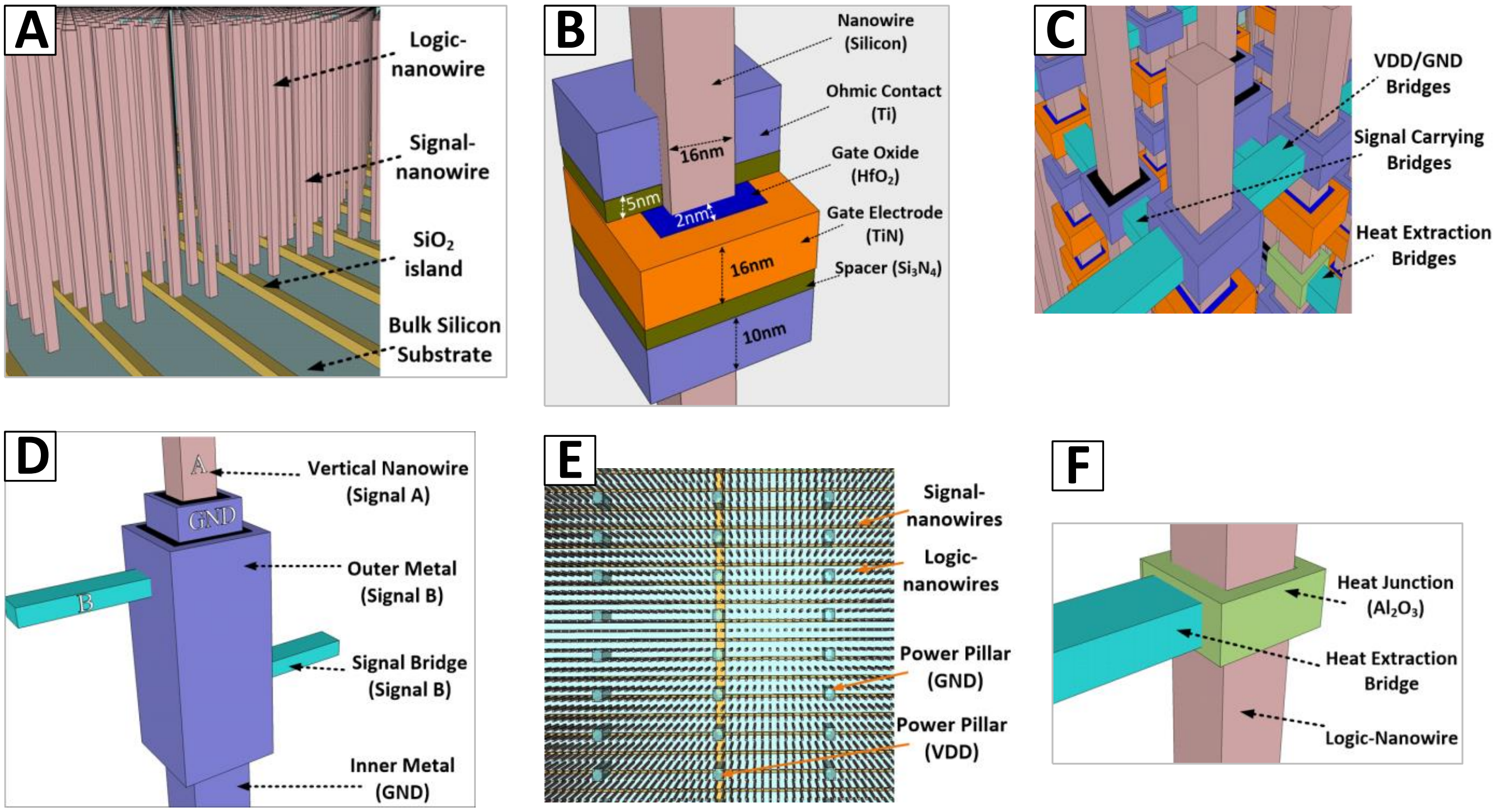

**Fig. S1.1 | Core fabric components.** A) Arrays of regular single crystal vertical Si nanowires, B) vertical Gate-All-Around Junctionless nanowire transistor, C) nanowire linking Bridges, D) Coaxial routing structures, E) sparse large area Heat Dissipating Power Pillars, F) Heat Extraction Junctions

### *1.3 Bridges*

Bridges are unique to the Skybridge fabric; they enable high degree of connectivity in 3-D with minimum area overhead, and also play a key role in heat extraction. Based on their roles, Bridges can be classified into two categories: *signal carrying Bridges* and *heat extraction Bridges*.

The primary role of signal-carrying Bridges is to form links between two adjacent nanowires, and carry Input/Output/Global signals (Fig. S1.1C). Depending on the circuit implementation, Bridges can be placed at different nanowire heights, and can propagate relatively long distances in the layout by *hopping* nanowires; *Coaxial routing structures* are used in conjunction with Bridges to facilitate this nanowire hopping. These routing features provide flexibility, and allow dense 3-D interconnection minimizing interconnect congestion. Section 4 details the 3-D connectivity benefits of Skybridge.

In addition to their usage as signal carrying links, the Bridges also facilitate heat extraction. *Heat extraction Bridges* provide thermally conductive paths for heat transfer from the heat source. They are used in conjunction with *Heat Extraction Junctions (HEJs)* and *large area Heat Dissipating Power Pillars (HDPPs)* to maximize heat extraction and dissipation. Subject to the thermal profile of the nanowires, HEJs and Bridges can be connected to any heated region in the logic-nanowire. Fig. S1.1F shows an example of a Bridge connected to a HEJ in the logic gate output region (see Section 5 for thermal modeling and heat extraction results for 3-D circuits).

### *1.4 Coaxial Routing Structures*

*Coaxial routing* refers to a routing scheme, where a signal routes coaxially to another inner signal without affecting each other. This routing is unique for Skybridge, and is enabled by the vertical integration approach. Fig. S1.1D shows an example: signal ‘A’ is carried by the vertical nanowire, whereas the signal ‘B’ is routed by Bridges; the Coaxial routing structure allows signal ‘B’ to hop the nanowire and continue its propagation. This coaxial routing is achieved by specially configuring material structures, insulating oxide and contact metal. By controlling the thickness of the insulating oxide, and by choosing low workfunction metal as Contact Metal, proper signal isolation can be achieved. A thick layer of $SiO_2$ as insulating oxide and Titanium (Ti) as Contact metal is well suited for this purpose. Workfunction difference between Ti and n-doped Si is such that there is no carrier depletion; moreover a thick layer of $SiO_2$ ensures no electron tunneling between the Contact metal and silicon nanowire.

Using multiple coaxial layers can provide noise isolation and route multiple signals. Coupling noise in dense interconnect networks and in dynamic circuits is a well-known phenomenon. By configuring the Coaxial routing structure to incorporate a GND signal for noise shielding, coupling noise can be mitigated. Fig. S1.1D also illustrates this concept; the GND signal in between signal A and B acts as noise shield, and prevent coupling between these two signals. More details on noise mitigation can be found in Section 2.2.2.

*1.5 Heat Extraction Junctions*

*Heat Extraction Junction* (HEJ) is an architected feature (Fig. S1.1F) used to extract heat from a heated region in logic-nanowire without affecting the underlying logic operation. An HEJ is a thermally conductive but electrically isolated junction. When combined with Bridges, the HEJs provide flexibility to be connected to any heated region in the logic-nanowire to prevent hotspot development.

These junction properties of an HEJ are achieved by carefully architecting material requirements. A sufficiently thick layer (6nm) of $Al_2O_3$ is used for this purpose – $Al_2O_3$, a good insulator with excellent thermal conduction property (thermal conductivity 39.18 $Wm^{-1}k^{-1}$ [18]).

*1.6 Heat Dissipating Power Pillars*

Large area *Heat Dissipating Power Pillars* (HDPPs) serve both the purpose of reliable power supply and heat dissipation. Depending on electrical and thermal requirements, these pillars are placed intermittently throughout the layout and are connected by Bridges. They occupy large area, and are specially designed to have low electrical resistance, and maximum heat conduction. As shown in Fig. S1.1E, HDPPs occupy a 2 x 2 nanowire pitch and would typically be placed on the periphery of circuit layouts. The 4 nanowires used in HDPPs are all metal silicided, and the region is filled with Tungsten (W) to maximize thermal conductance and minimize electrical resistance.

HDPPs that carry GND signals are connected to Bulk silicon at the bottom, whereas HDPPs carrying VDD signals are isolated from the bulk with $SiO_2$ islands (Fig. S1.1E). For heat extraction purposes, Bridges connect to HDPPs (GND) on one end and to HEJs on the other; this configuration ensures that the heat extraction Bridges are at reference temperature for maximum heat extraction. Details on HDPPs, and thermal analysis can be found in Section 5.

## 2. DEVICE, CIRCUIT STYLE AND MEMORY

Section Summary

This section details Skybridge's device, circuit style and volatile memory elements. We show the Vertical Gate-All-Around Junctionless transistor geometry, and TCAD simulated device characteristics. We present details on the 3-D compatible circuit style, and show different approaches to designing for high performance and low power at ultra-high density. We also introduce Skybridge's volatile memory approach equivalent with the CMOS SRAM.

Overview

The manufacturing compatibility and the ability to efficiently implement logic and memory functionalities in 3-D without incurring detrimental connectivity overhead are key requirements for realizing circuits in 3-D. The CMOS circuit style is not suitable for this purpose, since it requires customizations in complementary device doping, sizing and placements for functionality; such an implementation in 3-D would result in significant connectivity bottleneck, and escalate manufacturing complexities.

In Skybridge, 3-D circuit and connectivity requirements are met by synergistically exploring device, circuit and architectural aspects without compromising on manufacturability. A dynamic circuit style that is amenable to implementations in 3-D is chosen for realizing arbitrary logic and volatile memory circuits. This dynamic circuit style uses only single type uniformly sized Junctionless transistors. It is easily mapped onto arrays of regular vertical nanowires without requiring any customizations in terms of doping, sizing or incompatible routing; formation of active components is primarily by layer-by-layer material depositions. As discussed, to meet 3-D inter-circuit connectivity requirements, Skybridge has intrinsic routing features: signal nanowires, Bridges and Coaxial structures.

The dynamic circuit style, along with the 3-D integration scheme allows various choices to design for either high performance or low power, or a balance of both, at a very high density. The tuning knobs for Skybridge circuit implementations are cascading choices and compound gates, dual rail vs. single rail implementations, and fan-in. In the following, we present more on these choices, and discuss trade-offs with example circuits. We also show how coupling noise due to ultra-dense 3-D integration, is mitigated through optimizing circuit clocking scheme and architecting fabric features. The discussion begins with analysis of active device components, and follows by details on logic circuit styles and volatile memory design.

*2.1 Vertical Gate-All-Around Junctionless Transistor*

N-type vertical Gate-all-around (V-GAA) Junctionless nanowire transistor were chosen as active devices in the Skybridge fabric. V-GAA Junctionless transistors do not require abrupt doping variations within the device; as a result complexities related to precision doping in 3-D and high temperature annealing are eliminated. Stacking of transistors for circuit implementation requires only material deposition steps on pre-patterned vertical nanowires.

In V-GAA Junctionless transistors, channel conduction is modulated by the workfunction difference between the heavily doped channel and the gate. Due to this workfunction difference, the n-type devices used in Skybridge are normally OFF, and the channel carriers are depleted (note, p-type Skybridge fabrics would follow similar mindset as our n-type version). With the application of gate voltage, carriers start to accumulate and the channel conducts. Source/drain contact formation is done by metal-Si Ohmic contacts; there is no need for raised S/D structures [19]. We have carried out extensive process and device simulations to characterize the V-GAA Junctionless devices based on specific material and sizing in Skybridge. We have also experimentally demonstrated the Junctionless device concept; a p-type Tri-gated Junctionless device in 2-D was fabricated and characterized recently [7] in our group.

The 3-D Synopsys Sentaurus Process simulator [9] was used to create the device structure emulating actual process flow. In the process simulation, the substrate was initially doped to have 1e19 dopants/$cm^3$ doping concentration; the doping step was followed by vertical nanowire patterning using anisotropic etching, followed by sequential anisotropic material deposition steps to complete the V-GAA Junctionless transistor formation. The resulting device structure had 16nm long Si channel, 2nm of $HfO_2$ as gate oxide, 10nm thick TiN as gate electrode, 10nm thick and 5nm long $Si_3N_4$ as spacer material, and 10nm thick, 10nm long Ti as contact material (Fig. S1.1B). 3-D Sentaurus Device simulations [10] were performed on this device to characterize its behavior, while taking nanoscale effects into account. Silicon bandstructure was calculated using the Oldslotboom model [10], charge transport was modeled using hydrodynamic charge transport [10]; quantum confinement effects were taken into account by using density gradient quantum correction model [10]. Electron mobility was modeled taking into account effects due to high doping, surface scattering, and high-k scattering. The simulated device characteristics are shown in Fig. S2.1. This device had an On current of 27µA, Off current 0.1nA; subthreshold slope was found to be 78mV/dec, and threshold voltage (Vth) was 0.35V. These simulated device characteristics were used to generate a behavioral device model for HSPICE circuit simulations.

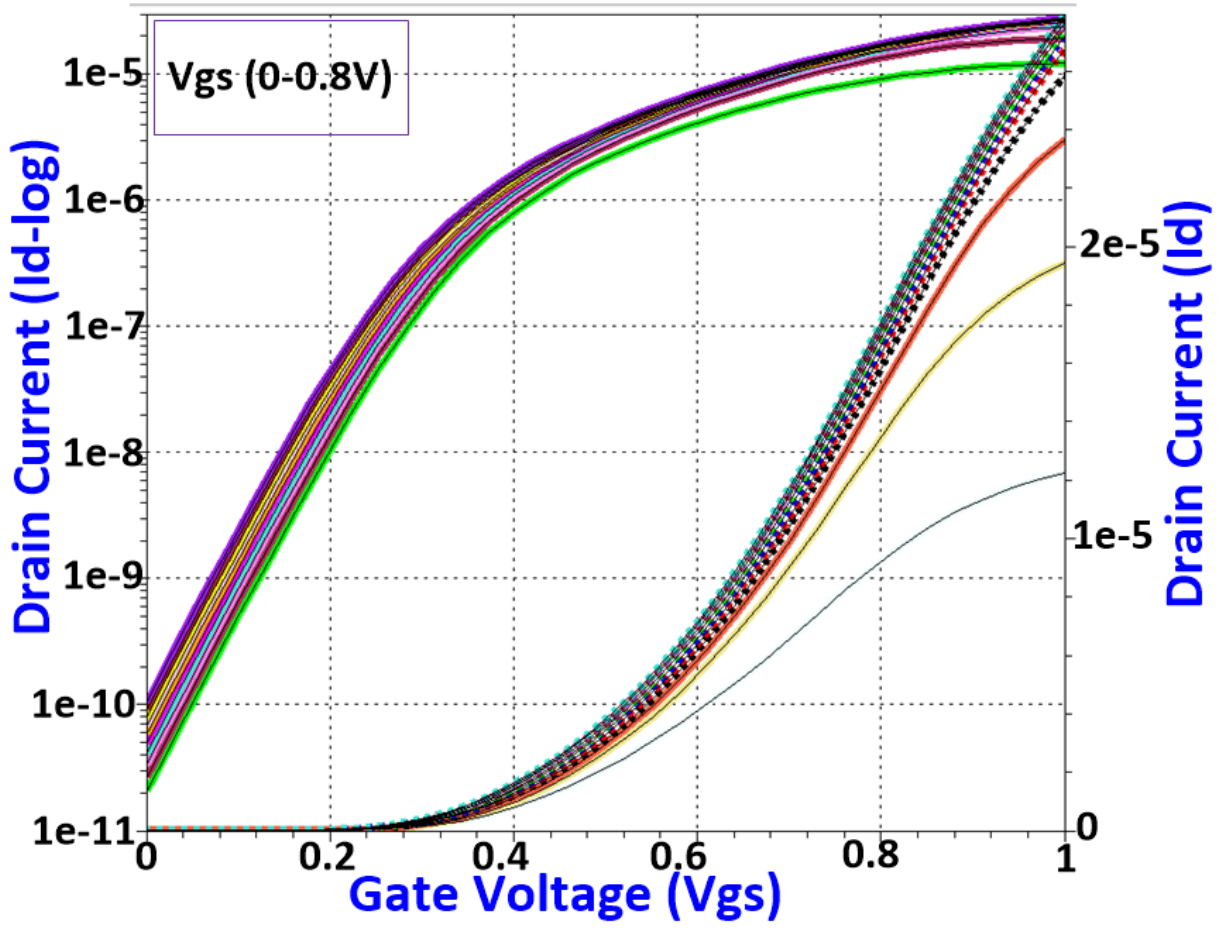

**Fig. S2.1 | 3-D TCAD simulation results.** Id-Vgs characteristics in log (left) and linear (right) scale for V-GAA Junctionless transistor. 16nm channel length, width and thickness; doping: As dopant, 1e19 dopants/cm3; 2nm $HfO_2$ gate dielectric; 10nm thick TiN gate electrode. Simulation shows 27µA $I_{on}$, 0.1nA $I_{off}$, SS 78mV/dec.

### *2.2 Skybridge's Circuit Style*

As outlined before, Skybridge circuits follow a dynamic circuit style that is compatible with 3-D integration requirements. The circuit style allows various design choices including cascaded NAND-NAND or single stage AND-of-NAND compound implementations for logic gates with dual rail or single rail inputs; these can be also combined in a hybrid logic style with high fan-in support. These design choices are generic and can realize any arbitrary logic; moreover, they provide flexibility to optimize Skybridge circuit designs for power or performance, or a balance of both at a very high density. In the following discussions we analyze each circuit style supported, and discuss their trade-offs. Other circuit implementations may be possible.

Fig. S2.2 illustrates the cascaded NAND-NAND and compound dynamic logic gate implementations. An example of cascaded dynamic logic is shown through XOR gate design in Fig. S2.2A, corresponding HSPICE simulated behavior and physical layout are shown in Figs. S2.2B and S2.2C. In cascaded dynamic logic style, complex logic is implemented in two stages using NAND-NAND logic. The output of one NAND stage is propagated to another NAND stage to complete logic behavior; both stages are micro pipelined for seamless signal propagation. The dynamic NAND gates in Fig. S2.2A operate with only n-type uniform V-GAA Junctionless transistors; dynamic circuit behavior is controlled by precharge (*PRE1*, *PRE2*), evaluate (*EVA1*, *EVA2*) and hold (*HOLD1*, *HOLD2*) clock phases. During precharge, the output node is pulled to VDD, and during evaluate period it is either pulled to GND or remains at VDD depending on the input pattern. During the hold phase, the output of current stage is propagated to next stage. In order to have full voltage swing in the output node, the pull up transistor's gate voltage is regulated to have higher voltage than VDD. Cascaded dynamic logic has the potential to achieve high performance, since the load capacitance at output is small for each NAND stage. More details on other types of cascaded dynamic circuits and their analysis can be found in our previous work [20][21][24].

Compound dynamic logic is another variation of dynamic logic style that is unique for the Skybridge fabric. The compound circuit style is designed such that maximum density benefits can be achieved in 3-D implementations. This also alleviates fine-grained clocking requirements. In a single stage, complex logic gates such as XOR, AND-of-NAND gates, etc. can be realized. An example of compound dynamic logic is shown in Figs. S2.2D-F. As shown in Fig. S2.2D, circuit operation is controlled by precharge (*PRE*), evaluate (*EVA*) control signals, and there is no need for cascading of stages; outputs of NAND gates are shorted to achieve AND-of-NANDs logic behavior. Fig. S2.2E shows HSPICE simulated waveforms that validate the compound logic behavior. Like cascaded NAND-NAND designs, this compound logic style is also generic for any logic function.

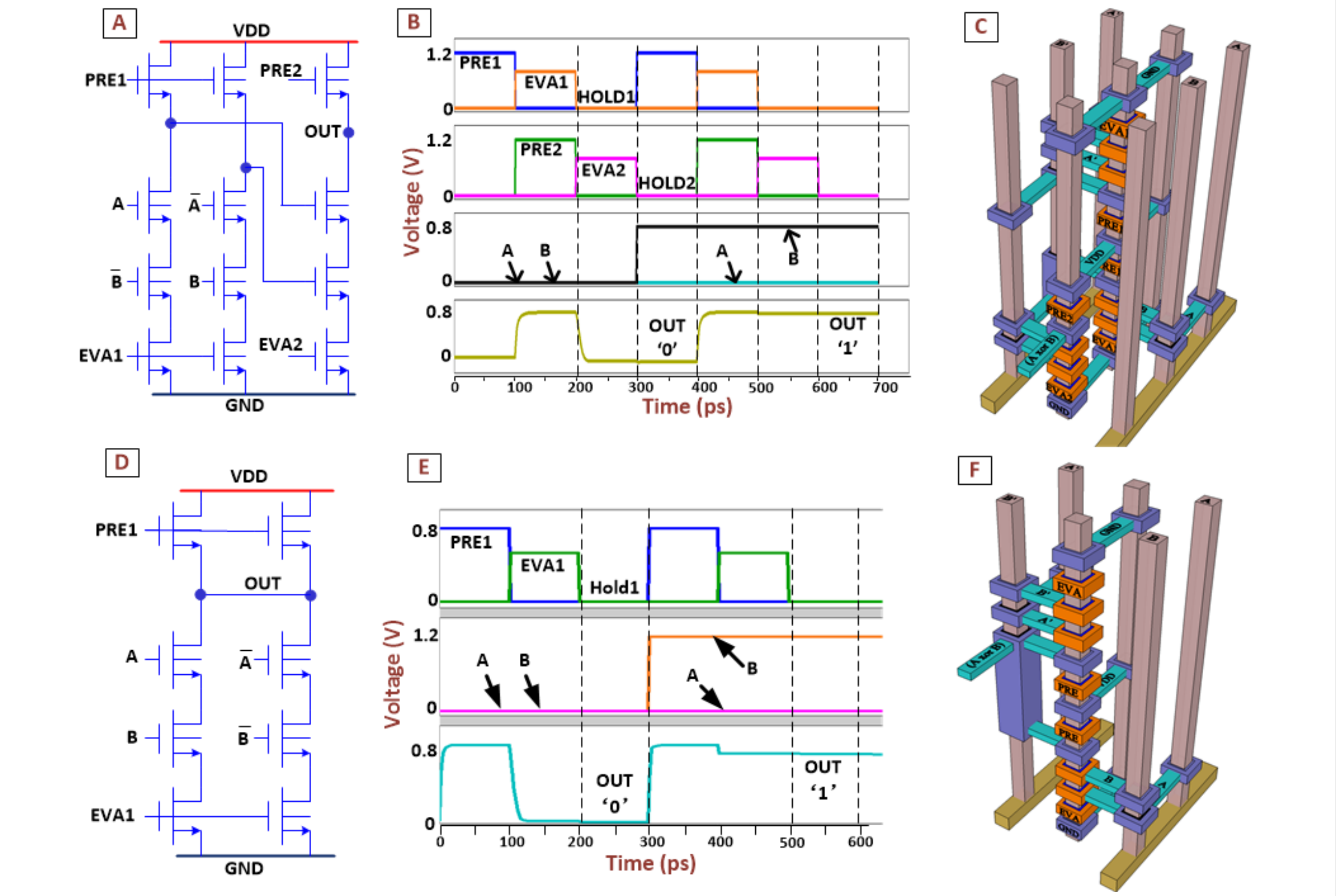

**Fig S2.2 | Cascaded NAND-NAND and Compound dynamic circuit styles for XOR gate.** A) Cascaded circuit style with two logic stages, each stage is controlled by separate PRE and EVA clock signals; B) HSPICE simulated waveforms for the XOR in (A); C) physical layout of cascaded XOR, occupying 3 logic nanowires, and 6 signal nanowires; D)compound dynamic circuit style; logic computation in one stage; two NAND gate outputs are combined in AND of NAND logic; E) HSPICE validations; F) physical layout of XOR gate in (D), only one logic nanowire is occupied for circuit implementation; 4 peripheral nanowires are used signal routing, which are shared with other circuits.

As evident from the physical layouts in Fig. S2.2C and Fig. S2.2D, Skybridge's 3-D implementation achieves tremendous density benefits. Cascaded NAND-NAND logic based XOR implementations require three logic nanowires (Fig. S2.2C), whereas a compound XOR implementation uses only one logic nanowire (Fig. S2.2F); the signal nanowires are shared with other logic gates. The compound dynamic style achieves maximum density by eliminating signal and clock routing overheads of cascaded logic, but lacks slightly in performance compared to cascaded logic since the load capacitance is higher due to output sharing. Our Skybridge designs for arithmetic circuits (Section 6) and microprocessor (Section 7) follow typically a hybrid logic style, where both the benefits of cascaded NAND-NAND and AND-of-NAND compound logic are combined for maximum density and performance.

These above circuit styles support both dual-rail and single-rail implementations, and thus allow flexible design choices for logic. In dual-rail logic, all true and complimentary signals are used as inputs, and the circuit is configured to generate both true and complimentary outputs at the same stage (Figs. S2.3A, S2.3B). On the contrary, single-rail logic uses only a combination of inputs required to generate true/complimentary output, a separate inverter stage is used to generate the opposite signal. Fig. S2.3C illustrates single-rail implementation, and Fig. S2.3D shows HSPICE simulation results. The clocking schemes are different for single-rail and dual-rail circuit styles. Single-rail logic uses two overlapping clock sequence *PRE1, EVA1, HOLD1* and *PRE2, EVA2, HOLD2* (Fig. S2.3D). In dual-rail logic, only one sequence of clock phases is used: *PRE*, *EVA*, *HOLD* (Fig. S2.3B), since all operations are performed in one stage. Single-rail logic is suitable to be used in Cascaded NAND-NAND circuit style, whereas dual-rail logic is more suitable for Compound AND-of-NAND circuit style.

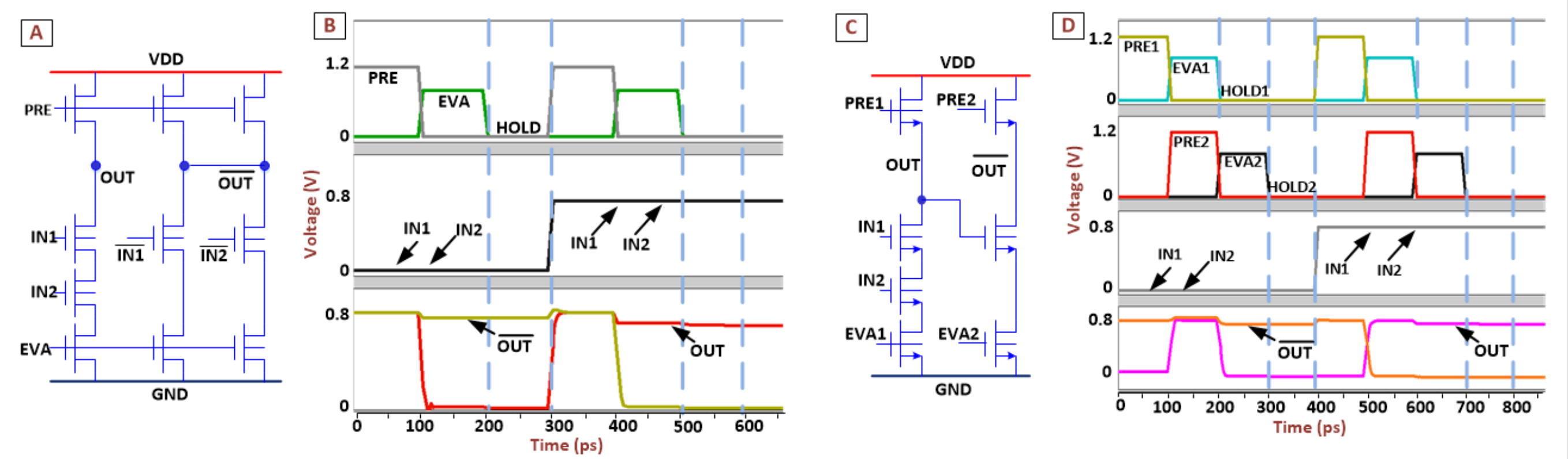

**Fig. S2.3 | Dual rail vs Single rail logic for Skybridge circuits.** A) Example of dual rail logic using 2 input NAND gate; both true and complementary signals are generated at the same stage; B) Simulated waveform of the NAND gate in (A); C) Single rail implementation of the same 2 input NAND gate using two clock stages; complementary output is generated in the second stage NAND gate; D) HSPICE validations of the single rail circuit in (C).

Both dual-rail and single-rail designs have associated trade-offs; in order to optimize circuit performance dual-rail logic is used, whereas single-rail logic results in lower power and higher density. In addition to aforementioned choices, Skybridge's unique dynamic circuit styles and fabric integration provides opportunities for more compact circuit implementations with high fan-in to maximize density. In the following we elaborate on fan-in choices for Skybridge circuits.

### *2.2.1 High Fan-In Support*

High fan-in logic is a well-known driver for compact circuit designs. Since they have fewer transistors and interconnects. Therefore, they are advantageous for both improving density and power consumption. However, high fan-in circuits are not widely used due their detrimental impact on performance compared to low fan-in cascaded designs. The performance degradation is particularly severe in CMOS, where the circuit style requires complementary devices, and the devices have to be differently sized, which adds to load capacitance, and thus lowers the performance. Generally, CMOS circuits are limited to only 4 or 2 fan-in based designs. In contrast, Skybridge's circuit style with only single type uniform transistors and 3-D layout implementation, allows high fan-in logic without corresponding typical performance degradation.

To evaluate the feasibility of high fan-in logic in Skybridge,

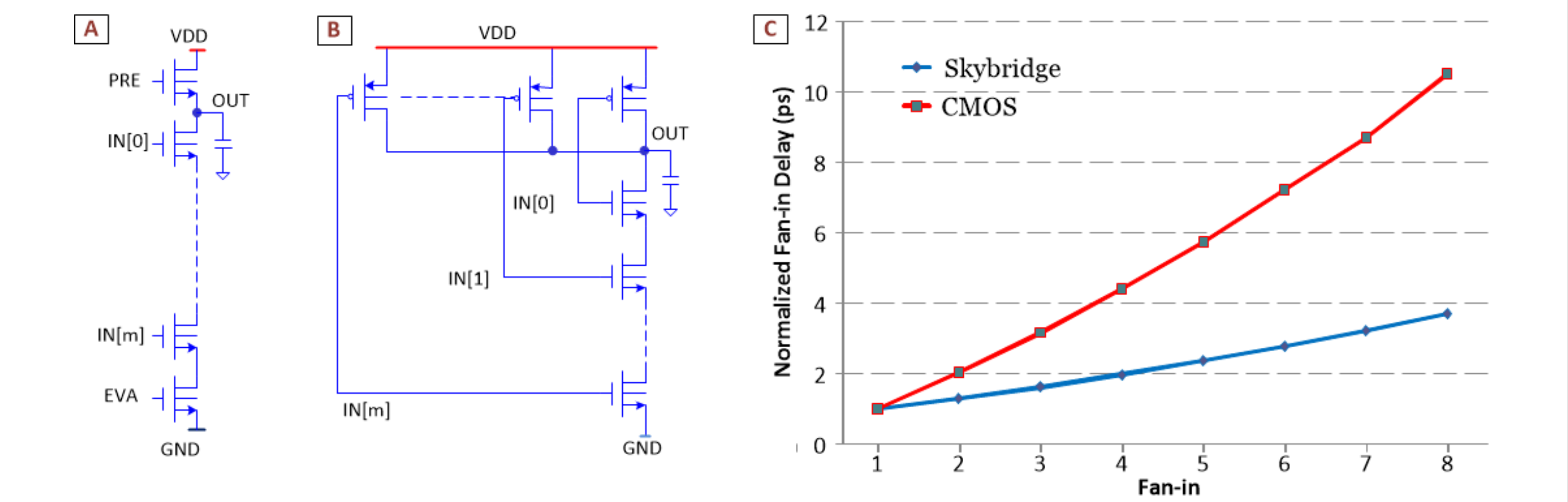

**Fig. S2.4 | Comparative analysis of high fan-in implications.** A) Skybridge NAND gate with 'm' number of fan-ins; B) CMOS NAND gate with 'm' number of fan-ins; C) fan-in sensitivity: CMOS delay increases sharply with increasing fan-in, Skybridge's delay increases almost linearly with high fan-in; the difference is primarily due to the higher load capacitance of CMOS circuit; CMOS uses complementary devices, higher fan-in results in higher parasitic capacitances.

we have carried out fan-in sensitivity analysis using a NAND gate as an example circuit. For Skybridge HSPICE simulations, TCAD generated V-GAA Junctionless device characteristics (Fig. S2.1) were used. Equivalent CMOS designs were simulated for comparison using 16nm tri-gated high-performance PTM device models [26]. The outputs of both Skybridge and CMOS NAND gates were connected to load capacitances that are equivalent to fan-out to 4 inverters in respective designs. The worst-case delay was captured during the falling edge of the output node.

As shown in Fig. S2.4A and S2.4B, Skybridge's NAND gate uses all n-type transistors, whereas the CMOS NAND gate uses both n- and p-type transistors. The total capacitance at the output node of Skybridge's NAND gate is from two adjacent transistors and from 4 inverter fan-out load capacitance. Inverter implementation in Skybridge is equivalent to one fan-in NAND gate with three transistors; one transistor is gated with input signal, and other two are gated with control clock signals. As a result, the load capacitance at the output node in Fig. S2.4A is from 4 n-type transistor gate capacitances and interconnects. On the other hand, the total capacitance at the output node of CMOS NAND gate in Fig. S2.4B is from adjacent transistors, which increases with fan-in, and from 4 inverter fan-out load capacitance. In a CMOS inverter, same input is driven to both n- and p-type devices; in addition, p-type devices are sized to be twice that of n-type. Hence the load capacitance in CMOS is from 4 n-type and 4 double sized p-type transistors, and interconnects.

The impact of higher capacitance at output node is evident from results in Fig. S2.4C. These results are normalized to one fan-in delay for respective designs. As shown in Fig. S2.4C, CMOS delay increases rapidly with higher fan-in, as more transistor parasitic capacitances are added to the total capacitance. On the contrary, Skybridge's delay increases almost linearly and the impact is less prominent, since the load capacitance remains same; the linear increase in delay is mainly due to increased resistance of additional transistors in the discharge path. By optimizing V-GAA Junctionless device characteristics, this delay can be improved further.

In the Section 6, we show high fan-in circuit implementations for large-scale designs. The benchmarking results indicate significant benefits can be obtained for Skybridge designs compared to CMOS.

### 2.2.2 *Noise Mitigation*

While the dynamic circuit style provides opportunities for efficient circuit implementations in 3-D, it is not immune from coupling noise. In dynamic circuits, the output is not driven during the hold phase; hence it is susceptible to coupling noise due to '1' to '0' and '0' to '1' transitions in cascaded logics [24]. In a dense 3-D integration, coupling noise from interconnects can also affect the circuit functionality.

In order to mitigate coupling noise affects, Skybridge has intrinsic architected features that provide noise shielding. The coaxial routing capability (Section 1.4), which is normally

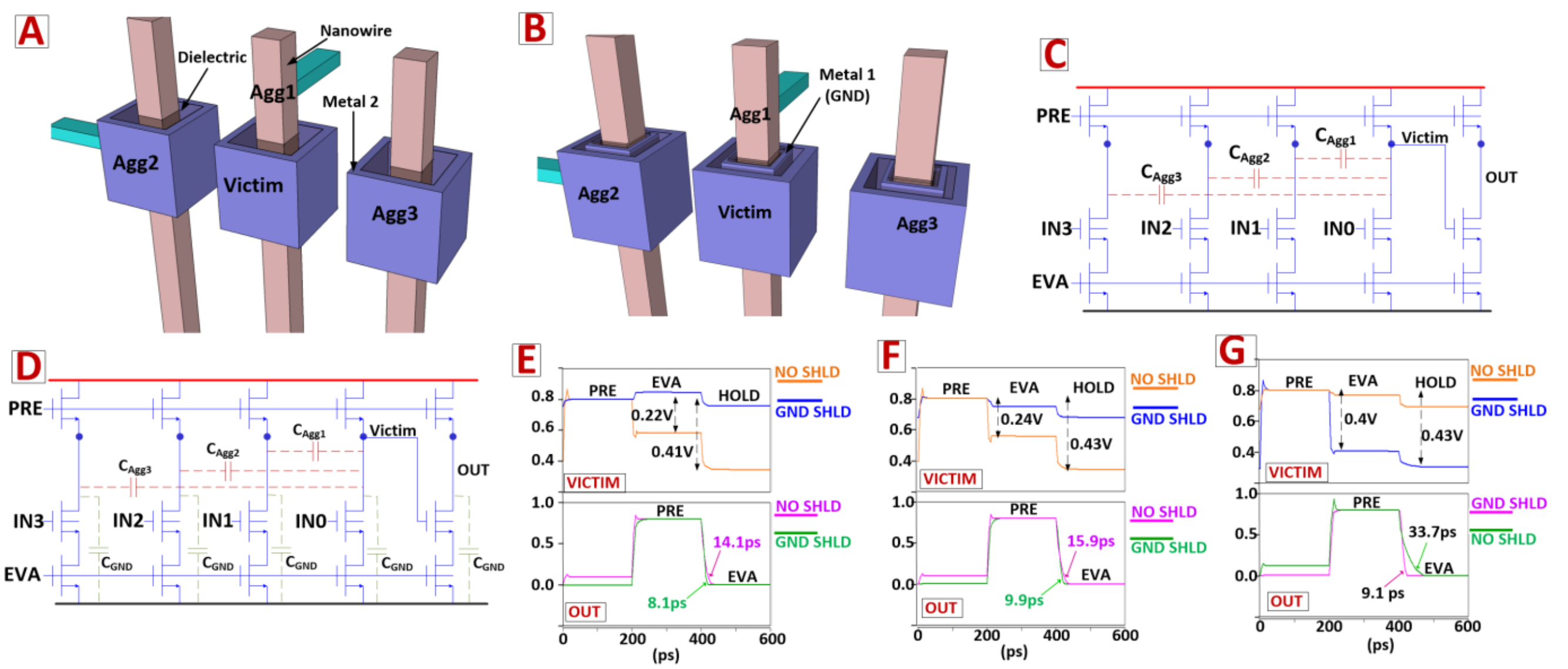

**Fig. S2.5 | Analysis of coupling noise.** A)worst case noise scenario; a victim signal is carried through outer metal shell in the middle nanowire, signals in inner nanowire, and in adjacent metal layers are transitioning from '1' to '0' while the victim signal is floating at '1'; B) layout with GND shielding layer to protect against coupling noise; C) the circuit depicting worst case scenario; D) the circuit schematic when GND shielding layers are incorporated; E) when one aggressor is active (Agg1 switching); F) when two aggressors are active (Agg 1 and 2 switching); G) when three aggressors are active (Agg 1,2 and 3 switching)

used for signal routing, is specially configured to incorporate a noise-shielding layer. A GND signal is routed in between inner nanowire and outer metal2 shell. The key concept of noise shielding using GND signal is to increase the overall capacitance at the floating nodes, thereby reducing the impact of coupling capacitance. This approach ensures coupling noise mitigation during logic cascading, and signal propagation in dense interconnect network. In addition to the noise shielding layer, the Skybridge circuit style uses a clocking control scheme that is known to provide noise resilience [24].

To evaluate the effectiveness of Skybridge's noise shielding approach, we have performed detailed simulations accounting for worst-case scenarios. The scenarios considered, are depicted in Fig. S2.5A. Worst case scenario 1 considers the case when a signal carried through outer metal layer is floating, and is affected by a driven signal that is routed through the inner nanowire; the nanowire signal in this case is aggressor 1. Worst case scenario 2 and 3 considers coupling from adjacent metal2 layers that carry driven signals; they are denoted as aggressor 2 and aggressor 3 (Figs. S2.5A). In all scenarios the victim signal is input to another NAND gate with single input; the switching activity of this NAND gate degrades floating node's stability even further. The corresponding circuit that emulates these worst-case scenarios is shown in Fig. S2.5C. The modified circuit schematic after incorporation of GND shielding layer is shown in Fig S2.5D, and its physical representation is shown in Fig. S2.5B. Simulation results are shown in Figs. S2.5E-G. Skybridge simulations use 3-D TCAD simulated V-GAA Junctionless device characteristics for HSPICE simulations, and takes into account interconnect parasitics from the actual 3-D layout. Capacitance calculations for Coaxial routing structures use the methodology in [25] and assume average routing lengths from a Skybridge microprocessor design (Section 7).

In all scenarios, the victim signal (carried through metal2) is kept floating at '1', and the aggressor signals (carried through inner nanowire, and adjacent metal2 lines) are transitioning from '1' to '0'. For clarity, only the results during transitions are shown in Figs. S2.5 E-G. As shown in Fig. S2.5E, for scenario 1, due to interconnect coupling from aggressor 1, the floating voltage drops from 0.8V to 0.58V; during the evaluation phase of cascaded stage, it drops further to 0.39V. The situation worsens for scenario 2 and 3, and in the worst-case the voltage drops to 0.39V. The performance degradation due to low input voltage is obvious, and in the worst case it reduces by 416% (Fig. S2.5G). The GND shielding approach increases the noise margin significantly with none to small degradation in performance. For scenario 1, the GND shielding recovers the noise margin completely and there is no performance degradation; for scenario 2 and 3 the noise impact is minimal, in the worst case the voltage drops by 0.08V, and the performance degradation from nominal is 12%.

#### 2.2.3 *Mitigation of Performance Impact Due to Long Interconnects*

Long interconnect RC delay is a critical factor that impact overall performance of nanoscale integrated circuits. Typically in CMOS, this issue is addressed by custom sizing of transistors to increase signal drive strength. In Skybridge, the 3-D circuit style and the fabric integration scheme provides several options to minimize this performance impact without any device customization. One such option is insertion of Dynamic buffers; dynamic buffers allow partitioning of long interconnect into small segments, and allow seamless signal propagation in a pipelined design, without impacting the overall throughput. Dynamic buffers are one fan-in NAND gates that are gated by complementary inputs. All Skybridge circuit design is such that both true and complementary values are present in the output. These dynamic buffers were used extensively in our arithmetic circuits and microprocessor designs (Section 6 and 7). Other choices for performance improvement are through fan-in optimization and logic replication. Both these choices can be used to boost drive current, and as a result to reduce long interconnect delay. By reducing fan-in of the driver circuit, the total resistance at the output node can be reduced, which in turn can increase the drive current at the output. Similarly, by replicating the driver logic in neighboring nanowires and by shorting the outputs, the drive current in long interconnect can be increased to reduce delay. In addition to these choices, CMOS like repeaters can be used to reduce the delay for very long interconnects that are used for semi-global and global signals. These repeaters can be placed with other mixed-signal analog power and clock generation circuits in sparse locations of the die. Such repeater requirement is significantly less for Skybridge large-scale designs; our analytical estimation shows the repeater count to be up-to 100x less than CMOS (Section 4). All these choices for performance optimizations provide flexibility to optimize Skybridge circuits in an application specific manner.

### 2.3 *Skybridge's Volatile Memory*

In addition to logic, ability to incorporate high performance volatile memory is a key requirement in integrated circuits. In Skybridge, the volatile memory implementation conforms to the 3-D integration requirements, and follows the aforementioned dynamic circuit styles. In this memory, two cross-coupled dynamic NAND gates are used to store true and complimentary values, and a separate read logic is employed to perform read similar to our previous design for 2-D fabrics [13]. The 8T-NWRAM schematic, HSPICE validation, and 3-D layout are shown in Figs. S2.6A-S2.6C.

As shown in Figs. S2.6A and S2.6B, the memory operation is synchronized with the input clocking scheme and the control signals. In order to write '1' or '0', the clock signals (*xpre*,

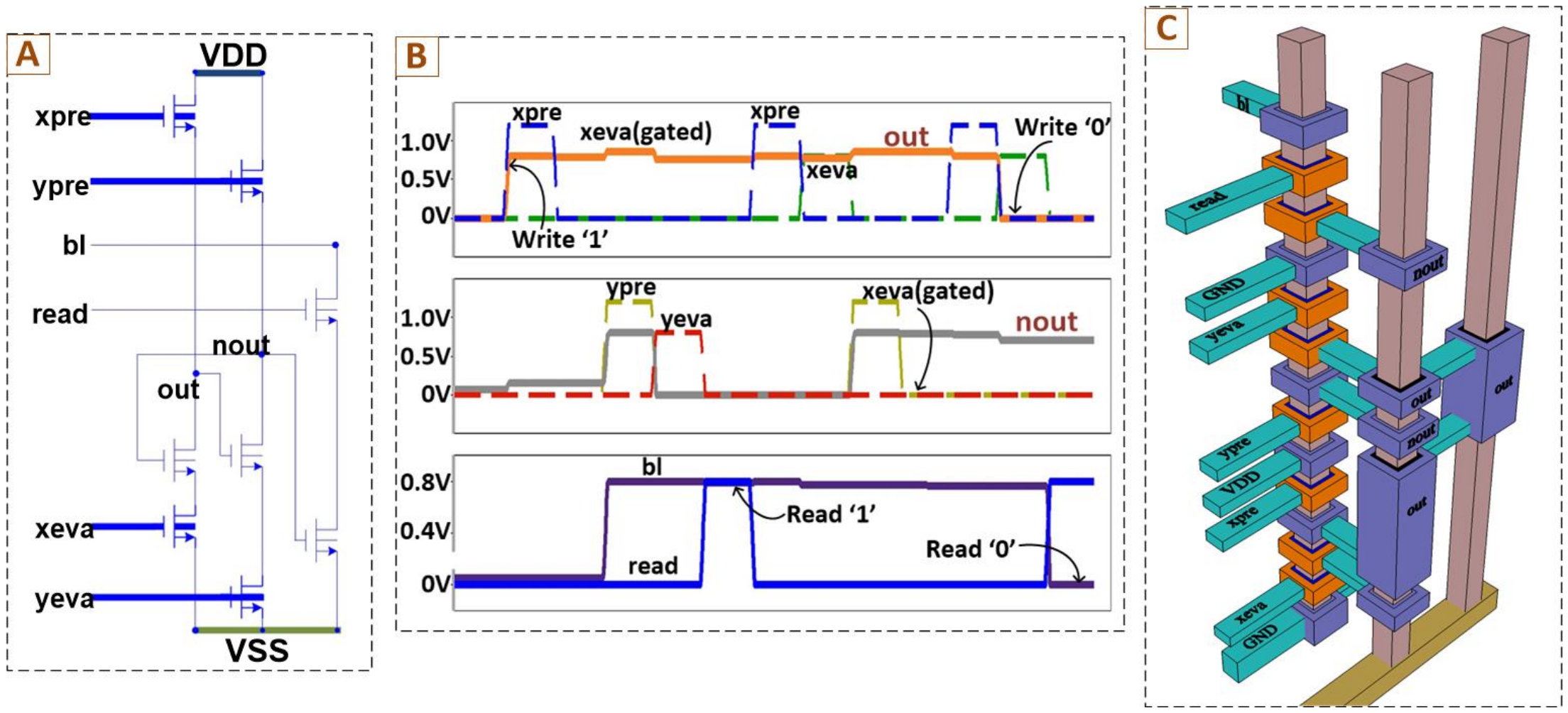

**Fig. S2.6 | Volatile memory design in Skybridge.** A) 8T-NWRAM circuit schematic; volatile memory implementation with two cross-coupled dynamic NAND gates, a separate read logic for read operation; B) HSPICE results showing write and read operations, and C) 8T-NWRAM ‘s physical layout.

*xeva*, *ypre*, *yeva*) are selectively turned ON. For example, to write ‘1’ in node *out*, *xpre* and *xeva* signals are turned ON, and this is followed by *ypre*, *yeva* signals. Once the node *out* is pulled to ‘1’, the complementary node gets pulled to ‘0’ during the *ypre*, *yeva* clock phases. A gated read logic is employed for memory read, and the operation is synchronized with the *read* signal. During the read operation, *bl* is initially precharged, and is subsequently discharged or remains at precharged voltage depending on the *nout* state, when the read signal is ON.

A key feature of this NWRAM is that it is not dependent on precise sizing of complementary transistors for memory operations as it is in the CMOS SRAM; as a result, device sizing-related noise concerns prevalent at nanoscale are mitigated. Furthermore, the read logic is separated from the write logic mitigating bit-flipping concerns during read operations. In addition, during periods of inactivity, all control signals are switched OFF, which reduces leakage power. At certain intervals, the clock signals are switched ON again to restore the stored values but there is no need for read-back and write for this periodic restoration.

The Skybridge layout of this volatile memory is shown in Figure S2.6C; noticeably, all 8 transistors required for memory operation are stacked in only one nanowire, whereas two adjacent nanowires are used for signal propagation, which can be shared by other memory cells. The ultra-dense implementation with reduced interconnections has huge implications on reducing active power and improving performance. Moreover, the Coaxial routing structures used for intra-cell routing provide additional storage capacitance, which is beneficial for prolonging bit storage without restoration, and thus help in reducing leakage power consumption. Benchmarking results are shown in Section 6.2.2.

## 3. FABRIC EVALUATION METHODOLOGIES

Section Summary

This section presents an overview of the methodologies used for interconnect estimation, thermal analysis, 3-D circuit functionality verification and benchmarking.

Overview

A comprehensive methodology, from the material layer to system, was developed to evaluate the potential of Skybridge vs. CMOS. A detailed methodology was followed to analyze connectivity implications of large-scale designs. 3-D interconnect modeling was done for a 10 M logic gate based design with Skybridge specific parameters; equivalent estimation was done for CMOS designs at 16nm technology node for comparison. Thermal analysis of Skybridge circuits was carried out using fine-grained models that account for thermal properties of materials, nanoscale dimensions and 3-D layout. All logic and memory circuit simulations followed an extensive bottom-up simulation methodology that included detailed effects of material choices, confined dimensions, nanoscale device physics, 3-D circuit style, 3-D interconnect parasitics, and 3-D coupling noise. For benchmarking purposes, equivalent CMOS designs were implemented using state-of-the-art CAD tools, and were scaled to 16nm using standard scaling rules.

### *3.1 Overview of Methodology for 3-D Interconnect Modeling, Wire Length Estimation and Repeater Count Distribution*

Predictive models [14][15] for estimation of interconnect distribution in 2-D and 3-D fabrics were employed. Parameters for these models such as Rent’s parameters, average fan-out and gate-pitch were extracted from the microprocessor and arithmetic circuits designed for Skybridge and CMOS. In addition, typical CMOS parameters from literature [14] were also considered for another level of comparison. This resulted in the full interconnect distribution for Skybridge and 2-D CMOS. In order to identify the boundaries between interconnect hierarchical levels, delay criterion was used (see Section 4). The number of repeaters for each hierarchical level was then estimated based on the optimal interconnect segment length for repeater insertion and the number of interconnects for a given length (from the interconnect length distribution). The optimal segment length for a given hierarchical level was determined based on interconnects resistance and capacitance parameters. Fig. S3.1A provides an overview, and Section 4 for details on the predictive models used.

### *3.2 Overview of Methodology for 3-D Thermal Analysis*

To analyze the thermal profile of 3-D circuits, and to quantify the effectiveness of Skybridge's heat extraction features, we have done circuit-level thermal evaluation using detailed modeling and simulation for the worst-case static heat scenario. The thermal modeling was done at transistor level granularity, and was extended for Skybridge circuits. In this model, each heat conducting region (e.g., channel, drain/source, contacts etc.) is represented with equivalent thermal resistance, and the thermal resistance value is determined from the actual thermal conductivity of material used, and material dimensions (see Section 5 for material properties). The effect of nanoscale confined dimensions on thermal conductivity is captured in thermal resistance calculations. For Skybridge circuits the same model was used to calculate thermal resistance of all active circuit components, accurately reflecting material dimensions and 3-D layout. HSPICE thermal simulations were done by analogous representation of thermal resistance and heat source in electrical domain. Worst case static heat scenario was considered for these simulations. Analysis was done on 8 fan-in based Skybridge circuits. Several conditions were simulated including heat conduction with and without Skybridge's heat extraction features at different gate temperatures. Fig. S3.1B illustrates the methodology used for thermal modeling. More details about thermal modeling and analysis can be found in Section 5.

### *3.3 Overview of Methodology for 3-D Circuit Evaluation*

As mentioned earlier, Skybridge circuit evaluation followed a bottom-up simulation methodology. Detailed simulations were done at device, core circuit and system levels. V-GAA Junctionless device behavior was characterized using 3-D TCAD Process and device simulations. Process simulation was done to create the device structure emulating the actual process flow; process parameters (e.g., implantation dosage, anneal temperature, etc.) used in this simulation were taken from our experimental work on Junctionless transistor [7]. Process simulated structure was then used in Device simulations to characterize device behavior. Detailed considerations were taken to account for confined device geometry, nanoscale channel length, surface and secondary scattering effects (see Section 2.1 Process and Device simulation results).

For circuit simulations, the TCAD simulated device characteristics were used to generate an HSPICE compatible behavioral device model (Fig. S2.1C). Regression analysis was performed on the device characteristics, and multivariate polynomial fits were extracted using DataFit software [23]. Mathematical expressions were derived to express the drain current as a function of two independent variables, Gate-Source ($V_{GS}$) and Drain-Source ($V_{DS}$) voltages. These expressions are then incorporated into sub-circuit definitions for voltage-controlled resistors in HSPICE [22]. Capacitance data from TCAD simulations is directly integrated into HSPICE using voltage-controlled capacitance (VCCAP) elements and a piece-wise linear approximation. The regression fits for current together with the piece-wise linear model for capacitances and sub-circuits define the behavioral HSPICE model for the V-GAA Junctionless transistor. This modeling methodology is similar to our prior work on horizontal nanowire device modeling [24].

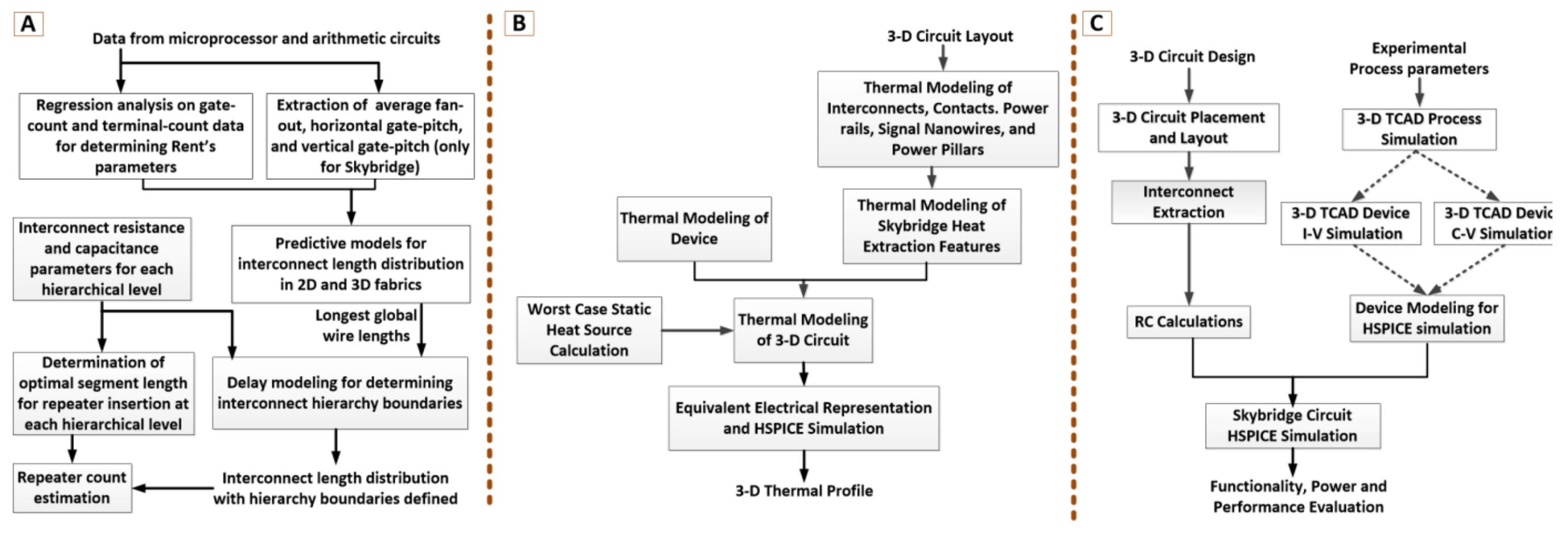

**Fig. S3.1 Skybridge fabric evaluation methodology.** A) Methodology for predicting the interconnect length distribution in Skybridge and 2-D CMOS; B) Thermal evaluation methodology for worst-case static heat scenario. Thermal resistance modeling is done for each circuit component, such as device, interconnect, power rail, etc. and combined to assemble thermal resistance network for the 3-D circuit; electrical equivalent circuit model is then used for HSPICE evaluations; C) Skybridge circuit evaluation methodology; the bottom-up approach uses TCAD Process and Device simulated device characteristics in HSPICE simulations. Interconnect parasitics and noise effects from 3-D layout are also captured in these simulations.

In addition to accurate device characteristics, Skybridge circuit simulations also accounted for 3-D layout specific interconnect parasitics and coupling noise effects (Fig. S3.1C) considering actual dimensions and material choices. Circuit mapping into Skybridge fabric and interconnection were according to manufacturing assumptions and followed fabric's design rules and guidelines (see Section 9). Coupling noise considered was due to cascading of logic stages, and signal propagation through dense 3-D interconnect network. V-GAA Junctionless transistors used for fabric evaluation had 16nm channel length. All manufacturing assumptions and design rules followed ITRS guidelines for 16nm technology node [34]. Capacitance calculations for Coaxial routing structures were according to the methodology in [25], and resistance calculations were according to the PTM interconnect model [33]. The PTM model [33] was also used for metal routing RC and coupling capacitance calculations.

For benchmarking CMOS implementations, of arithmetic circuits and a microprocessor, state-of-the-art CAD simulation tools (Synopsys Design Compiler, Cadence Encounter, and Synopsys HSPICE) were used. Behavioral design, physical layout, placement, interconnect extraction, and HSPICE simulations were performed at 45nm technology node. Extracted results were then scaled to 16nm technology using standard scaling rules [11][12]. A separate methodology was used for SRAM comparisons, since SRAM designs are specific to foundry processes and vary widely in literature. We have considered scaled design rules [13] and used PTM's 16nm high performance transistors for HSPICE SRAM simulations. These design rules may likely be optimistic for CMOS SRAM scaling, since manufacturing complexities escalate at sub 20nms, which leads to larger designs for variation tolerance.

## 4. INTERCONNECT DISTRIBUTION IN SKYBRIDGE AND COMPARISON WITH CMOS

<u>Section Summary</u>

This section details the methodology used to predict the interconnect length distribution for the Skybridge fabric. The implications are discussed and compared with CMOS.

<u>Overview</u>

Conventional integrated circuits (ICs) using CMOS technology implement logic with gates placed in a two-dimensional array on a die. In terms of connectivity between the gates, this scheme limits the degree of connectivity resulting in very long interconnects for large-scale systems. In nanoscale technologies, the interconnect delay is the dominant component as it scales quadratically with interconnect length. To mitigate this, repeaters are used to break long interconnects into shorter segments; as a result the total delay scales linearly with respect to length. While this method ameliorates the performance impact, it introduces significant overhead in terms of area and leakage power dissipation due to repeaters. Increasing wire resistance and delay for nanoscale technologies exacerbates this problem leading to a significant increase in the number of repeaters required to maintain acceptable performance [2].

Skybridge offers fine-grained 3-D integration with vertical stacked gates that are connected with 3-D bridges, providing significant benefits due to a higher degree of connectivity. This results in higher gate density than 2-D technologies, resulting in shorter wires for signal propagation. In this section, we quantify the wiring benefits of Skybridge using predictive models for interconnect length distribution in a large-scale system, and analyze the impact on repeaters. We show that due to much shorter interconnection requirements, Skybridge drastically reduces the number of repeaters, which implies tremendous area and leakage power savings.

### *4.1 Skybridge Interconnect Hierarchy*

In Skybridge, we classify interconnects into three different tiers – *Local* interconnects for close-proximity communication between logic gates; *semi-global* interconnects for intermediate-range communication; and *global* interconnects for long distance communication across chip, clocking and power distribution (see Fig. S4.1). Local interconnects are used for short distance communication in the immediate neighborhood of a gate, implemented intrinsically with *bridges* (also see Section 1). *Semi-global* and *global* interconnects require wider aspect ratios and pitch than local interconnects to support intermediate-long distance communication. These are implemented using metal routing layers on top of the vertical nanowires.

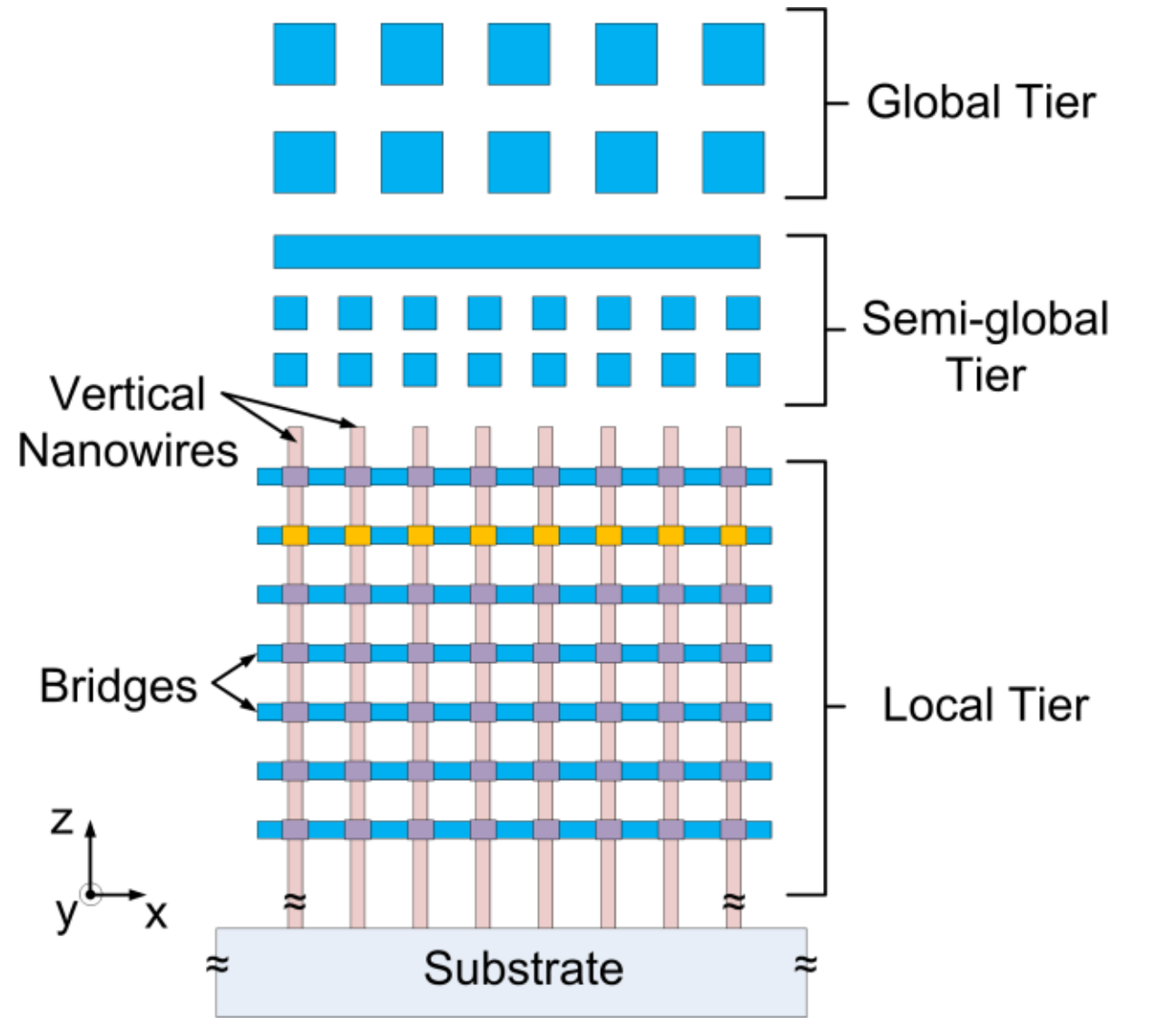

**Fig. S4.1 | Skybridge interconnect hierarchy.** Local interconnects are implemented using bridges for short distance communication. Semi-global and global interconnects are implemented using wide metal layers on top of the vertical nanowire array for intermediate-long distance communication across the IC.

### *4.2 Overview of Predictive Model*

We develop and use a predictive model for 3D integrated circuits tailored to the Skybridge fabric to estimate the interconnect length distribution for a large-scale design. This model uses Rent's rule to predict the number of interconnects of a given length in a large system. Such predictive models have been used in literature to derive system-level interconnect distribution for 2D CMOS [15] and stacked-die approaches [14]. We follow a similar high-level mindset in Skybridge by considering that all gates are distributed uniformly for a given design (see Fig. S4.2B), and the number of gates that can be vertically stacked on nanowires determines the number of gate layers in Skybridge, but adjust to the requirements and overall fabric architecture of Skybridge. By following Rent's rule for terminal count estimation and through the use of intrinsic 3D bridges and metal routing layers for connectivity, this predictive model is adjusted and applied to Skybridge. Rent's parameters, fan-out/fan-in and gate-pitch reflect Skybridge's circuit-style, architecture, connectivity style and high gate density. These parameters are shown in Table S4.1 for both CMOS and Skybridge. The subsequent sections detail the methodology used for extracting each of them and overall model.

The total wire-length distribution in an IC is determined by estimating (i) the number of interconnections $I(l)$ of length $l$ (using Manhattan routing measured in terms of gate-pitches) between a set of logic gate pairs, and (ii) the number of such logic gate pairs $M(l)$ separated by distance $l$. The total number of interconnections of length $l$ is then given by:

$$f(l) = \Gamma . I(l) . M(l), \qquad (1)$$

where Γ is a normalization constant. For 2-D CMOS with $L_{max}$ as the maximum interconnect length, the number of gate-pairs $M_{2D}(l)$ separated by distance $l$ is estimated by the following equation:

$$M_{2D}(l) = \frac{l^3}{3} - l^2 L_{max} + \frac{l^2 L_{max}}{2}, \; 1 \le l < \frac{L_{max}}{2} \qquad (2)$$

$$= \frac{(L_{max}-l)^3}{3}, \; \frac{L_{max}}{2} \le l < L_{max}$$

For 2-D CMOS, the longest interconnect $L_{max}$ is the one that spans from one corner of a square IC to the opposite corner using Manhattan routing. If the total number of gates under consideration is $N_{tot}$, and assuming that they are distributed uniformly throughout the IC, $L_{max}$ is for 2-D CMOS is $2(\sqrt{N_{tot}}$ -1) in units of gate-pitches (gate-pitch is defined as the average separation between adjacent gates). This result can be extended to Skybridge, and is given by

$$M_{SB}(l) = \sum_{i=0}^{G_z-1}(G_z - i).M_{2D}(l - ip_z).u(l - ip_z), \quad (3)$$

where $G_z$ is the number of gates that can be accommodated vertically in Skybridge, $p_z$ is the vertical gate-pitch, and $u$ is a unit-step function. The maximum interconnect length spanning three dimensions using Manhattan routing is $2[\sqrt{(N_{tot}/G_z)} - 1] + (G_z - 1)p_z$. Here, we use $G_z$=2 for Skybridge.

**Supplementary Table S4.1 | Parameters for interconnect prediction models**

| | Rent's Parameters | | Average horizontal gate-pitch | Horizontal gate-pitch normalized to CMOS | Vertical gate-pitch $p_z$ | Average Fan-out |
|---|---|---|---|---|---|---|
| | **k** | **p** | **G.P.$_{avg}$ (nm)** | | **(nm)** | **f.o.** |
| **CMOS (parameters extracted from designed circuits)** | 3.416 | 0.473 | 803.87 | x1 | NA | 1.7 |
| **CMOS (parameters from literature)** | 4 | 0.66 | | | | 3 |
| **Skybridge** | 5.39 | 0.577 | 150 | x0.186 | 448 | 2.018 |

The number of interconnects of length $l$ gate-pitches, $I(l)$ is estimated using Rent's rule as described below. For a partitioned design, Rent's rule relates the number of logic gates $N$ within a sub-module or logic block to the number of external signals or terminals $T$ to that block as follows.

$$T = k.N^p \quad (4)$$

Here, $k$ is the Rent's coefficient defined as the average number of terminals per logic gate. Rent's exponent $p$ is an empirical parameter used to fit the observed data from circuits to the relationship above. Consider the group of gates shown in Fig. S4.2. Here for the gates under consideration in block A, there are several gates in block C that lie at a Manhattan distance of $l$ gate-pitches. By counting the number of terminals from logic block A to logic block C, we get the total interconnections from block A to block C. Using a partial Manhattan circle approximation [15] and taking the average fan-out (*f.o.*) into consideration, the total number of interconnects of length $l$ gate-pitches between blocks A and C is given as follows.

$$I(l) = \frac{\alpha k}{N_C}[(N_A + N_B)^p - (N_B)^p + (N_B + N_C)^p - (N_A + N_B + N_C)^p] \quad (5)$$

Here $k$ and $p$ are Rent's parameters, $\alpha = (f.o.)/(1+f.o.)$, and $N_A$ is set to 1. For 2-D CMOS, $N_{A\text{-}2D} = 1$, $N_{B\text{-}2D} = l(l-1)$ and $N_{C\text{-}2D} = 2l$. This can be extended to Skybridge as

$$N_{A-SB}(l) = 1, \quad (6)$$

$$N_{B-SB}(l) = N_{B-2D}(l) + \frac{1}{G_z}\sum_{i=1}^{G_z-1}[2(G_z - i).N_{B-2D}(l - ip_z).u(l - ip_z)], \quad (7)$$

$$N_{C-SB}(l) = N_{C-2D}(l) + \frac{1}{G_z}\sum_{i=1}^{G_z-1}[2(G_z - i).N_{C-2D}(l - ip_z).u(l - ip_z)]. \quad (8)$$

Substituting equations (6)-(8) in (5) gives the expression to estimate the number of interconnections between gates separated by a distance of $l$ gate-pitches. If $I_{total}$ is the total number of interconnects [27], then the normalization constant $\Gamma$ is then calculated as,

$$\Gamma = [\mathrm{I_{total}}]/(\sum_{l=1}^{\mathrm{L_{max}}} M(l).I(l)) \quad (9)$$

$$= [\alpha k N_{tot}(1 - N_{tot}{}^p)]/(\sum_{l=1}^{\mathrm{L_{max}}} M(l).I(l)).$$

### *4.3 Determination of Parameters for Skybridge*

#### *4.3.1 Rent's Parameters*

We use data from the designed Skybridge circuits to extract Rent's parameters. Using the definition of Rent's coefficient $k$

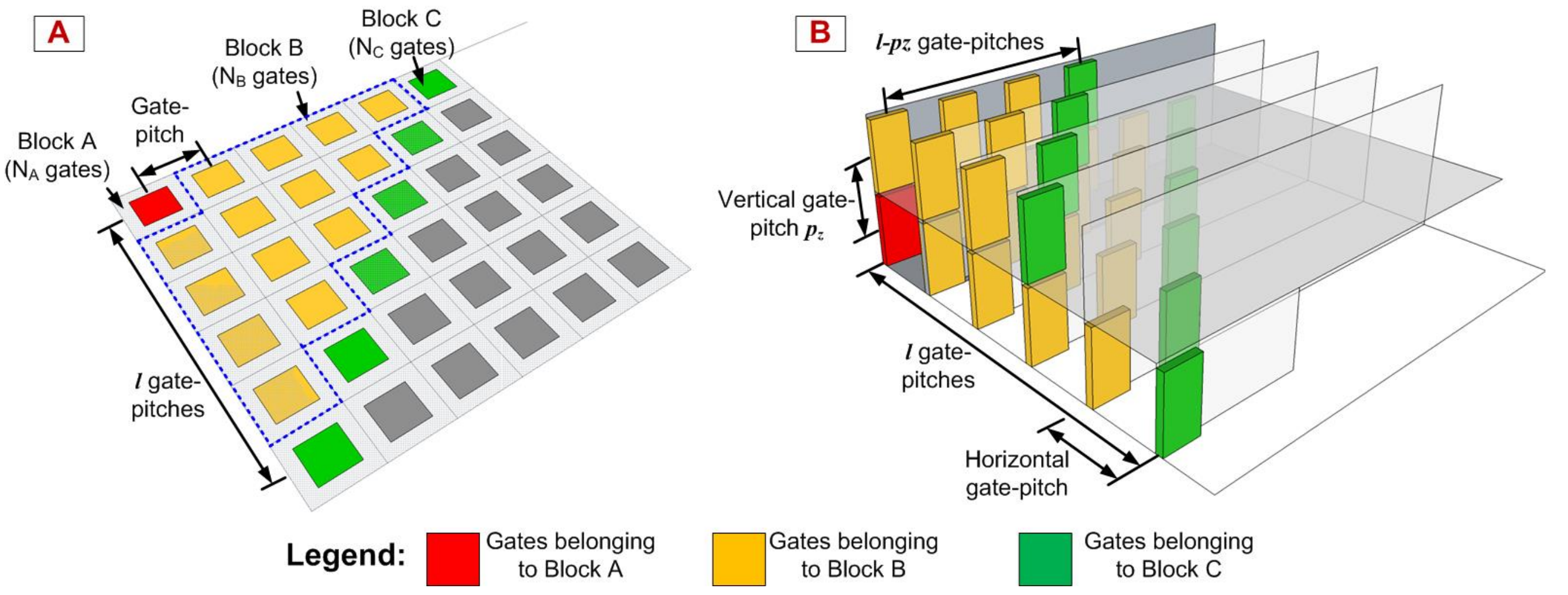

**Fig. S4.2 | Interconnect estimation.** Procedure to estimate number of interconnects between a gate-pair separated by $l$ gate-pitches in (A) 2-D CMOS integrated circuits, and (B) 3-D Skybridge integrated circuits. Here block A is the source gate and the destination gate belongs to block C. Block B contains all the gates lying between this gate pair. Gates are laid out both vertically and horizontally in Skybridge vs. only horizontal in CMOS.

(average number of terminals per gate), we enumerate the gates and their terminal count for all designed circuits, and calculate the average of the terminal counts. To estimate Rent's exponent $p$, we extract data-points by computing gate-counts $N$ and terminal-counts $T$ for sub-modules of circuits at various levels of hierarchy, and use regression-based curve fitting for equation (4). In case of multiple terminal counts for a given gate-count, which is possible since different circuits can have same number of logic gates but differ in the number of I/O terminals depending on the function being realized, we use the geometric mean of these data-points for the regression analysis since it has been statistically observed to track Rent's rule quite accurately [28]. The results of this analysis are shown in Table S4.1.

### 4.3.2 Average Gate-Pitch and Fan-Out

Gate-pitch is defined as the average separation between adjacent logic gates. The interconnect prediction model described earlier for a 3-D fabric like Skybridge takes horizontal and vertical gate-pitch as parameters. We determine the Skybridge average horizontal gate-pitch across all the designed logic circuits, by considering the number of gates and the area occupied by the circuits. For each module, if the footprint area is $A$ and it contains $N$ gates with a vertical stacking of 2 gates, the horizontal gate pitch ($G.P.$) assuming uniform distribution of gates is calculated as follows.

$$G.P. = \sqrt{2A/N} \tag{10}$$

For $m$ modules under consideration, the net average horizontal gate pitch across all logic circuits is calculated as

$$G.P._{avg} = \frac{1}{m}\sum_{i=1}^{m}\sqrt{\frac{2A_i}{N_i}}. \tag{11}$$

The vertical gate-pitch $p_z$ is determined by dividing the total height of nanowires by number of gates that can be stacked vertically (in this case 2). The average fan-out is determined by calculating the average fan-out of each module in the designed logic circuits, followed by taking the arithmetic mean across all modules. The parameters extracted using this method for Skybridge are shown in Table S4.1.

For comparison with 2-D CMOS, we use two sets of parameters for the predictive models for a comprehensive evaluation. The first set of parameters is extracted from designed circuits that were used to compare with Skybridge. In addition, typical values for Rent's parameters and average fan-out are taken from literature [15] for microprocessors, and both sets of parameters are used to derive CMOS interconnect distributions (see Table S4.1).

## 4.4 Interconnect Hierarchy, Delay Models and Repeater Count Estimation

### 4.4.1 CMOS Interconnect Hierarchy

CMOS integrated circuit interconnects are typically classified into three tiers – *global* interconnects responsible for long distance communication across the chip, clocking and power distribution; *semi-global* interconnects for intermediate range communication; and *local* interconnects for short range communication between gates. Each tier is characterized by wiring parameters such as wiring pitch (which determines load capacitance observed by drivers), and wire aspect ratio and material choice (determining the output resistance). These parameters affect the signal delay when using a particular tier to communicate between gates. *Local* interconnects are closest to the device layer and typically have a tighter pitch to allow dense packing of transistors. *Semi-global* and *global* wires are typically much wider with higher aspect ratios and wider pitch to allow for long distance communication across the chip and reduce delays due to propagation.

For long interconnects, the propagation delay starts to dominate the overall delay of a logic gate driving the interconnect. To mitigate this, these long interconnects are broken down into shorter segments driven by repeaters (static inverters, see Fig. S4.3A) [29]. This scheme results in delays that scale linearly with the length of the interconnect as opposed to being quadratic.

#### 4.4.1.1 Interconnect Delay Model

The equivalent RC circuit used to model the delay of each interconnect segment of length $l$ is shown in Fig. S4.3B. If $R_{tr}$

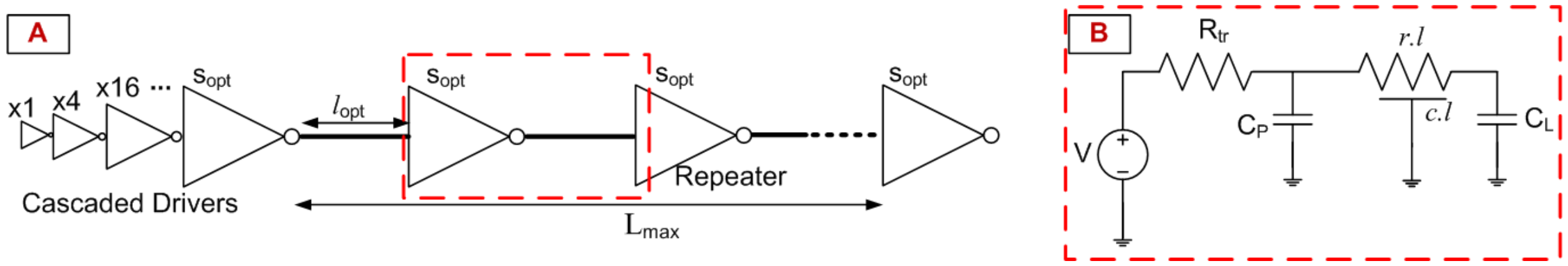

**Fig. S4.3 | Repeater insertion.** (A) Segmentation of long interconnects and repeater insertion in 2-D CMOS. Cascaded drivers are used to drive the large repeaters. Here, multiples indicate the size of the driver in units of minimum inverter size. (B) Equivalent RC circuit used to model the delay of each wire segment driven by a repeater.

**Supplementary Table S4.2 | Parameters for delay modeling**

| | a(x) | b(x) |
|---|---|---|
| **Propagation Delay (50% swing)** | 0.4 | 0.7 |
| **Fall/Rise Time (10% - 90% swing)** | 0.9 | 2.2 |

is the resistance of the driver transistor having a parasitic output capacitance $C_P$, $r$ and $c$ are the resistance and capacitance per unit length for the interconnect respectively, and $C_L$ is the load capacitance of the next stage then the delay of this segment is given by

$$\tau = b(x).R_{tr.}(C_L + C_P) + b(x).(cR_{tr} + rC_L).l + a(x).rcl^2 \ . \quad (12)$$

Here $R_{tr}$, $C_P$ and $C_L$ can be expressed in multiples of resistance $r_0$, parasitic output capacitance $c_p$ and input capacitance $c_0$ respectively of a minimum sized inverter. If the size of the driver is $s$ times the minimum inverter, then $R_{tr} = r_0/s$, $C_L = sc_0$ and $C_P = sc_p$. *a(x)* and *b(x)* are constants determined by the voltage swing being considered (see Table S4.2). CMOS static logic delay is typically characterized by the propagation delay, which considers a 50% output voltage swing (i.e. $x$=0.5). For a given set of wiring parameters, optimal interconnect length $l_{opt}$ and optimal repeater size $s_{opt}$ (in multiples of minimum-sized inverters) can be determined to minimize the overall delay by the following expressions [30].

$$l_{opt} = \sqrt{\frac{b(x).r_0.(c_0 + c_p)}{a(x).r.c}} \quad (13)$$

$$s_{opt} = \sqrt{\frac{r_0 c}{r c_0}} \quad (14)$$

The total delay of a full interconnect as a function of its length $l$ consisting of $n$ such segments is then simply $\tau_d(l) = n\tau$. Since the repeaters used are much larger than minimum-sized inverters, cascaded drivers are typically used where each succeeding driver size is progressively increased till the required size is reached. This is then used to drive the rest of the interconnect segment (see Fig. S4.3A). The transistor parameters ($r_0$, $c_0$, and $c_p$) were extracted for 16nm PTM FinFET models [26][32]. The interconnect resistance parameters (see Table S4.3) were taken from ITRS [34] and capacitance parameters were derived using PTM Interconnect RC models [33] which takes into account both ground and coupling capacitances.

**Supplementary Table S4.3 | Wiring parameters for 16nm technology [34]**

| 16nm Node | Effective Resistivity (µOhm-cm) | Wire Aspect Ratio | Wire Pitch (nm) | Maximum delay as a fraction of clock period (β) |
|---|---|---|---|---|
| Global Wires | 5.26 | 2.34 | 152 | 0.9 |
| Semi-Global Wires | 6.96 | 2 | 76 | 0.25 |
| Local Wires | 6.96 | 2 | 38 | 0.25 |

*4.4.1.2 Interconnect Classification*

The interconnect distribution can be classified into different tiers by estimating the longest interconnect for a given tier. This is determined based on the maximum allowed delay expressed as a fraction $\beta$ of clock period [31]. Since *global* signals are expected to have large delay while propagating over large distances, they are typically allowed to use 90% of the clock period ($\beta$=0.9) for signal propagation alone. *Local* and *semi-global* wires typically are allowed to use 25% of the clock period ($\beta$=0.25) for propagation delay, while accommodating delay due to intermediate logic stages during the remaining time. The longest global wire $L_{max\text{-}global}$ can be determined from the interconnect distribution as the length $l$ for which $f(l)$ = 1. Using this as baseline, the longest interconnects in *local* and *semi-global* tiers can be estimated using the delay criterion as follows.

$$\frac{\tau(L_{max-local})}{\tau(L_{max-global})} = \frac{\beta_{local}.T_{clock}}{\beta_{global}.T_{clock}} = \frac{\beta_{local}}{\beta_{global}}$$

$$\tau(L_{max-local}) = \frac{\beta_{local}}{\beta_{global}}\tau(L_{max-global}) \quad (15)$$

$$\tau(L_{max-semi-global}) = \frac{\beta_{semi-global}}{\beta_{global}}\tau(L_{max-global}) \quad (16)$$

*4.4.1.3 Repeater Count Estimation*

In any given tier, the interconnects whose lengths are between $l_{opt}$ and $L_{max}$ will have optimally sized repeaters inserted for minimizing propagation delay. The number of segments can be computed for a given interconnect length $l$ for that tier, which in turn yields the number of repeaters required *R(l)*. Using the interconnect distribution *f(l)* which estimates the total number of interconnects of a given length $l$, the total number of repeaters in a given tier $i$ can be estimated as follows.

$$R_i = \sum_{l=lopt-i}^{Lmax} f(l).R(l) \tag{17}$$

### 4.4.2 Skybridge Interconnect Hierarchy

In Skybridge, we similarly define different tiers for interconnects (see Fig. S4.1). *Semi-global* and *global* interconnects require wider aspect ratios and pitch than local interconnects to support intermediate-long distance communication. These are implemented using metal routing layers on top of the vertical nanowires. To minimize signal delays over long interconnects, *semi-global* and *global* wires are segmented and static CMOS inverters drive each segment. These repeaters can be implemented through coarse-grained heterogeneous integration using islands of CMOS inverters. *Local* interconnects, implemented with *bridges* (described in Section 1), are not segmented since fine-grained integration with CMOS would incur overhead in area. This scheme limits the maximum length of *local* interconnects in Skybridge, and results in more interconnects being assigned to the *semi-global* tier. The wire parameters for *semi-global* and *global* tiers are assumed to be the same as that for 16nm CMOS (see Table S4.3).

#### 4.4.2.1 Interconnect Delay Model

Interconnection scheme for different tiers in Skybridge is shown in Fig. S4.4. Delay modeling is similar to the method described earlier, where delay for each segment is calculated using equation (12). For Skybridge dynamic circuits, the switching model considers the fall-time at the output, i.e. 10% to 90% voltage swing. This fall-time delay should be accommodated within the *evaluation* clock period (see Section 2 for dynamic clocking scheme in Skybridge) for correct functionality of cascaded logic gates. The corresponding parameters *a(x)* and *b(x)* for Skybridge are shown in Table S4.2.

#### 4.4.2.2 Interconnect Classification

The delay criterion used to classify interconnects in Skybridge is different from CMOS. Since Skybridge uses dynamic circuits, the entire *evaluation* clock period is devoted to the total delay of a gate driving an interconnect load. Thus for all tiers, β = 1 which implies that the delay of the longest interconnect in any tier is the same.

$$\tau(L_{max-local}) = \tau(L_{max-semi-global}) = \tau(L_{max-global}) \tag{18}$$

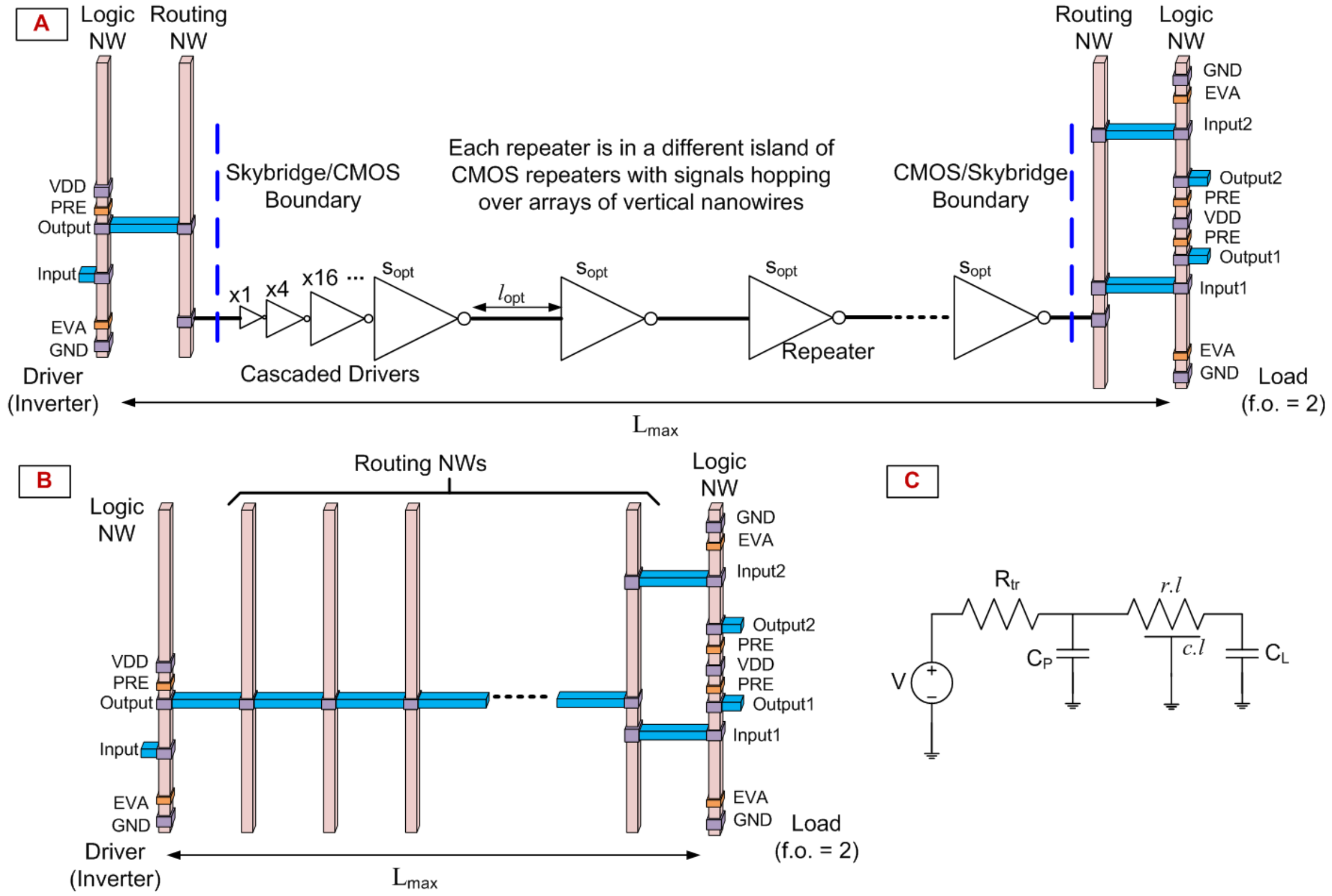

**Fig. S4.4 | Interconnection scheme in Skybridge.** (A) Global/semi-global interconnects using static CMOS repeaters; and (B) Local interconnects. Here, there are no repeaters used for local interconnects. (C) Equivalent R-C model to estimate interconnect delay.

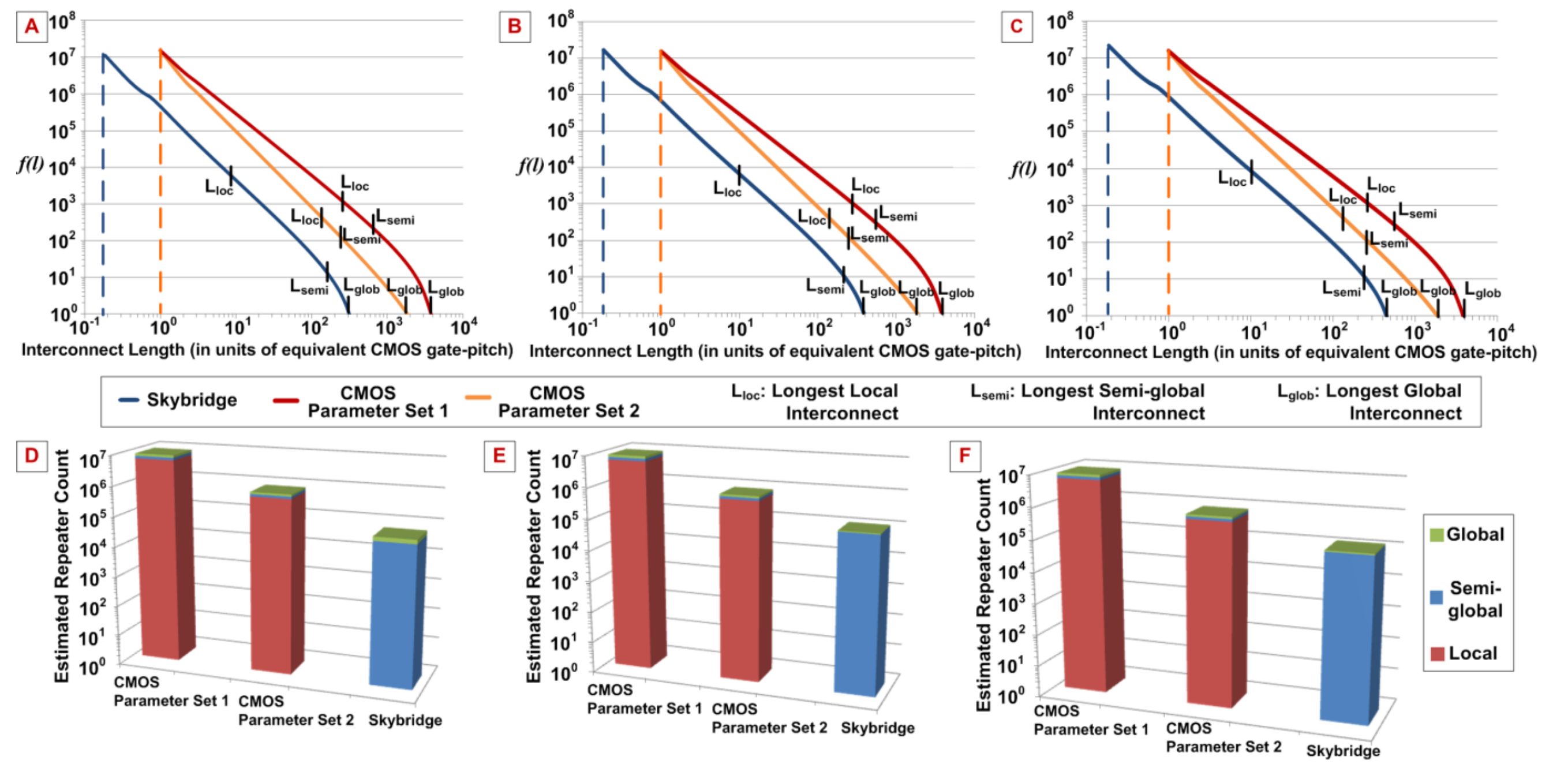

**Fig. S4.5 | Interconnect distribution (A) – (C) and estimated repeater counts (D) – (F) in Skybridge vs. 2D CMOS.** Here, number of gates $N_{CMOS} = 10^7$. Number of gates in Skybridge are (A) $N_{SB} = 0.5 \times 10^7$ and corresponding repeater counts in (D); (B) $N_{SB} = 0.75 \times 10^7$ and corresponding repeater counts in (E); and (C) $N_{SB} = 10^7$ and corresponding repeater counts in (F). Parameters for Skybridge: $k$=5.39, $p$=0.577 (Rent's parameters), average fan-out = 2.018. For CMOS, Parameter Set 1: $k$=4, $p$=0.66, average fan-out = 3; and Parameter Set 2: $k$=3.416, $p$=0.473, average fan-out = 1.7.

Similar to the methodology described before, the longest global interconnect is determined from the distribution as the length $l$ for which $f(l) = 1$. Then using the delay criterion for Skybridge, the maximum local and semi-global interconnect lengths are determined.

#### 4.4.2.3 *Repeater Count Estimation*

The methodology for estimating the total number of repeaters for a given integrated circuit is the same as described in Section 4.3.1.3. In Skybridge, since *local* interconnects are not segmented, there are no repeaters for this tier. Only *semi-global* and *global* tiers use repeaters for long wires whose length exceeds the optimal segment length.

### 4.5 *Evaluation and Benchmarking with CMOS*

Using the models described in this section, we compare the system-level implications of Skybridge with 2-D CMOS. In particular, we look at the effect of reduced interconnect lengths in Skybridge on repeater count. Since Skybridge supports high fan-in circuits (see Section 2.2.1 on fan-in), the atomic logic gates are much more expressive than CMOS and are expected to result in fewer gates for a given high bit-width function. Thus, we analyze three different scenarios for comparison with CMOS. If $N_{CMOS}$ is the total number of gates considered for a 2-D CMOS circuit, the number of gates for Skybridge implementation ($N_{SB}$) is varied between $0.5N_{CMOS}$, $0.75N_{CMOS}$ and $N_{CMOS}$. The results are shown in Fig. S4.5.

Here, we can see that the longest interconnect in Skybridge is significantly shorter when compared to 2-D CMOS, specifically up to 10x shorter. We also see that there is up to 2 orders of magnitude reduction in the total number of repeaters required in Skybridge. This implies tremendous reduction in terms of area overhead of repeaters as well as leakage power savings for larger designs such as multi-core processors.

## 5. THERMAL MANAGEMENT IN SKYBRIDGE FABRIC

<u>Section Summary</u>

This section presents thermal management details in Skybridge fabric. Through transistor level modeling we analyze thermal profiles in Skybridge circuits, and show the effectiveness of Skybridge's intrinsic heat extraction features.

<u>Overview</u>

Thermal management is a crucial issue at nanoscale. As transistors are reaching ultra-scaled dimensions, heat dissipation paths are reducing, thus giving rise to self-heating in transistors. The situation worsens for 3-D designs, where multiple transistors are stacked vertically, and thermal resistance from heat source to sink increases. In Skybridge nanoscale thermal issues are addressed through architected heat extracting features being built-in as core fabric components. This integrated mindset is a significant departure from traditional CMOS approaches, where heat extraction from active circuit is addressed only as after-thought (i.e., during operation, and at system level).

The intrinsic heat extraction features of Skybridge fabric are: (i) selective placement of power rails (i.e., VDD and GND) to control heat flow direction, (ii) *Heat Extraction Junctions (HEJs)* to extract heat from a heated region in a circuit, (iii) sparsely placed large area *Heat Dissipating Power Pillars (HDPPs)* for heat dissipation to sink.

(i) In Skybridge, logic and memory functionality is achieved in vertical nanowires, where transistors are stacked and metal contacts are established at selective places in nanowires for output and power rails (i.e., VDD and GND). The placement of power rail contacts has huge thermal implications, since it determines the current and heat flow direction in a vertically implemented fabric. For example, in a vertically implemented dynamic NAND gate if the VDD is placed on the top and GND is placed at the bottom, electrons will flow from GND towards VDD and generate heat along its path. In turn the generated heat will flow from top (i.e., hot region) to bottom (i.e., cool region) towards reference temperature. In this fabric, the power rails are positioned vertically such that heat flow towards substrate is maximized. Since, each logic nanowire pillar accommodates two dynamic NAND gates, and one power rail can be shared between two gates, the VDD contact is positioned in the middle and GND contacts are made at the top and at the bottom. This configuration allows heat transfer from VDD to bottom GND and towards heat sink in the bulk and allows the bottom of the nanowires to be at the same temperature as the substrate.

(ii) HEJs are specialized junctions that are used to extract heat from a logic nanowire without perturbing its operation. HEJs are connected with Bridges to transfer heat to the bulk through HDPPs. The Bridges that carry heat are different from other generic signal carrying Bridges, since these always carry only one type of electrical signal (GND) and serve the purpose of heat extraction only. HEJs in conjunction with Bridges allow flexibility to selectively extract heat from a 3-D circuit layout without any loss of functionality or performance.

(ii) HDPPs are intrinsic to Skybridge fabric, and are used for both power supply (i.e., VDD and GND signals) and heat dissipation. These pillars are large in area (2nw pitch x 2nw pitch) and have specialized configuration with metal silicidation and fillings particularly to facilitate heat transfer. The top GND and middle VDD contacts in each logic nanowire connect to these large area pillars through Bridges. The power pillars are different in-terms of dimension, layout and material configuration from signal pillars, which carry input/output/clock signals from different logic/clock stages.

In the following, we present details on thermal characteristics of Skybridge fabric, and show effectiveness of its architectural features. Fine-grained thermal modeling approach is presented for 3-D circuits, and is followed by detailed evaluation.

### *5.1 Thermal Modeling and Analysis*

In order to characterize the thermal profile during operating conditions heat modeling was done for circuits at transistor-level granularity as outlined in Section 3.2. This fine-grained modeling is especially important due to nanoscale dimensions of active devices; at this scale, confined dimensions and scattering affects drastically reduce thermal conductivity of silicon channel, which leads to rapid self-heating. From a circuit perspective, such fine-grained modeling allows detail understanding about heat generation in circuits, and implications of materials and architectural choices for heat dissipation.

#### *5.1.1 V-GAA Junctionless Transistor*

In this section we show thermal modeling of a single n-type GAA Junctionless transistor. Material and geometry considerations of this device are reassessed from thermal perspective. Fig. S5.1A shows cross-section of n-type GAA Junctionless transistor, where heat generation is mainly due to electron-phonon interaction in the drain region. During ON state, free electrons accelerate from the source region towards the drain. Here they scatter due to interactions with other electrons, phonons, and impurity atoms causing the lattice temperature to increase [17]. Depending on the material considerations and geometry of the transistor, this temperature gradient can either dissipate quickly without any impact or slowly dissipate and cause transistor ON current degradation.

In order to estimate temperature gradient within transistor region, an electrical analogy of thermal model can be used [16]. An approximation of generated heat, Q (Watts) can be:

$$Q = Ids * Vds \qquad (19)$$

In eq. (19), Ids is drain-source current, and Vds is drain-source voltage. The relationship between heat (Q) and temperature-gradient (ΔT) is:

$$\Delta T = \frac{L}{K * A} * Q \qquad (20)$$

In eq. (20), $L$ is the length of heat conduction path, $k$ is thermal conductivity and $A$ is cross-section area of heat conduction path. $Q$ and $T$ are analogous to current ($I$) and voltage ($V$) respectively in electrical domain, and thermal resistance is analogous to electrical resistance. This allows us to model the thermal circuit as an equivalent electrical circuit for analysis under various operating conditions.

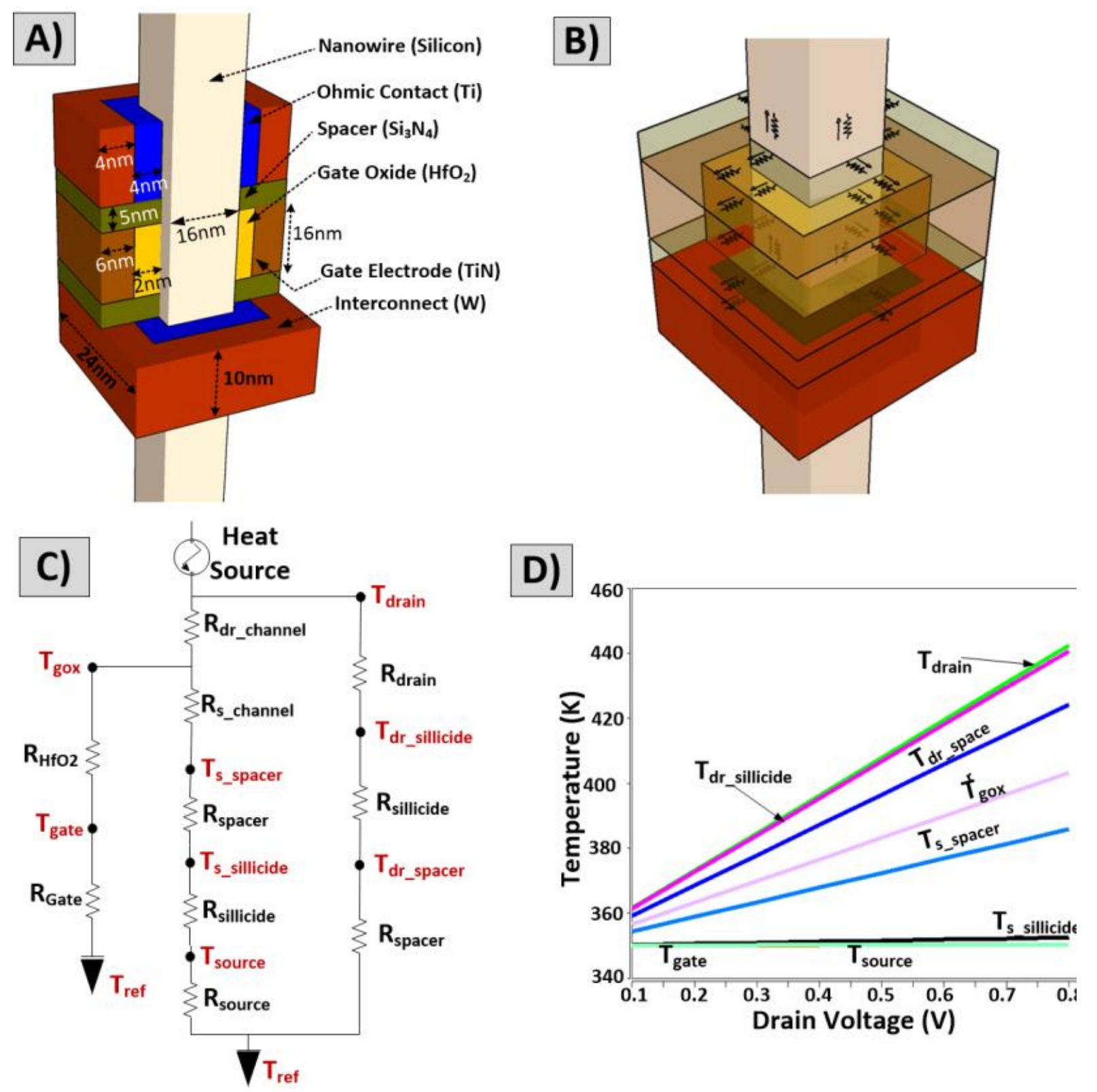

**Fig. S5.1 | Thermal modeling and simulations of V-GAA junctionless transistor.** A) V-GAA Junctionless transistor cross-section; B) heat dissipation paths are shown; heat source being the drain region; C) heat resistance model for a single transistor; drain side of the channel acts as heat source, heat is dissipated through the contacts in drain, source and gate; D) thermal simulation results for a single transistor; temperature profile at various transistor regions with the increase in drain voltage.

Material considerations and nanoscale effects are captured in thermal conductivity parameter *k*, whereas geometry considerations are accounted in (*L*/*A*) portion of eq. (20). Surface scattering, trap states and confinement effects reduce channel conductivity significantly at nanoscale. Pop. et. al., reported [17] thermal conductivity of 10nm thin silicon layer to be as small as 13 $Wm^{-1}K^{-1}$, which is one order of magnitude less than bulk silicon (147 $Wm^{-1}K^{-1}$). Table S5.1 lists different materials used in GAA Junctionless transistor and circuit thermal modeling. Material specifications (i.e., 2-D dimensions, thermal conductivity), in the heat flow path are also mentioned in Table S5.1, which is visually depicted in Fig. S5.1B.

**Supplementary Table S5.1 | Properties of materials used in transistor modeling**

| Region | Material | Dimension (L x W x T) nm | Thermal Conductivity $Wm^{-1}K^{-1}$ |
|---|---|---|---|
| Drain Electrode | Ti | 10 x 16 x 12 | 21 [36] |
| Drain-Si | Sillicide | 10 x 16 x 16 | 45.9 [37] |
| Spacer | $Si_3N_4$ | 5 x 16 x 18.5 | 1.5 [39] |
| Channel | Doped Si | 16 x 16 x 16 | 13 [17] |
| Gate Oxide | $HfO_2$ | 16 x 18 x 2 | 0.52 [40] |
| Gate Electrode | TiN | 10 x 16 x 6 | 1.9 [43] |
| Heat Junction | $Al_2O_3$ | 4x16x18.5 | 30 [18] |
| Interlayer | C doped $SiO_2$ | | 0.6 [38] |
| Bridge | W | 43.5x58x 16 | 167 [44] |

Thermal model of GAA Junctionless transistor was developed using an equivalent thermal resistance network considering the heat conduction path and device geometry, based on the methodology discussed in [16] for multigated transistors. The resistance network built from the thermal conduction paths in Fig. S5.1B and with corresponding material parameters (Table S5.1) is shown in Fig. S5.1C. As illustrated, there are three paths to reference temperature through contacts at drain, gate and source regions. Following the transistor's underlying self-heating principle the heat source is placed on the drain side of the channel. From the heat source, heat travels either through the silicide, spacer and contact at the drain, or through the channel towards the gate contact, or through the channel towards the source contact. Heat flow is depended on the least resistance path to reference temperature. This resistance network model and device characteristics from TCAD simulations (VDD = 0.8V and ON current = $3.2x10^{-5}$ A; Section 2.1) were used for HSPICE simulations. Fig. S5.1D shows the simulation result for a single isolated transistor. For this simulation, routing resistance from contact to bulk was considered to be negligible. The reference temperature was assumed to 350K. As shown in Fig. S5.1D, the temperature is highest at the drain side and gradually lowers towards the source; the trend is same for varying drain voltages. However the slope of change in temperature is different in various regions due to effective thermal resistance in each dissipation path.

### *5.1.2 Thermal Model & Analysis of Skybridge Circuits*

In order to understand thermal constraints present in realistic scenarios and to validate thermal extraction capabilities in Skybridge, we have performed detailed thermal circuit modeling using thermal resistance networks. HSPICE simulations were carried out to characterize static thermal behavior of the circuit during worst case operating condition.

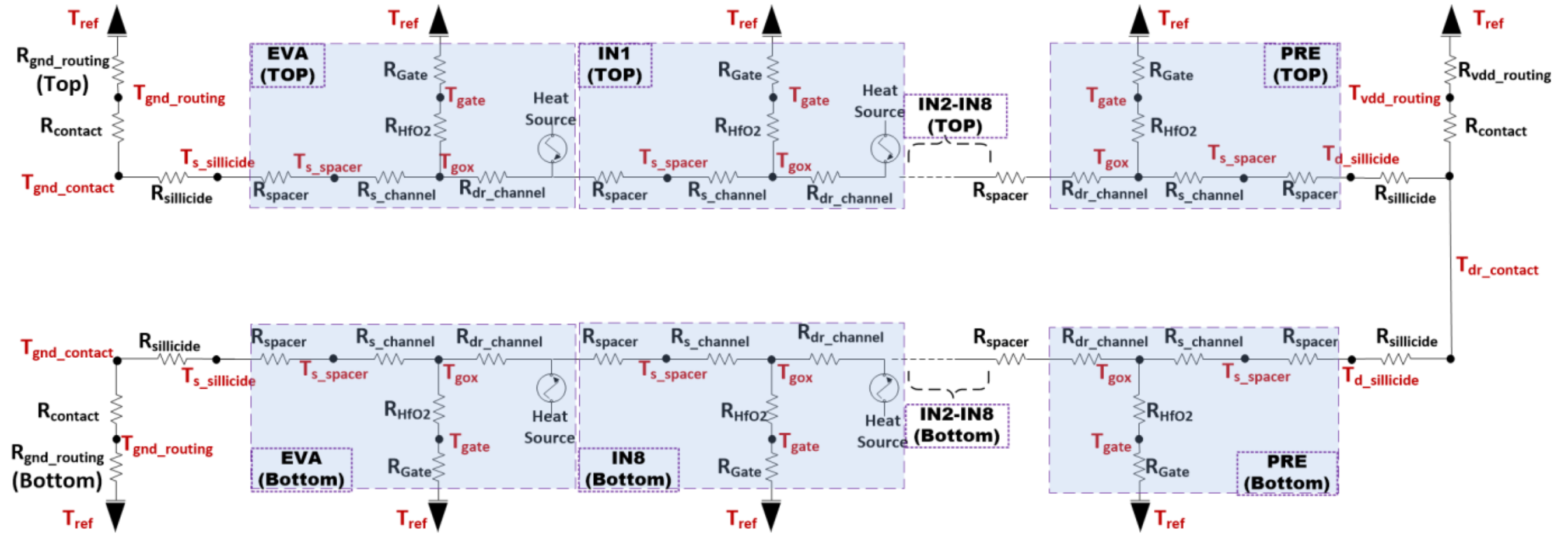

**Fig. S5.3 | Thermal modeling of circuits.** 2 sub-circuit representation in single nanowire is shown; the thermal resistance network is built based on vertical GAA Junctionless transistor model (Fig. S5.1C) and nanowire transistor stack schematic (Fig. S5.2). Each Ohmic contact to nanowire is represented by nanowire silicidation resistance, Ohmic contact resistance and routing resistance. Average routing distance from each metal electrode (i.e., gate electrode, Ohmic contact, power rail contact) to heat sink was assumed from 8bit Skybridge carry look-ahead adder (CLA) circuit.

Fig. S5.2 shows example sub-circuits with two independent 8-input dynamic NAND gates implemented in single nanowire. GND contacts are on the top and bottom of the nanowire and VDD is in the middle. The placement of these power rail contacts dictates the dissipation paths. Additional heat dissipation paths are through the transistor gate regions, through interlayer dielectric, and through doped silicon nanowire (see Fig. S5.2). Gate input Bridges along with gate contacts contribute significantly in heat extraction, if the contact itself (i.e., source of gate input) is in reference temperature. If the gate input is at different temperature, heat dissipation through gate may vary.

The 3-D thermal resistance network for the nanowire in Fig. S5.2 is shown in Fig. S5.3. As depicted, metal contacts, silicided nanowire, transistors, Skybridges, signal and power pillars are all represented by thermal resistances. The modeling of thermal resistance follows similar methodology described in Section 5.1.1. Design rules for 3-D circuit layout and transistor are same as in Section 9.1 and Section 2.1.

HSPICE simulations were carried out for worst case thermal profile. For the sub-circuits in Fig S5.2, the worst case scenario is during the *EVA* phase of operation when all the transistors are 'On' and each of them act as a static heat source. Heat source (i.e., power in electrical analogy) at the drain side of each transistor in the NAND gate was determined by dividing maximum heat ($I_{on}$ x VDD) with number of ON transistors. This is overly pessimistic, since in a dynamic circuit multiple transistors are stacked, and the state of each transistor's drain/source diffusion capacitances determines the current flow. As a result the current in drain regions are much lower than this worst static case.

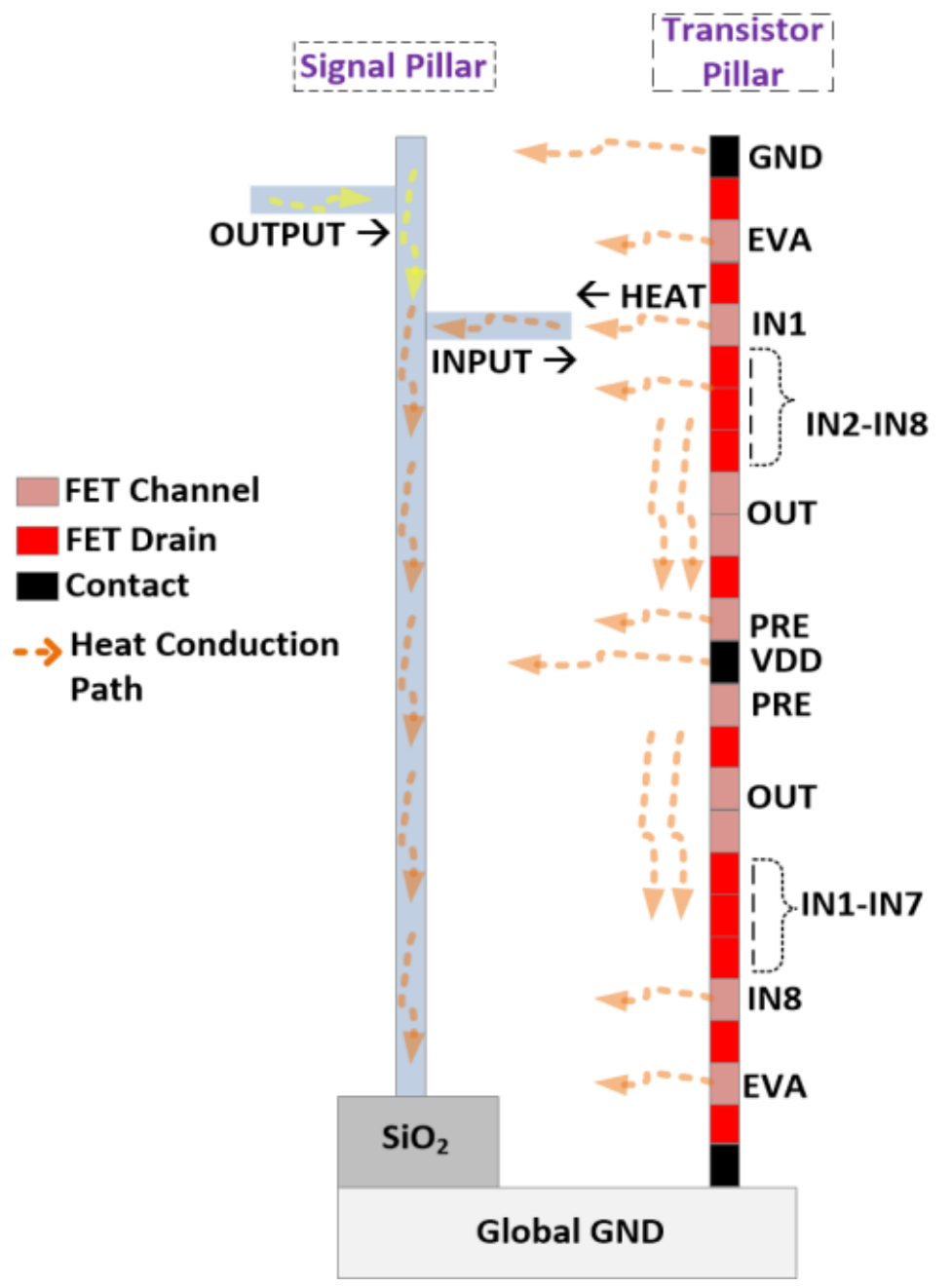

**Fig. S5.2 | Heat dissipation paths in circuits.** 2 dynamic NAND gate (8 fan-in and Pre and Eva transistors) are implemented in vertical nanowire; NAND gates share VDD contact in the middle; heat dissipation is through the nanowire, power rail contacts (VDD and GND), through gate electrodes and through interlayer dielectric. A signal nanowire is shown. Bridges carry signal from the signal nanowire to inputs; heat flows opposite to the direction of incoming signal through the gates depending on the temperature of gate input Bridges and signal nanowires.

As mentioned earlier, the gate contact plays an important part in heat dissipation. In our HSPICE simulations, we model different scenarios for gate input temperature: (i) at maximum, (ii) half of the maximum, and (iii) reference. Maximum temperature in gate contact represents the scenario when there is no heat conduction through the gate (i.e., thermal resistance in the gate is inifinite); half of the maximum scenario refers to the condition that the heat conduction through the gate is half of the best case scenario, when the gate is at reference temperature and contributes fully as major heat dissipation path. Simulation results are shown in Fig. S5.4. The best case results are obtained for scenario (iii), when there are multiple heat dissipation paths. For the top-most transistor, the temperature in the drain region is as high as 4307K in scenario (i); however with more heat dissipations through the gate, the temperature reduces drastically to 667K (scenario (ii)) and to 480K (scenario (iii)). Fig. S5.4 also shows the trend that temperature decreases towards the bottom of the transistor stack.

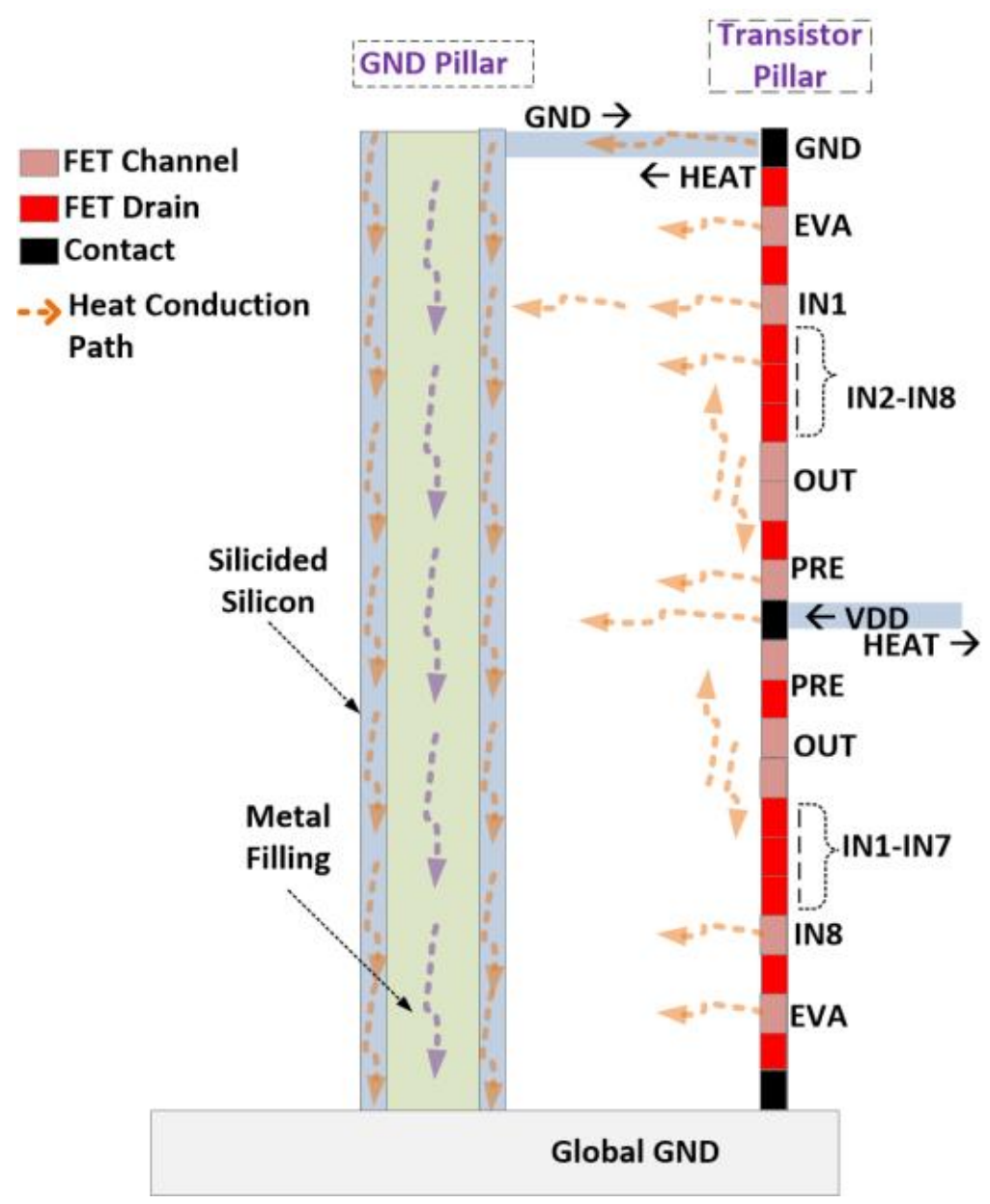

**Fig. S5.5 | Incorporation of Heat Dissipating Power Pillar (HDPP):** An intrinsic feature in Skybridge fabric to mainly facilitate heat extraction. HDPPs are connected to logic-nanowire through Bridges at the top (GND) and the middle (VDD) of the nanowire. HDPPs are configured (132 nm x 132nm area, 4 sillicided nanowire pillars, metal filling (W)) to maximize heat dissipation.

### *5.2 Skybridge's Heat Extraction Features*

#### 5.2.1 *Heat Dissipation Power Pillars (HDPPs)*

Skybridge's heat extraction features maximize heat dissipation by providing thermally conductive paths. HDPPs, when connected to power rails provide such paths. The HDPPs are intermittent power pillars that serve both the purpose of local power supply and heat dissipation. These pillars are specially designed to maximize heat conduction; they occupy 2x2 nanowire pitch, (132nm x 132nm) area in our current fabric design; within this area there are 4 silicided pillars (16nm x 16nm) each. The rest of the volume has Tungsten (W) filling to maximize heat conductance (Fig. S5.5).

For the example sub-circuits in Fig. S5.2, we have connected the power rails contacts at the top, middle and bottom to HDPPs and characterized thermal effects. The configuration is visually depicted in Fig. S5.5. The average routing distance was assumed to be 10 nanowire pitches, which is half the width of an 8-bit carry look ahead adder (CLA) layout in Skybridge; the 8-bit CLA is representative of large scale circuit design. Simulation results are shown in Fig. S5.6. Clearly, for scenario (i), large area power pillars have huge impact in heat dissipation, since they provide extra heat conduction paths to reference temperature other than the silicon nanowire; the temperature reduces to 2433K in scenario (i), which is a 43% reduction from 4307K. For scenario (ii) and (iii) the change in temperature is less obvious, since the gate contacts constitute major heat dissipation paths. Noticeably, the trend in change in temperature across various transistors is different in this case. Peak temperature from the top of the transistor stack gradually decays at the middle when contacts are made to VDD pillars, and then there is slight increase again and ultimately it decays to the reference temperature. In the middle of the nanowire, contacts to VDD pillar provide less heat resistance path, and as a result the temperature drops sharply; further down the nanowire, as we go away from the power rail contacts, temperature increases slightly. These results indicate that HDPPs play a prominent role in heat extraction from circuits. Based on this understanding, we have added new architectural features to maximize heat extraction from logic-nanowire pillars and to dissipate it through HDPPs.

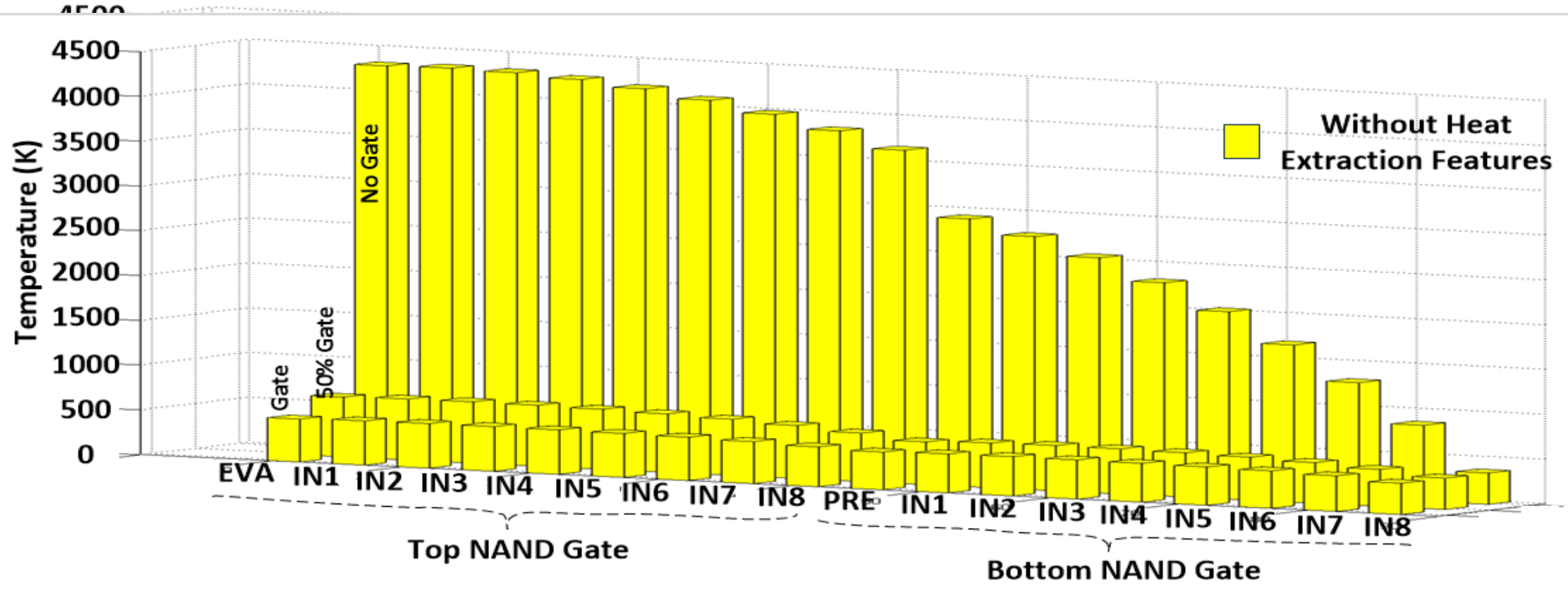

**Fig. S5.4 | Thermal simulation results of Skybridge circuits without heat extraction features.** temperature profile of each transistor in the logic-nanowire in Fig. S5.2 is shown. Thermal profile of shows the importance of heat dissipation paths, for the scenario when no heat extraction through gate is considered, temperature is as much as 4307K, in the EVA transistor. When heat extraction through gate contact is considered, temperature reduces drastically to 667K and 480K for 50% and 100% gate extractions respectively.

### 5.2.2 *Heat Extraction Junctions (HEJs)*

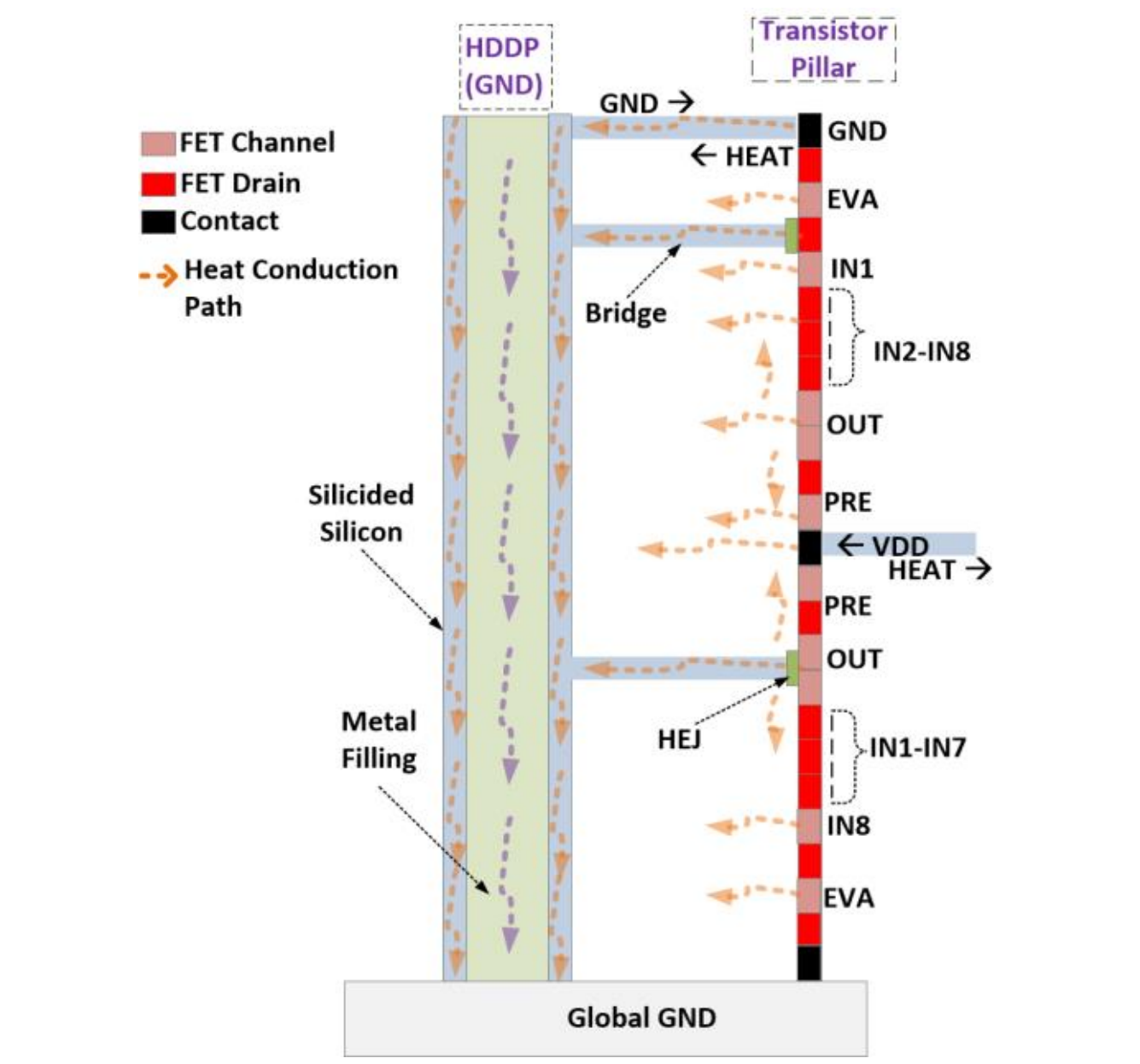

**Fig. S5.7 | Heat Extraction Junctions (HEJs):** HEJs for heat extraction and dissipation through Bridges and a HDPP is shown. HEJs are placed at selective places in the logic-nanowire; they extract heat without perturbing the electrical signal. $Al_2O_3$ is used as Junction material for excellent thermal conduction and electrical insulation.

Heat Extraction Junctions (HEJs) are specialized junctions that are used solely for heat extraction in a logic nanowire without perturbing its electrical operation. HEJs facilitate heat transfer to Bridges and HDPPs. The heat extracting Bridge connects to an HEJ on one side and to HDPP (GND) pillar on the other;

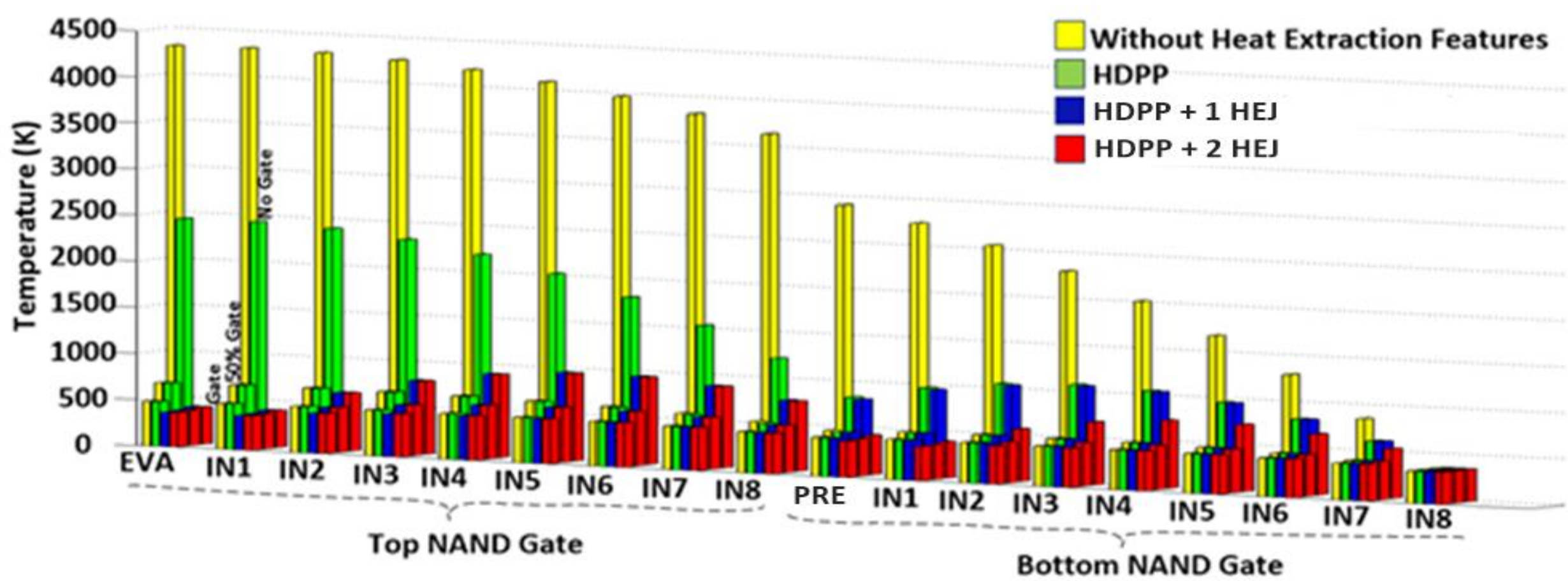

**Fig. S5.8 | Impact of HEJs, Bridges and HDPPs for heat extraction.** Two cases are simulated: with 1 HEJ and with 2 HEJs per logic nanowire connected to Bridges and HDDPs for heat management. In the case of 2 HEJs per nanowire, they are connected to two output regions of dynamic NAND gates. For the case with no heat dissipation through gate, the temperature decreases from 4307K to 400K when 1 HEJ is used in topmost Eva transistor, and from 2909K to 426K in the bottom Eva transistor for 2 HEJs. Improvements are also observed for the cases when the gate electrode is at half of the maximum temperature (1 HEJ: from 667K to 376K) in the topmost Eva transistor and (2 HEJ: from 479K to 398K) in the middle Eva transistor; in case of the gate electrode at reference temperature, temperature drops from 479K to 367K for 1 HEJ at the topmost Eva transistor, and from 422K to 389K for 2HEJs at the middle Eva transistor.

this ensures that the heat extraction Bridges are at reference temperature initially to facilitate heat transfer from the hot region towards cool region. Fig. S5.7 illustrates this concept. $Al_2O_3$ meets the material requirements for such HEJ since it has excellent thermal conductance (39.18 $Wm^{-1}K^{-1}$ [18]), and is a good electrical insulator. The thickness for $Al_2O_3$ was chosen to be 6nm, which is sufficient to prevent any electrostatic control from Bridge contacts to silicided silicon. The HEJs can be placed at any point on the logic-nanowire and can be connected with Bridges for heat extraction; this allows certain degree of freedom and enables custom design choices for hotspot mitigation.

Fig. S5.8 shows simulation results that indicate the effectiveness of the HEJs when combined with Bridges and HDPPs. Two conditions are illustrated: (a) one HEJ connected to the drain region in the topmost transistor in the logic nanowire, and (b) two HEJs are connected to two most heated regions in the logic-nanowire (two topmost transistors in each NAND gate). In these simulations, power rail contacts were assumed to be connected to HDPPs in the same way as was discussed in the previous sub-section. The routing distances for Bridges were assumed to be 10 nanowire pitches.

As illustrated in Fig. S5.8, radical improvement in temperature profile is achieved when all the fabric heat extraction features are active. Up to 90% reduction in temperature is achieved when only one HEJ is used in the logic nanowire. For the scenario when there is no heat extraction through gate contacts, HEJ, Bridges and HDPPs jointly reduce the temperature from 4307K to 400K in the topmost transistor, and the average temperature drops from 2977K to 793K, a 73% reduction. The average temperature reduces further, 78% when two HEJs are used in conjunction with Bridges and HDPPs. Substantial improvements are also observed when gate contacts contribute to heat dissipation. For the scenarios when gate contacts are at half of the maximum temperature and at reference temperature, the average temperature reduces by 12% and 4.5%, and 15.4% and 6.5% for heat extractions with one HEJ and two HEJs, respectively. These results validate the effectiveness of Skybridge's heat extraction features. The simulation results indicate that even with 1 HEJ per logic nanowire, the average temperature for the worst-case heat generation can be reduced to acceptable temperatures below the breakdown voltage of Junctionless transistors. These transistors were shown to operate even at temperatures as high as 500K [45]. In addition, depending on design requirements, modifications can be done with placement of HDPPs and number of HEJs in circuits to reduce the average temperature even further.

# 6. CIRCUIT DESIGN EXAMPLES AND BENCHMARKING

<u>Section Summary</u>

In this section, various circuit design examples in Skybridge are shown. We present arithmetic circuits at different bit-widths and show how they scale. Benchmarking results against projected scaled CMOS designs are also provided.

<u>Overview</u>

In this section we detail on arithmetic circuit implementations using carry look-ahead adders and array multiplier circuits. These arithmetic circuits combine compound and cascaded dynamic logic styles in dual rail logic for optimum performance at low power and ultra-high density. The density benefits are maximized by using high fan-in logic. Connectivity requirements are met by utilizing the fabric's routing features. The effect of coupling noise due to dynamic circuit style and dense interconnections is mitigated through the noise shielding approach introduced in Section 2.2.2.

In order to study the scalability aspects of Skybridge designs, we have implemented arithmetic circuits at 4, 8 and 16-bit-widths, and benchmarked against CMOS designs at 16nm. The benchmarking was done by accounting for detailed effects of material structures, nanoscale device physics, circuit style, 3-D circuit layout, interconnect parasitics and noise coupling. In addition to arithmetic circuits, we have also benchmarked Skybridge's volatile RAM against projected scaled CMOS SRAM designs at 16nm. The results show tremendous benefits can be obtained for Skybridge designs; for example, the 16-bit CLA design achieves 60x density, 10x power and 54% performance benefits over equivalent CMOS designs, and Skybridge's NWRAM achieves 4.6x density, 4.3x active power and 50x leakage power benefits at comparable performance over CMOS SRAMs.

## *6.1 Circuit Design Examples and Scalability Aspects*

### *6.1.1 Basic Arithmetic Circuits*

Adders and multipliers are core arithmetic computing blocks in ALUs, and are often extended to implement other arithmetic operations such as complement, subtraction and division. Some of the circuits presented here are also used for the Skybridge microprocessor design (Section 7).

#### *6.1.1.1 Carry Look-Ahead Adder*

CLA is well-known parallel adder for fast computation. A block diagram of a 4-bit CLA is shown in Fig. S6.1A; it consists of propagate-and-generate, carry, buffer and summation blocks. The propagate-and-generate block is used to produce intermediate signals $P_i$ and $G_i$ (where i = 0 to 3), which are used for calculating Sum and Carry respectively; the logic expressions used are $P_i = (A_i \oplus B_i)$ , $G_i = A_i \bullet B_i$. The carry block is used to compute intermediate carry signals and final carry output. The logic expression for carry generation is $C_i = G_{i-1} + P_{i-1} \bullet C_{i-1}$, where 'i' is from 1 to 4. The buffer block is used to buffer a signal and maintain signal integrity. The sum block generates the final sum output using the intermediate $P_i$ and $C_i$ signals; the logic expression is $S_i = A_i \oplus B_i \oplus C_i = P_i \oplus C_i$.

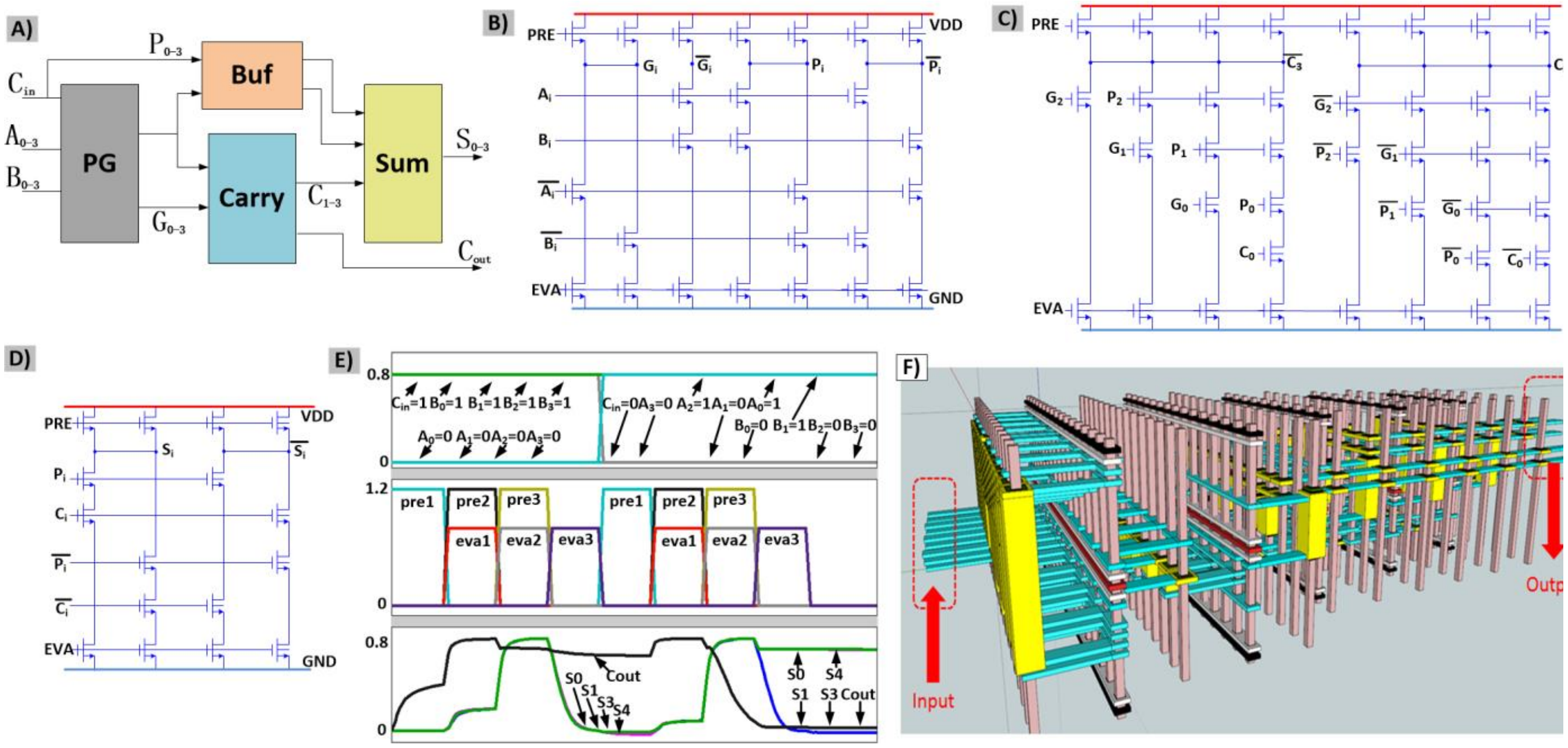

**Fig. S6.1. 4-bit carry look-ahead adder (CLA)**. A) Overall block diagram of 4-bit CLA; it contains propagate and generate (PG), carry, buffer and sum blocks; B) circuit schematic of PG block; both true and complementary values are generated in the compound dual rail logic; C) schematic for carry block using the same circuit style, inputs from PG block is used; D) schematic of sum block, inputs from both PG and carry blocks are used; E) HSPICE simulated waveforms validating the expected adder behavior; F) physical layout of a CLA in the Skybridge fabric.

The Skybridge specific implementations of these logic blocks use both compound and cascaded dual-rail dynamic logic styles (see Section 2 for details). The circuit schematics are shown in Figs. S6.1B-D. As shown in Figs. S6.1B, and S6.1D, the XOR logic for computing $P_i$ and $S_i$, and their complementary signals, is done using compound dynamic gates. The $C_i$ and $\sim C_i$ computations also use dynamic compound gates in AND-of-NANDs logic, as shown in Fig. S6.1C. The generated intermediate signals are propagated to the next stage of compound gates through cascading. HSPICE simulation results validating the CLA circuit behavior are shown in Fig. S6.1E.

The physical implementation of a CLA is shown in Fig. S6.1G. The circuit mapping into Skybridge follows the guidelines summarized in Section 9.

*6.1.1.2 Array Multiplier*

Array based multipliers are widely used for fast parallel multiplications. The core concept is illustrated in Fig. S6.2A: multiplication is achieved by a series of additions. The hardware implementation of the algorithm uses adder units for these iterative additions. The block diagram for the multiplier is shown in Fig. S6.2B. As illustrated, the multiplication is performed with the help of AND logic, half adder and full adders. AND operation is performed simply by using a compound gate with two inverted inputs (to perform AND-of-NANDs). The half adder and full adder implementations follow ripple carry logic, and are implemented using XOR and NAND gates. Implementation of these logic units use similar compound circuit implementations as in CLA. The result of each addition is cascaded to other adder units to generate the total multiplication output. HSPICE simulated waveforms for this multiplier circuit are shown in Fig. S6.2C; the two operands illustrated for the 4-bit multiplication are 0011 and 0111, yielding 00010101. The physical layout of this multiplier can be seen in Fig. S6.2D.

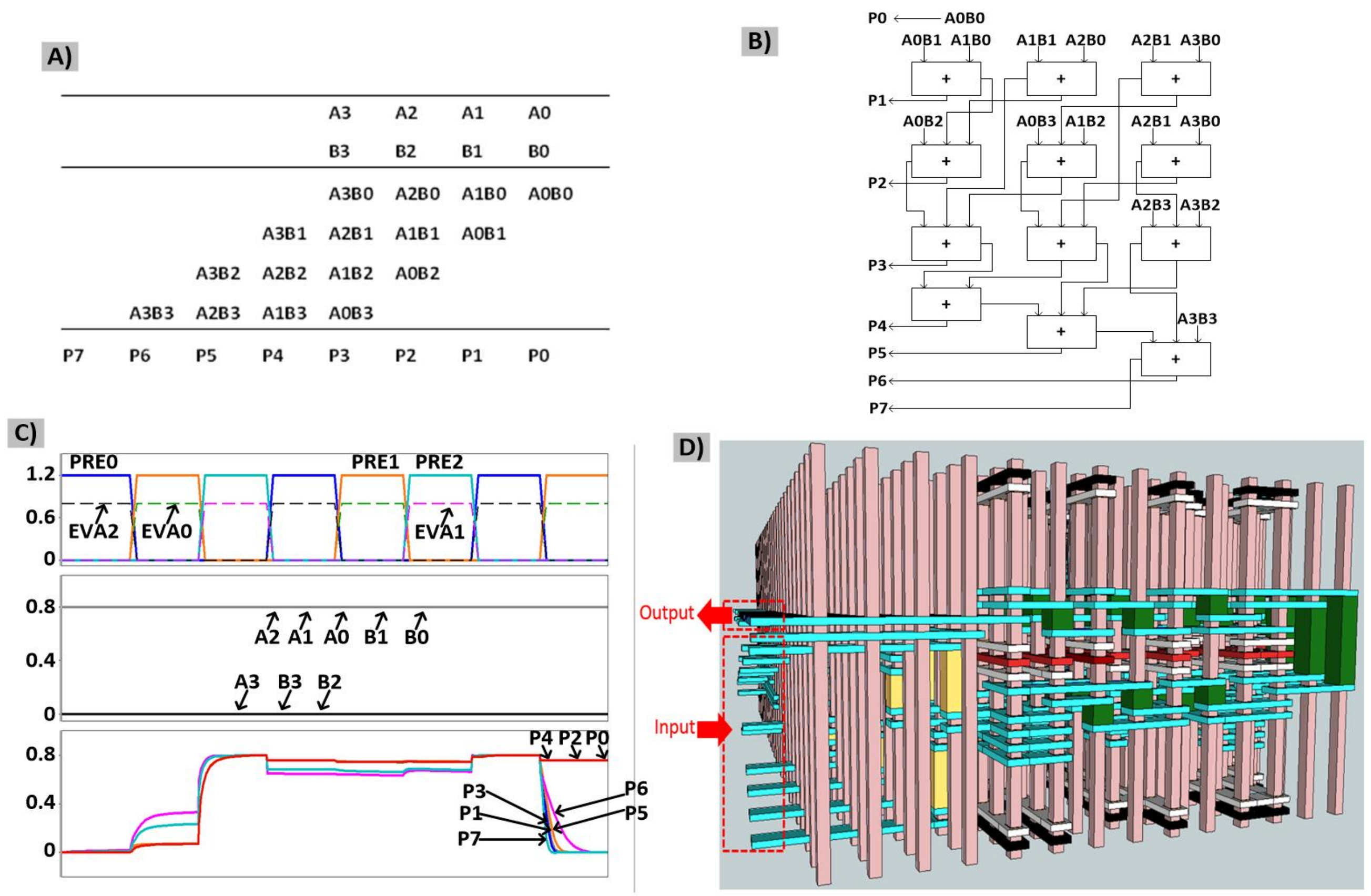

**Fig. S6.2 | 4-bit array multiplier.** A) 4-bit array multiplication algorithm; B) block diagram of the array multiplier; in order to do iterative additions half adder and full adders are used, the multiplication is completed in 9 stages; in this figure, the flow is from the top towards bottom; C) HSPICE validations of the multiplier; multiplication between 0011 and 0111 results in 00010101; the final result is generated at the 9th clock phase; D) physical layout of a 4-bit multiplier in Skybridge.

*6.1.2 High Bit-Width Arithmetic Circuits*

In order to evaluate the potential of Skybridge designs at higher bit-widths, we have extended the 4-bit CLA designs to 8- and 16-bit CLAs. An additional objective was to evaluate the impact of high fan-in on key design metrics such as density, power and performance.

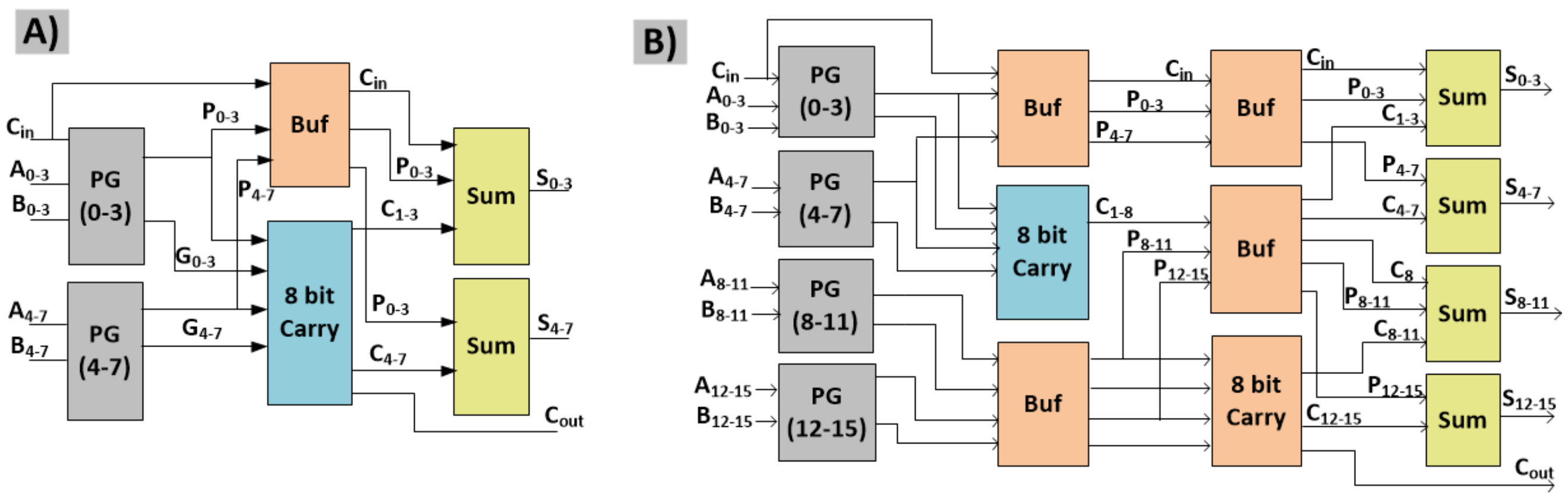

**Fig. S6.3 | High bit-width arithmetic examples: 8-bit and 16-bit CLAs.** A) 8-bit CLA block diagram; it consists of 4-bit propagate and carry (PG), 4-bit buffer, 8-bit carry and 2 4-bit sum units. PG blocks generate intermediate signals for parallel addition, buffer is used for signal synchronization, and for signal propagation; sum and carry blocks generate sum and carry respectively; B) 16-bit CLA block diagram; it consists of 4 4-bit PG, 4 4-bit buffer, 2 8-bit carry and 4 4-bit sum blocks.

8-bit and 16-bit CLA block diagrams are shown in Fig. S6.3. Both designs use 4-bit PG and Sum blocks as core building blocks. The implementations of these 4-bit blocks remain the same irrespective of the bit-width choices. However, the carry block's complexity increases with bit-width, since $C_i$ is calculated using logic expression: $C_i = G_{i-1} + P_{i-1} \bullet C_{i-1}$. For higher orders of $C_{out}$, the complexity increases exponentially. As a result, two carry blocks cannot be used in the same clock stage without cascading in 8-bit CLA design; such partitioning of the carry block will result in throughput degradation.

However, for a 16-bit CLA design, two 8-bit carry blocks were used. A single 16-bit carry block in a single clock stage would result in 17 fan-in circuits, which would cause severe degradation of overall performance (details on fan-in sensitivity can be found in Section 2.2.1). The maximum fan-ins assumed are 4, 9 and 9 for 4-bit, 8-bit and 16-bit CLAs respectively.

*6.2 Benchmarking*

*6.2.1 Benchmarking of Arithmetic Circuits*

The benchmarking of Skybridge circuits vs. CMOS followed the bottom-up evaluation methodology described in Section 3.3. The benchmarking results for arithmetic circuits are shown in Table S6.1.

As evident from the results, Skybridge designs achieve significant benefits across all metrics. Table S6.1 shows that the 4-bit array multiplier Skybridge design has 39.3x density and 4x power advantage at comparable performance vs. the CMOS multiplier. The 4-bit Skybridge CLA is 24.6x denser and has 12X reduced power; whereas 8 and 16-bit CLA designs that use 8 fan-in are 48x and 60.5x denser, respectively, and consume 12x and 10x less power, respectively, in comparison to equivalent 16-nm CMOS designs. The active power results show almost linear dependence to throughput. The 16-bit Skybridge design is 54% higher performance vs. the CMOS version. Due to the Skybridge fabric and circuit style, the load capacitance that each gate output sees is reduced, and as a result high fan-in designs are possible and beneficial in Skybridge circuits. Our 16-bit results show better overall results with higher bit-widths vs. CMOS.

**Supplementary Table S6.1 | Scalability potential of Skybridge designs**

| | Throughput (Ops/sec) | | Power (µW) | | Area (µm$^2$) | |
|---|---|---|---|---|---|---|
| | CMOS | SB | CMOS | SB | CMOS | SB |
| **4-Bit Multiplier** | 5.0e9 | 5.1e9 | 42.3 | 172 | 50 | 1.27 |
| **4-Bit CLA** | 9.9e9 | 10.4e9 | 235 | 19.4 | 18.7 | 0.76 |
| **8-Bit CLA** | 4.5e9 | 5.7e9 | 287 | 23.5 | 64.7 | 1.34 |
| **16-Bit CLA** | 2.4e9 | 3.7e9 | 297 | 27.8 | 130.2 | 2.15 |

*6.2.2 Benchmarking of Volatile Memory*

Cell-based evaluation of Skybridge volatile RAM vs. scaled 16nm high performance 6T-SRAM is shown in Table S6.2. The Skybridge RAM has 4.6x density, 4.24x active power and 50x leakage power benefits, and operates at similar frequency as the high performance SRAM (Table S6.2). These benefits of Skybridge RAM are achieved due to 3-D integration and innovative circuit style. The density benefits are obvious from the Skybridge RAM 3-D layout (Fig. S2.6C), since only one logic-nanowire is used for memory implementation, which is equivalent to one transistor area. The dense implementation also implies intra-cell routing is less, which is advantageous to reduce active power. The active power in this RAM is further reduced compared to SRAM, due to its fundamental operating style. The write operation in Skybridge RAM is synchronized with clock, and only true or complementary value is written at a certain time as opposed to SRAM where both values transition at the same time leading to higher switching activity, and as a result more active power compared to Skybridge RAM. The leakage power in Skybridge RAM is significantly less, since the RAM design uses dynamic circuit style with multiple transistors stacked in series forming high resistance path from storage node to GND. Moreover, the Skybridge RAM's restoration scheme ensures that during periods of inactivity all control signals can be switched off, which reduces leakage power further (Details on Skybridge RAM operation can be found in Section 2.3). Despite reduced intra-cell routings of Skybridge RAM, the performance results were found to be similar to that of high performance scaled SRAM; this is mainly because, the SRAM design uses highly optimized strained 16nm MOSFET that has higher ON current than unstrained V-GAA Junctionless transistor used for Skybridge RAM. By optimizing transistor characteristics, higher performance can be achieved from Skybridge RAM design. We also expect similar or higher benefits to be maintained across all metrics for large-array memory designs.

**Supplementary Table S6.2 | Memory comparison: Skybridge 8T-NWRAM vs. CMOS 6T-SRAM**

| | Delay (ps) | Active Power (µW) | Leakage Power (nW) | Area (µm$^2$) |
|---|---|---|---|---|
| **CMOS 6T-SRAM** | 20 | 1.4 | 8.2 | 0.065 |
| **Skybridge 8T-NWRAM** | 20.2 | 0.33 | 0.164 | 0.014 |

## 7. DESIGN OF A SKYBRIDGE PROCESSOR

Section Summary

In this section, a Skybridge processor design is shown. A simple 4-bit WIre Streaming Processor (WISP-4) was built at the transistor level, functionally verified at the circuit level, and benchmarked against equivalent CMOS processor at 16nm. In addition to the larger bit-width arithmetic circuits presented in earlier sections, this allowed to exercise the Skybridge inter-circuit connectivity/routing concepts.

Overview

The WISP-4 processor design uses a load-store architecture, which is common in modern RISC processor designs. It is composed of blocks such as program counter (PC), read-only memory (ROM), register file, buffers, decoders, multiplexers and arithmetic logic unit (ALU), and is capable of performing memory access and arithmetic operations. WISP-4 was designed with five stages of pipeline, and each stage is micro-pipelined with internal clock signals driving Skybridge dynamic circuits. Design of all logic and memory circuits for processor follow the Skybridge's circuit styles (see Section 2). Circuit placements and layouts are in accordance to the Skybridge fabric design rules and guidelines (see Section 9).

Using the bottom-up evaluation and benchmarking methodology discussed in Section 3.3, extensive simulations were carried out to validate the WISP-4 design, and to evaluate its potential against equivalent CMOS implementation. Benchmarking results show that Skybridge's WISP-4 processor is at-least 30x better in terms of density and 2.48x better in terms of performance/watt vs. projected scaled 16nm CMOS. These results are in fact pessimistic since Skybridge is expected to improve for higher bit-widths and complexity vs CMOS due to shorter interconnects and higher possible fan-in.

*7.1 WISP-4 Architecture*

The architecture of WISP-4 is shown in Fig. S7.1. It has five pipeline stages: Instruction Fetch, Decode, Register Access, Execute and Write Back. During Instruction Fetch, an instruction is fetched from ROM and is fed to instruction decoder. In Instruction Decode, the fetched instruction is decoded to generate control signals, and to buffer the register addresses and data. In the next stage, buffered data is stored in register file and prepared for sequential execution in the Execute stage. After ALU operations in the Execute stage, results are stored in the register file during Write Back. The synchronization of pipeline stages is maintained through micro pipelining of logic blocks at each stage; this is possible, since all logic block implementation is through the Skybridge logic style, which uses clock signals as control inputs.

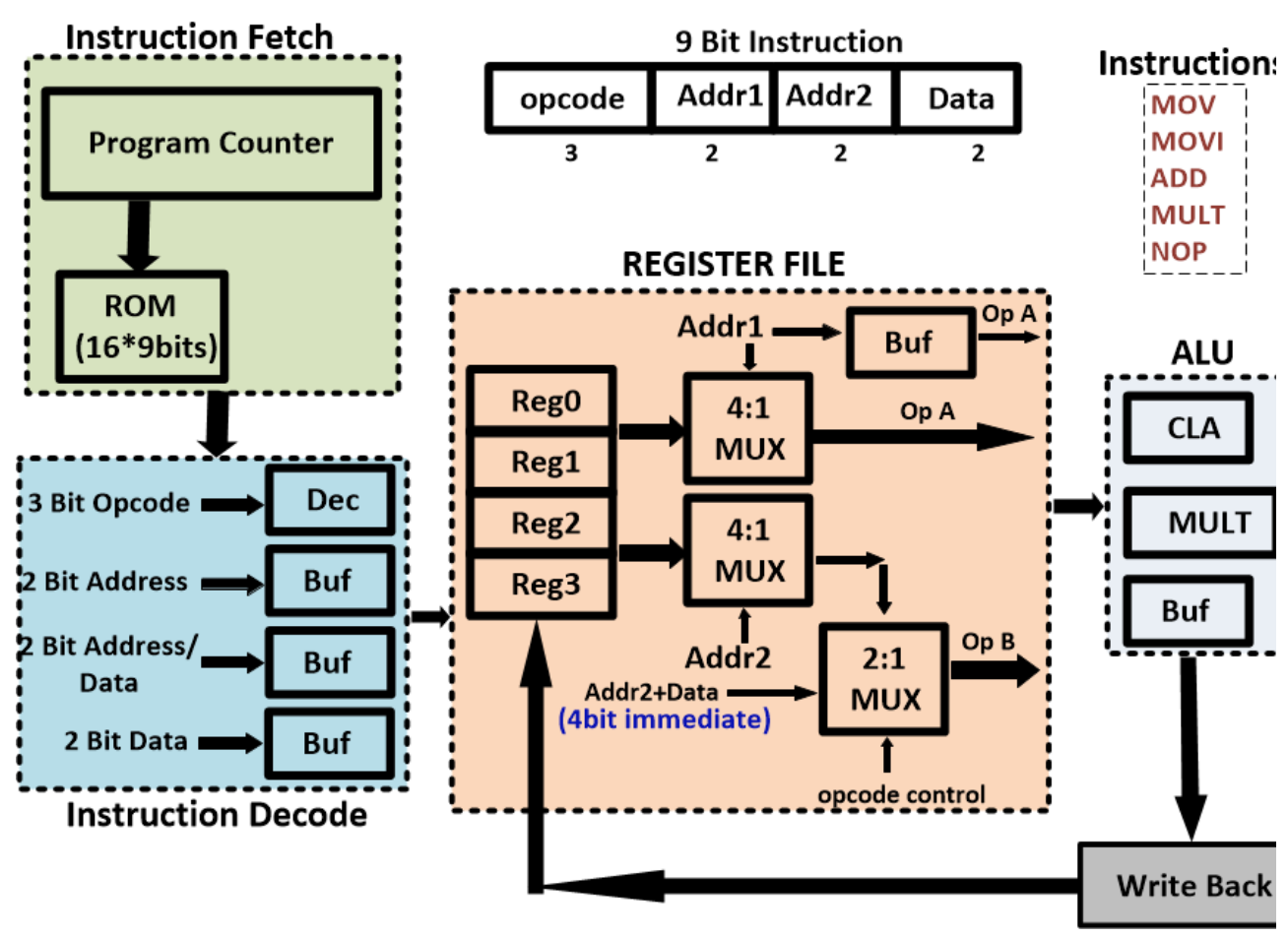

**Fig. S7.1 | Skybridge 4-Bit Wire Streaming Processor (WISP-4).** Block diagrams showing the WISP-4 organization; it has 5 pipelined stages: Instruction Fetch (IF), Instruction Decode (ID), Register Access, Execute and Write Back. 5 instructions are supported: move (MOV), move immediate (MOVI), addition (ADD), multiplication (MULT) and stall (NOP).

The instruction fetch unit consists of a program counter (PC) and a ROM (Fig. S7.2A). The PC is a 4-bit binary up counter that is used to continuously increment the instruction address every clock cycle. This implementation uses a 4-bit CLA; one of its inputs is constant '1', and another is the result of previous calculation. The result of PC is fed to a 4:16 decoder to select one of the 16 rows from the instruction ROM. The ROM stores a set of instructions to be executed and has a total capacity of 16x9bits in this prototype. The output of ROM is a 9-bit instruction and contains 3-bit operation instruction (opcode), two 2-bit source/destination register addresses or 4-bit data (see Fig. S7.1).

As shown in Fig. S7.2B, the instruction decode unit consists of a 3:8 decoder and buffers to decode operation type in an instruction (opcode), and to buffer the address and data. Five operations are supported in the current design: MOV, MOVI, ADD, MULT, NOP. MOV (move) and MOVI (move-immediate) opcodes are used to move or store data in registers. ADD and MULT opcodes are used for addition and multiplications respectively. NOP stands for no operation, and is used for stalling the pipeline.

The Register file consists of registers, 2:1 and 4:1 multiplexers, and buffers. Registers are used to store operands, and multiplexers are used to generate control signals for ALU. Buffers are necessary for synchronization of data between stages.

The ALU in WISP-4 consists of a CLA, array multiplier, buffer, and 2:1 multiplexers. The block diagram of ALU is shown in Fig. S7.2D. 4-bit CLA and multiplier units are used for addition and multiplication on 4-bit operands. The buffer unit is used for data buffering and to write back in the next stage. 2:1 multiplexers select the output of ALU, which is stored in the register file during Write Back stage.

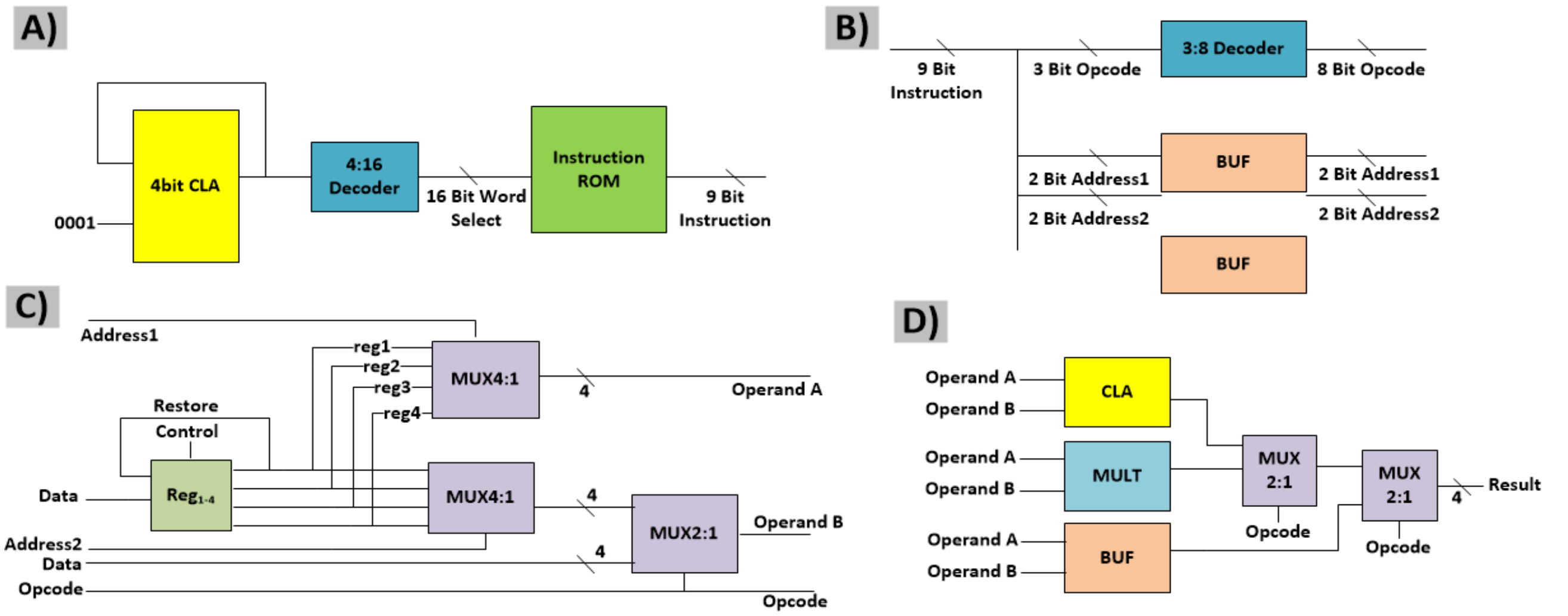

**Fig. S7.2 | Block diagram of each pipeline stages.** A) Instruction Fetch stage contains 4-bit CLA for program counter, 4:16 decoder to decode ROM address and 16*9 ROM to store instructions; B) Instruction Decode stage contains a 3:8 decoder to decode opcode and two 2-bit buffers for buffering address and data; C) Register Access stage has four 4-bit registers to store operands, two 4:1 multiplexers and one 2:1 multiplexer for operand selection; D) Execute stage contains arithmetic units: 4-bit CLA and multiplier for addition and multiplications, a buffer for data buffering, and two 2:1 multiplexers for result selection.

Circuit-level implementation of these processor units follows the Skybridge circuit style. Both Compound and cascaded dynamic logic styles are combined for efficient implementations. 4-bit CLA and multiplier circuits and HSPICE validations were shown in Section 6; in this section we show the core supporting circuits.

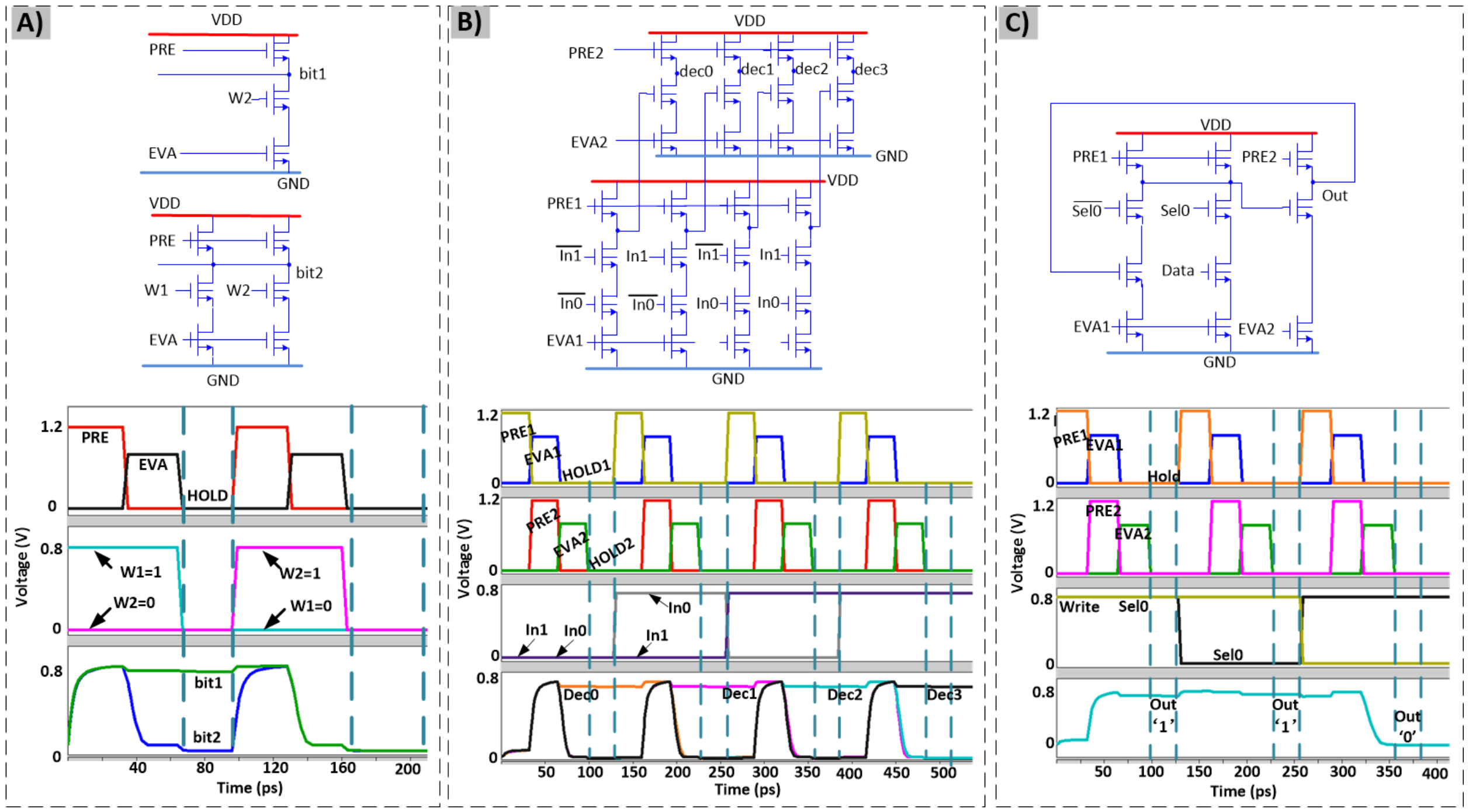

**Fig. S7.3 | 2-bit ROM, 2:1 decoder and a latch.** A) 2-bit ROM implementation using Skybridge's circuit style. The circuit is preconfigured to produce '0' or '1' output at selected locations; the schematic (top) is configured to produce '1' at bit1 location when W1 is selected, '0' at bit2 when W2 is selected. HSPICE results are shown in the bottom figure; B) 2:4 decoder schematic and HSPICE results are shown; cascaded logic style is used for this; output of first stage is propagated to the second stage for inversion operation; C) A latch implementation; latch operation is controlled by Sel0 and Data inputs; HSPICE simulation results are shown in the bottom subfigure.

Fig. S7.3 shows 2-bit ROM, 2:4 decoder, and a latch. The ROM is pre-configured to generate either '1' or '0' output at selected locations. For example, to emulate permanent storage of '1' and '0' in word1, bit1 and word2, bit2 locations, 3 dynamic one input NAND gates are used. As shown in Fig. S7.3A (top), the bit1 location is associated with a NAND gate that has only word2 (*W2*) as input; whereas, the bit2 location is associated with shorted outputs of two NAND gates, whose inputs are word1 (*W1*) and word2 (*W2*) respectively. All NAND gates shown in Fig. S7.3A (top) are controlled by the same PRE, EVA control signals. During *W1* select, *W2* is '0', therefore *bit1* read-out value is '1', and during *W2* select both *bit1* and *bit2* read-out values are '0' as expected. Fig. S7.3A(bottom) shows the HSPICE simulated waveform validating ROM behavior.

The 2:1 decoder implementation uses a cascaded dynamic logic style; output of first stage is propagated to second stage for inverted final output. Fig. S7.3B shows the circuit schematic and related HSPICE simulation results. The dynamic latch implementation is shown in Fig. S7.3C. It uses a 2:1 multiplexer and a NAND gate for required functionality; depending on the input (*Data*) and select signal (*Sel0*), either new data is latched or old data (*out*) is retained through the feedback logic. Fig. S7.3C (bottom) shows the HSPICE simulations for this latch, validating circuit operation.

The WISP-4 design lays the foundation for processor implementations in Skybridge fabric. This design can be easily extended to higher bit-width similar to the ones shown in Section 6 can be used. In addition, Skybridge's volatile RAM can be used to realize high performance on-chip caches.

### *7.2 Processor Evaluation and Benchmarking*

The processor implementation follows the 3-D circuit mapping guidelines and design rules outlined in Section 9. The bottom-up simulation methodology that takes into account device, circuit, and 3-D layout details was used. Effects of coupling noise were taken into account. CMOS equivalent WISP-4 implementation was completed using state-of-the-art CAD tools and scaling to 16nm was done using standard design rules [11][12]. Details about methodology can be found in Section 3.3.

As shown in Table S7.1, the Skybridge WISP-4 design significantly outperforms the equivalent CMOS version. At-least 30x density, 2.94x power and 18.6% performance benefits are obtained. Higher benefits are expected for higher bit-width implementations. The scalability of Skybridge circuits was shown through arithmetic circuits in Section 6.

**Supplementary Table S7.1 | Skybridge vs. CMOS comparison for microprocessor**

| WISP-4 Processor | **Throughput (Ops/sec)** | **Power (μW)** | **Area ($\mu m^2$)** |
|---|---|---|---|
| **CMOS** | $4.3 \times 10^9$ | 886 | 289 |
| **Skybridge** | $5.1 \times 10^9$ | 301 | 9.52 |

## 8. EXPERIMENTAL VALIDATIONS, WAFER SCALE MANUFACTURING, AND SENSITIVITY TO MANUFACTURING IMPERFECTIONS

<u>Section Summary</u>

Skybridge is an entirely new 3-D technology that is targeting beyond CMOS IC design, addressing scaling at the end of the CMOS roadmap. During the several years we pursued validation, we focused to *address and validate beyond-CMOS scaling aspects* rather than just prototyping at a much coarser scale than this technology targets. As compared to integrated circuit directions/concepts that would alter one aspect of CMOS, e.g., replace the channel material or change the device, but otherwise follow circuit and fabric assembly mindset as in CMOS (and thus can be manufactured largely within an existing process flow), this task is much more daunting. Taping out a Skybridge chip is arguably beyond reach on the short term; even as we show it being practical, no foundry is set up for this process yet and the costs would be initially very high, somewhat similar to when a foundry transitions to a new subsequent nanoscale technology node.

So, how can a 3-D technology like Skybridge, that is comprehensive across multiple layers of abstraction, be trusted at this stage? We believe that such a new fabric validation/direction across multiple layers of abstraction must combine the following key aspects to be credible, as we also show here that it should:

- <u>Demonstrate active device behavior at nanoscale experimentally</u>. We develop and show the junctionless behavior at 30-nm nanowire scale and show key process steps involved in a cleanroom prototype.

- <u>Not exceed industry's requirements for wafer scale integration</u> – whether lithography, doping, deposition, or alignment related. All our wafer-level process steps are proven based on CMOS and have been already shown in industry in volume projection (e.g., [49][50][54]), and/or follow ITRS guideline. In fact, we show that even if we significantly relax the manufacturing requirements we would still get very significant benefits, as compared to a projected 16nm CMOS node for which manufacturing is not yet known (per the ITRS [46]).

- Address practical issues associated with the wafer-scale manufacturing pathway upfront. We study extensively the key manufacturing related aspects of the Skybridge fabric including (i) aspect ratio implications, (ii) nanowire profile implications, (iii) what happens when lithography requirements are relaxed vs. CMOS, and (iv) implications of conservative/pessimistic material depositions resulting in sparser vertical arrays, etc.

- Establish large-scale benefits credibly. For large-scale circuits our validation follows a very detailed bottom-up methodology that includes both *process and device simulation in addition to extensive circuit-level simulation*. This captures experimental process parameters, materials, nanoscale geometry, nanoscale carrier and heat transport, precise 3-D circuit placement, 3-D layout, and 3-D interconnect parasitics.

- Show that benefits are not tied to ultimate/aggressive manufacturing assumptions. We show that significant benefits can be maintained even for overly pessimistic manufacturing scenarios: for example, a 16-bit CLA adder design in Skybridge has 29.5x density benefits over 16-nm CMOS even when the minimum pitch for nanowire array definition is considered to be almost *three times larger* than the ITRS projection for 16nm CMOS.

Overview

Contrary to CMOS, which relies on advanced lithography-dependent device miniaturization for technology scaling, Skybridge offers a paradigm shift. In Skybridge, scaling is primarily achieved by 3-D integration, where transistors are integrated vertically; 3-D circuit implementation, connectivity and thermal management requirements are carefully architected to reduce manufacturing complexities. Lithographic precision in this fabric is required only for the uniform nanowire array pattern definition; transistor channel length is determined by gate material deposition, which is lower cost, and known to be controlled to few Angstrom's precision.

In addition, Skybridge manufacturing uses only a single layer of crystalline silicon for vertical transistor channels, and same alignment markers for all the mask registration steps, which alleviate the challenges associated with the high temperature crystallization of amorphous silicon [4], and inter-layer misalignments [3][5] that are critical for stacked CMOS approaches [3][4][5]. Table S8.1 shows key manufacturing requirements for CMOS, stacked CMOS and Skybridge, highlighting Skybridge's advantages.

To validate core aspects of Skybridge fabric and to account for worst case manufacturing imperfections in high volume wafer-scale assembly, we have carried out experimental demonstrations of the Junctionless device, developed an extensive evaluation framework using process, device, and circuit simulators for large-scale designs, established a wafer-scale manufacturing pathway, and evaluated implications of manufacturing imperfections. Our experimental validation of Junctionless transistor included: development of a process flow utilizing process and device simulations, experimental prototyping and characterizations. Metrology after key process steps showed accordance with initial simulation results; the fabricated junctionless device had an $I_{ON}/I_{OFF}$ ~$10^4$. Process steps for device validation involved substrate doping, nanowire patterning, and sequential material depositions, which are central for Skybridge wafer-level assembly also.

Characterization of large-scale designs, including high bit-width arithmetic circuits and a microprocessor, was done using detailed process, device, circuit and architectural simulations.

The wafer-scale manufacturing pathway for Skybridge fabric was developed based on known materials and established processes. In addition to our experimental demonstrations, all the other steps involved for Skybridge manufacturing including: high aspect ratio nanowire formation, anisotropic material deposition, 3-D photoresist structures for selective material deposition, and non-mechanical planarization were already demonstrated [47]-[58]. Furthermore, the manufacturing pathway was developed accounting for ITRS defined lithographic guidelines for critical dimension, minimum lithography spacing and alignment tolerance during mask-substrate registration at 16nm, which further attest to its feasibility. The sensitivity analysis for manufacturing imprecision was completed at device-level granularity accounting for worst case scenarios; variations in nanowire aspect ratio, nanowire profile, pattern definition and material deposition were considered, and fabric evaluation was done for key metrics such as density, power and throughput. Evaluation results indicated that significant benefits can be achieved even for overly pessimistic manufacturing scenarios; for example, 16-bit CLA design in Skybridge had 29.5x density benefits over CMOS even when the minimum pitch for nanowire array definition was considered to be three times larger than ITRS defined limit at 16nm, for which manufacturing solutions are known to date.

**Supplementary Table S8.1 | Manufacturing requirements and challenges: CMOS vs. Stacked CMOS vs. Skybridge**

| | CMOS | | Stacked CMOS [3][4][5] | | Skybridge | |
|---|---|---|---|---|---|---|
| | **Requirements** | **Challenges** | **Requirements** | **Challenges** | **Requirements** | **Challenges** |
| **Lithography** | Determining factor for scaling; defines channel length, contact, interconnect, and via | Aberrations in light source; variation prone; design rule explosion; costly | Same as CMOS | Same as CMOS | Precision only for nanowires; interconnect definition relaxed | Uniform nanowire pattern definition |
| **Doping** | High precision for complementary doping | Very difficult to maintain uniform doping gradients across the die | Same as CMOS | Same as CMOS | Doping required only once; Single type uniform across the die | --- |
| **Patterning** | Complex shapes: zigzag patterns and different dimensions | Increasing variation | Same as CMOS | Same as CMOS | High aspect ratio nanowires | Pattern density |
| **Deposition** | Interconnect, Via material filling | Processing temperature in gate-first process | Same as CMOS | Same as CMOS | Transistor, contact definition, and interconnect filling | --- |
| **3-D Photoresist Structures** | --- | --- | --- | --- | Used for selective deposition | Maintaining photoresist thickness |
| **Planarization** | CMP after each deposition layers | Corrosions in metal; rigidity | Same as CMOS | Same as CMOS | Etch-back oxide or novel material | --- |
| **Alignment and Registration** | Layer by Alignment, and registration offset at different layers | Lithographic precision dependent | Same as CMOS | Same as CMOS | Same alignment and registration across all layers | Lithography dependent |
| **Thermal Annealing** | --- | --- | For crystallizing each deposited Silicon layer [4] | High temperature affects material structures | --- | --- |
| **Through Silicon Vias** | --- | --- | Coarse grain [3] die-die TSVs; fine grain layer-layer TSVs[4] | Misalignment; uniform material filling; Relatively new process | --- | --- |
| **Thinning and Bonding** | --- | --- | Processed Wafer/Die thinning for bonding | Die-bond related issues [3]: Die stress, crack formation, Die to Die misalignment | --- | --- |

### *8.1 Experimental Validation of Junctionless Nanowire Transistor*

As outlined earlier, we have experimentally validated the Junctionless device concept and demonstrated key steps for Skybridge assembly. Our clean room device validation work involved co-exploration of process/device simulations, and experimental metrology to optimize process steps. Initial process parameters were derived by simulating the actual process flow; SRIM, Synopsis Sentaurus Process and Device simulators were used for this purpose. Direct pattering with Electron-beam lithography (EBL) was used for experimental prototyping.

#### *8.1.1 Process and Device Simulations*

In our process and device simulation framework, Stopping Range of Ions in Matter (SRIM) simulator [61] was used to extract ion implantation parameters, Sentaurus Process [9] was used to create device structures simulating the actual process flow and Sentaurus Device [10] was used to model carrier transport in these device structures. These simulations provided realistic insight on implications of materials, and process and device parameter choices for experimental prototyping.

In Junctionless transistors, channel conduction is modulated by the workfunction difference between the channel and the gate; the nanoscale dimension of the channel is fundamental for its operation. The surround gate structure in Skybridge's V-GAA Junctionless transistor offers maximum gate to channel electrostatic control for the 16nm diameter vertical nanowire channel. For demonstration purposes, to achieve similar device operation, we have used an SOI wafer, where the top device silicon layer was thinned to 15nm. The buried Oxide layer in SOI wafer ensured that there are no leakage paths, and sufficient gate control is achieved over the horizontal nanowire channel.

The same SOI wafer configuration was used in Process and Device simulations. The wafer had a 100nm thick top device layer (Si), 378nm middle buried oxide ($SiO_2$) layer and 500um bottom handle layer (Si). Fig. S8.1A shows Ion (B+) distribution plot obtained from SRIM on this SOI wafer. The acceleration voltage (28 KeV) used in SRIM simulations, obtained from stopping range table for Boron dopants and silicon substrate, was chosen such that the bottom 20nm of the top Si layer had maximum doping concentration. In order to identify the annealing temperature for substrate recrystallization and to create a device structure for simulations with realistic process assumptions, ion implantation parameters (acceleration voltage 28KeV, implant dosage 1e14 atoms/$cm^2$) obtained from SRIM was used in Sentaurus Process simulations. Several process conditions were simulated to identify parameters for annealing. Substrate annealing at 1000° C, for 60 minutes in $N_2$ ambient was found to be adequate for substrate recrystallization, and diffusion and activation of dopants. Fig. S8.1B shows uniform dopant distribution in the top silicon layer after annealing. The ion implantation process was modeled using Monte Carlo (TRIM) simulation model [9]. Diffusion and activation processes were modeled using Charged Cluster model [9].

The doped substrate was then used to create tri-gated junctionless nanowire FET device structures in Sentaurus Process. The device creation process involved steps that are similar to Skybridge's wafer-scale process flow (Fig. S8.2)- i) substrate thinning from 100nm to 15nm, ii) nanowire patterning, iii) mask to define gate region, iv) $HfO_2$ gate oxide deposition, v) gate material (Ti) deposition, and vi) Al source, drain contact formation.

The device structure created from Sentaurus Process was used to simulate electrical properties of junctionless nanowire transistor using Sentaurus Device simulator. Carrier transport was modeled using Hydrodynamic charge transport model with density gradient quantum corrections [10] to take into account quantum effects at nanoscale. Secondary scattering effects were also taken into account. Simulations were done for various device configurations; gate oxide, channel width and channel length were varied, doping concentration, channel thickness were kept the same at 1e19 dopants/cm$^3$ and 15nm respectively. Fig. S8.1C shows $I_d$-$V_{gs}$ characteristics for different gate oxides; 1nm $HfO_2$ shows superior characteristics with $I_{ON}/I_{OFF} \sim 10^7$ compared to 3nm $SiO_2$, 1nm $SiO_2$ and 3nm $HfO_2$, which is primarily due to stronger electric field due to thinner $HfO_2$ high-*k* dielectric. Fig. S8.1D and Fig. S8.1E shows simulated $I_D$-$V_{GS}$ characteristics for different channel widths and channel lengths. Clearly, nanowire FETs with narrower channels and longer gate lengths show better characteristics ($I_{ON}$ ~ 30uA, $I_{OFF}$ ~ 5pA) due to higher electrostatics of the metal gate over channel.

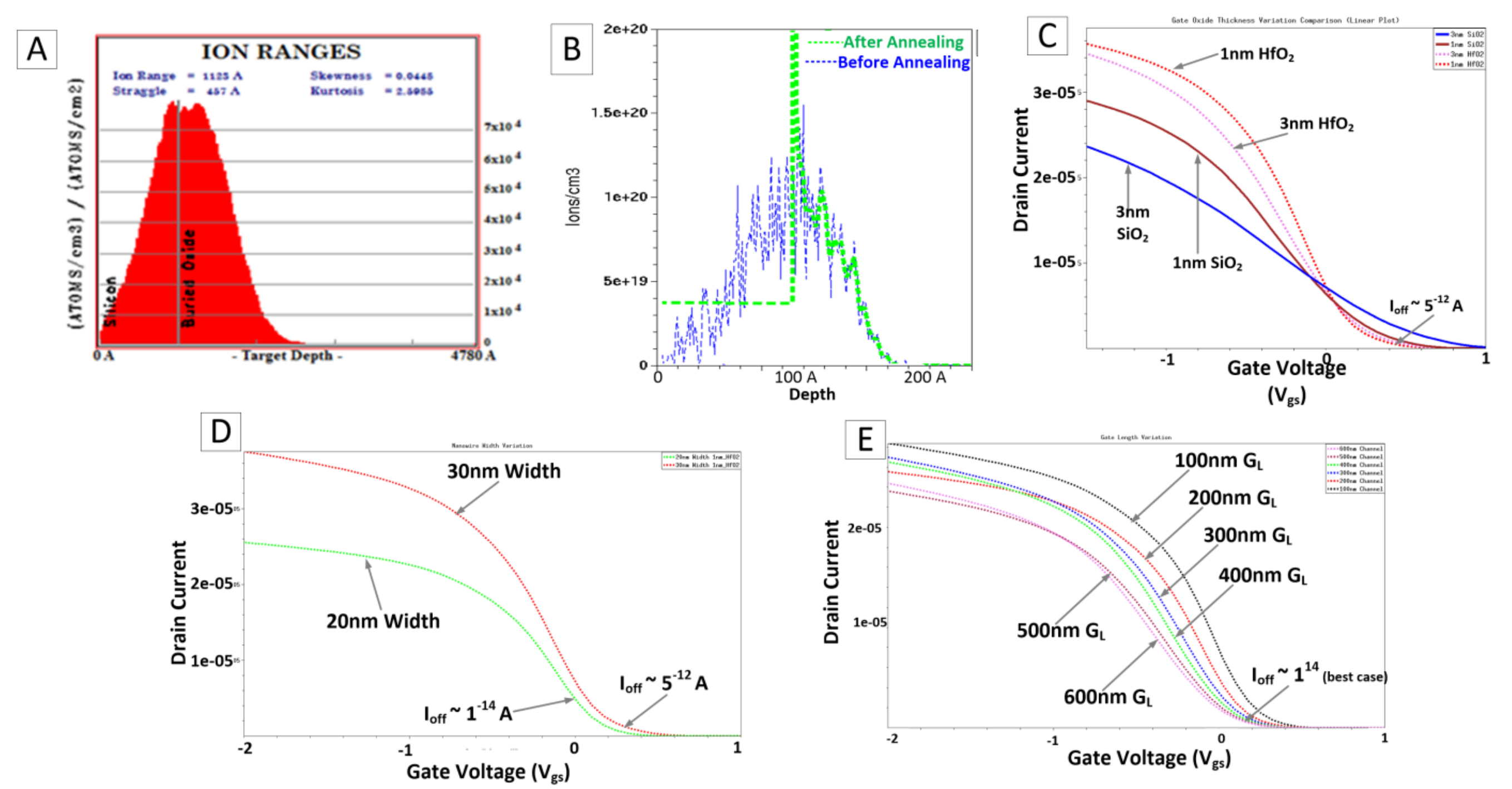

**Fig. S8.1 | Process and Device simulation results**. A) SRIM simulation plot showing ion distribution in SOI wafer for 28KeVimplant, B) Sentaurus process simulation plot showing ion distribution in SOI wafer before and after thermal annealing at 1000° C, D) $I_d$-$V_{gs}$ curve showing variations due to gate oxide choice, E) $I_d$-$V_{gs}$ curve showing impact of nanowire channel width, D) $I_d$-$V_{gs}$ showing the effect of gate length on drain current.

### *8.1.2 Experimental Process Flow for Device Fabrication and Metrology*

This prototyping was based on direct patterning of silicon nanowires on Silicon-on-Insulator (SOI) substrates using Electron-Beam Lithography (EBL). The prototyping approach used is shown schematically in Fig.S8.2. The starting material is an SOI wafer (Fig.S8.2A) where the top device layer is doped with P+ dopants (N-type devices would follow similar methodology). The ion implantation and annealing steps for uniform doping of Si device layer was carried out using simulated process parameters (Acceleration voltage: 28KeV, dosage: 1e14 dopants/cm$^2$, Implant tilt: 7 degrees, Annealing Temperature: 1000° C, Annealing Duration: 60min, Annealing Ambient: $N_2$). The implantation was such that initially the bottom 20nm of the top Si layer had maximum doping concentration in the order of 1e19 dopants/cm$^3$ (Fig.S8.2B). After doping step, the substrate was thinned down to 15nm with anisotropic RIE using $SF_6$+$CHF_3$ etch recipe (Fig.S8.2C). Using EBL and PMMA resist, contact pads and alignment markers were patterned, and were followed by Ti (5nm) and Au (25nm) deposition by E-beam Evaporator (Fig.S8.2D). Using these alignment markers, sub-30nm nanowire features were patterned between contact pad extensions, followed by Ni evaporation and liftoff to define Ni features on top of the substrate (Fig.S8.2E). The Ni features acted as an etch mask for defining nanowires on the SOI. Anisotropic RIE using $SF_6$ + $CHF_3$ mixture was then used to etch the surrounding Si, followed by Piranha (3:1 $H_2SO_4$:$H_2O_2$) treatment to remove Ni etch mask. This resulted in Silicon nanowires directly patterned on the SOI substrate (Fig.S8.2F). Nanowires at widths as small as 30nm, 20nm and 15nm have been successfully demonstrated using this approach. Atomic layer deposition technique was used for Halfnium oxide ($HfO_2$) deposition (Fig.S8.2G), followed by alignment, patterning, evaporation and liftoff to define metal gate (Fig.S8.2H). Material selection and thickness parameters for gate oxide and gate metal were as derived from process and device simulations.

Extensive metrology was done after each process step to verify expected results. Four point probe measurements were carried out to determine doping concentration in Silicon substrate after ion implantation and were found to be ~8 x $10^{18}$ dopants/cm$^3$, which was in range of the desired concentration ($10^{19}$ dopants/cm$^3$). Atomic Force Microscopy (AFM) measurements were done to determine the surface roughness and the silicon thickness after substrate thinning and pattern transfer steps. Substrate thinning and nanowire patterning results are shown in Fig.S8.3A and Fig.S8.3B. As shown in Fig.S8.3A, thinned Si substrate had less than 1nm of surface roughness variation after anisotropic etching of top SOI layer from 100nm to 15nm. Fig.S8.3B shows AFM image of a 15nm thick patterned Silicon nanowire on top of $SiO_2$ substrate.

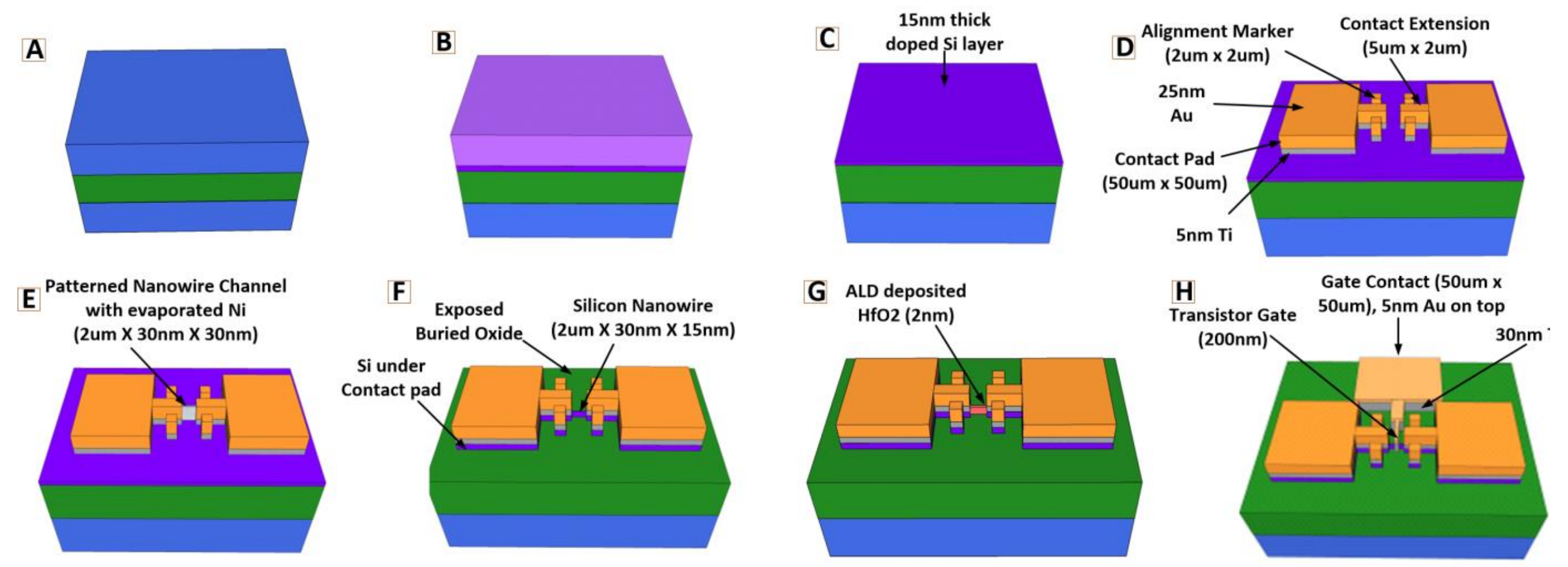

**Fig. S8.2 | Experimental process flow for nanowire junctionless device.** A) SOI wafer as starting wafer; 100nm Si device layer (top), 378nm buried Oxide layer (middle), and 500um Si handle layer (bottom). B) Ion implantation and annealing. C) Substrate thinning to 15nm using RIE. D) Contact pad and alignment marker formation. E) Patterning of Nickel feature. F) Nanowire pattern transfer. G) ALD $HfO_2$ deposition. H) Gate formation and gate material depositions.

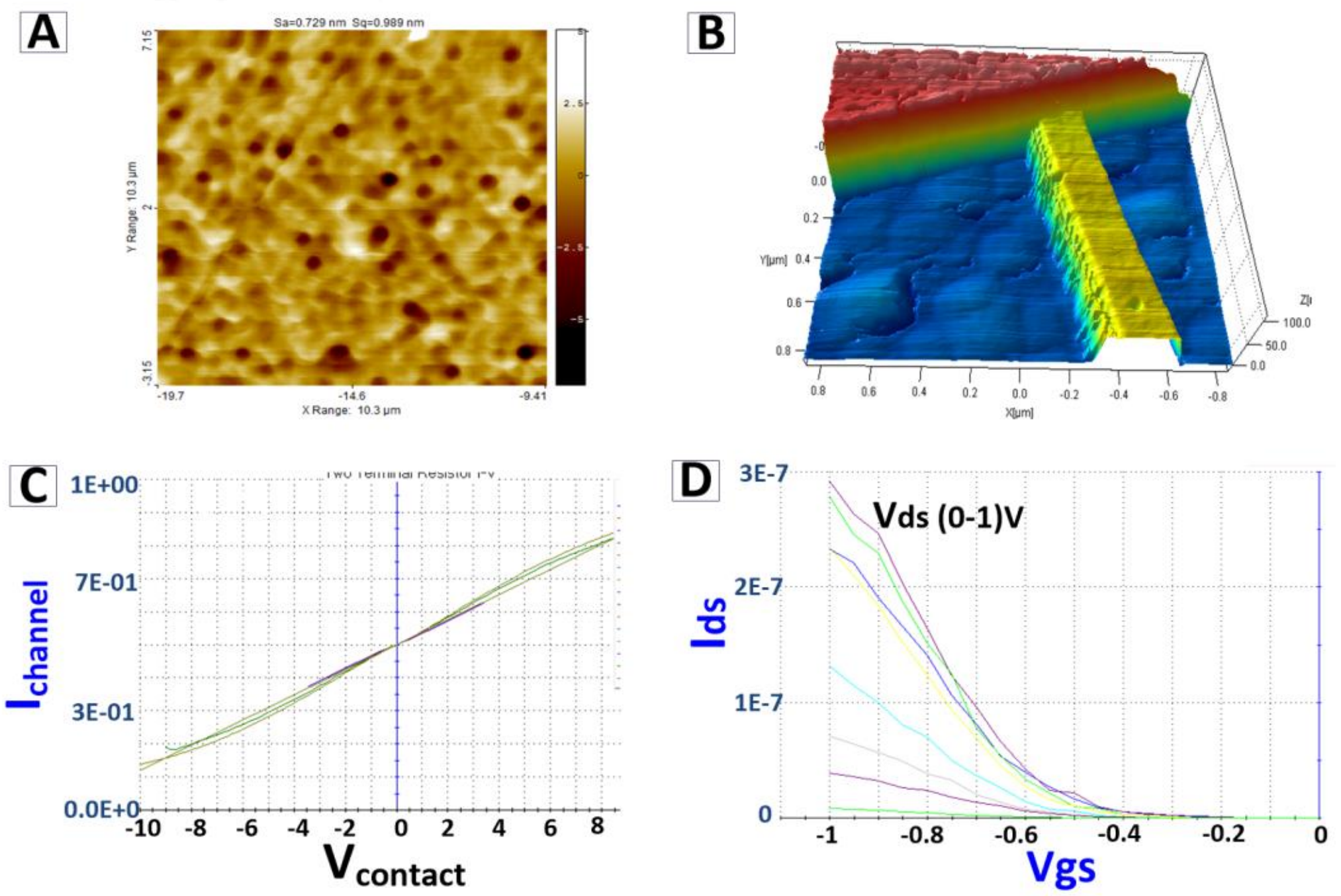

**Fig.S 8.3 | Experimental results.** A) AFM results: less than 1nm surface roughness after RIE thinning, B) 15nm thick Si nanowire on top of SiO2 substrate. C) I-V measurements of nanowire channel showing linear increase in current for wide range of voltages. D) Id-Vgs characteristics of fabricated p-type junctionless nanowire FET, the device is normally OFF at 0Vgs, turns ON fully at -1Vgs.

I-V measurements were carried out on individual junctionless nanowire FETs to characterize electrical properties. In order to determine ON current and contact resistivity in junctionless FETs, two point probe I-V measurements were done on nanowire channels, which were patterned in between source and drain contacts. Excellent Ohmic behavior was achieved from source/drain contacts (contact metal stack: 5nm Ti + 30nm Au), since the underlying substrate was heavily doped. Fig.S8.3C shows I-V characteristics of heavily doped nanowires with source/drain contacts, the gate voltage was varied from -10V to +10V and linear increase in current was observed. Ellipsometry was done to determine $HfO_2$ thickness after atomic layer deposition at 150° C. We were able to deposit and measure $HfO_2$ films down to 1nm, and the thickness was found to be uniform across the die.

Three point probe measurements were done on junctionless nanowire FETs. Dimensions for fabricated devices were 30nm wide and 15nm thick nanowire channel, 2nm thick $HfO_2$ gate dielectric, 200nm long gate and 50nm thick gate metal stack. A stack of 30nm Titanium layer and 20nm thick Gold layer served as gate metal stack. Fig.S8.3D shows $I_D$-$V_{GS}$ characteristics of p-type junctionless nanowire FETs when a metal gate stack was put on top of silicon nanowire channel. The $I_D$-$V_{GS}$ characteristics in Fig.S8.3D accurately depicts junctionless device characteristics, where the workfunction difference between Titanium/Au gate and P+ doped Silicon nanowire channel depletes the channel and the device is normally *OFF* at $V_{GS}$ = 0V. With the application of negative gate voltages ($V_{GS} < V_{TH}$), the carriers accumulated and the channel conduction was maximum. These devices had an $I_{ON}/I_{OFF} \sim 10^4$ and threshold voltage ~ -0.3 V.

### 8.2 *Bottom-Up Evaluation Using Process, Device and Circuit Level Simulations*

This has been discussed extensively in previous sections. The following overview is provided for completeness. Skybridge's circuit evaluation was done with a bottom-up methodology considering materials, processes, nanoscale geometries, nanoscale carrier and heat transport, 3-D circuit placement, 3-D layout and 3-D interconnects. A set of simulation tools: SRIM, Sentaurus Process, Sentaurus Device and HSPICE were used for this purpose. Sentaurus Process and SRIM simulation tools were used extensively to emulate actual wafer-scale process steps for Skybridge fabrication. Process parameters from experimental prototyping (Section 8.1) were used, and detailed process modeling accounting for material properties, nanoscale dimensions, and defects was done.

Sentaurus Process created device structures were used in Sentaurus Device simulations to characterize electrical properties. Carrier transport in these structures was modeled using advanced charge transport models with quantum corrections; secondary effects due to high doping, surface and coulomb scattering were also considered.

In order to capture the effect of experimental process parameters, material choices and nanoscale dimensions on large-scale circuit simulations, we developed a HSPICE compatible V-GAA Junctionless transistor model using TCAD Process and Device simulated device characteristics under various assembly assumptions. The device model was then used in circuit simulations in conjunction with extracted interconnect RC parameters from 3-D circuit layout. The 3-D layout design followed Skybridge specific design rules and guidelines that were derived from the manufacturing pathway, and are in accordance to ITRS defined lithographic guidelines [46].

This evaluation methodology was extended for all Skybridge circuit designs including: 4, 8 and 16-bit arithmetic circuits, Skybridge RAM and WISP-4 microprocessor.

### *8.3 Skybridge's Wafer-Scale Manufacturing Pathway*

This sub-section details the wafer-scale manufacturing pathway for Skybridge fabric. List of materials used in this manufacturing flow, dimensions, properties and applications in Skybridge are outlined in Table S5.1 and S9.1.

*Starting Wafer*

The starting wafer for Skybridge assembly is a customized highly doped silicon wafer. As shown in Fig. S8.4A, at the bottom of the wafer is bulk silicon, which can be connected to the package heat sink through backside metallization and bonding substrate; on top of bulk silicon are islands of $SiO_2$, which serve the purpose of electrically isolating the silicon nanowire pillars from the bulk; a layer of crystalline silicon is deposited on top and doped (concentration ~ $10^{19}$ dopants/$cm^3$; see Section 2.1 for doping requirements), which completes the wafer preparation process. Noticeably, doping is required only once *prior* to any processing steps.

*Nanowire Template Patterning*

Patterning of arrays of high aspect ratio vertical nanowires is the next step in the manufacturing flow. All the nanowires have similar aspect ratio, and they maintain uniform distances between each other. The nanowire patterning is done such that nanowires are patterned alternately on top of horizontal $SiO_2$ islands, and a group of nanowires are patterned on top of horizontal $SiO_2$ lands at sparse intervals (Fig. S8.4B, Fig. S1.1E). This is done to isolate input/output signal carrying pillars and large area VDD signal carrying pillars from shorting the bulk silicon and creating undesired latch-up conditions.

High aspect ratio uniform vertical nanowires with smooth surfaces can be achieved through different processes such as patterning with oxidation and etch back technique, Inductively Coupled Plasma (ICP) etching, etc. Several research groups have demonstrated high aspect ratio nanowires that are in line with Skybridge's requirements in the version evaluated in previous section (as we will show, these can be further relaxed while still providing significant benefits). Yang et al. in [48] have demonstrated 20nm wide, 1µm tall (1:50) nanowires using oxidation and etch back techniques, while in [47], Mirza. et al., demonstrated nanowires of various widths ranging from 30nm to 5nm with very high aspect ratios, the highest aspect ratio being 1:50, using plasma etching technique. In addition, these nanowires were shown to withstand processing conditions for Gate–All-Around (GAA) vertical transistor formation [48].

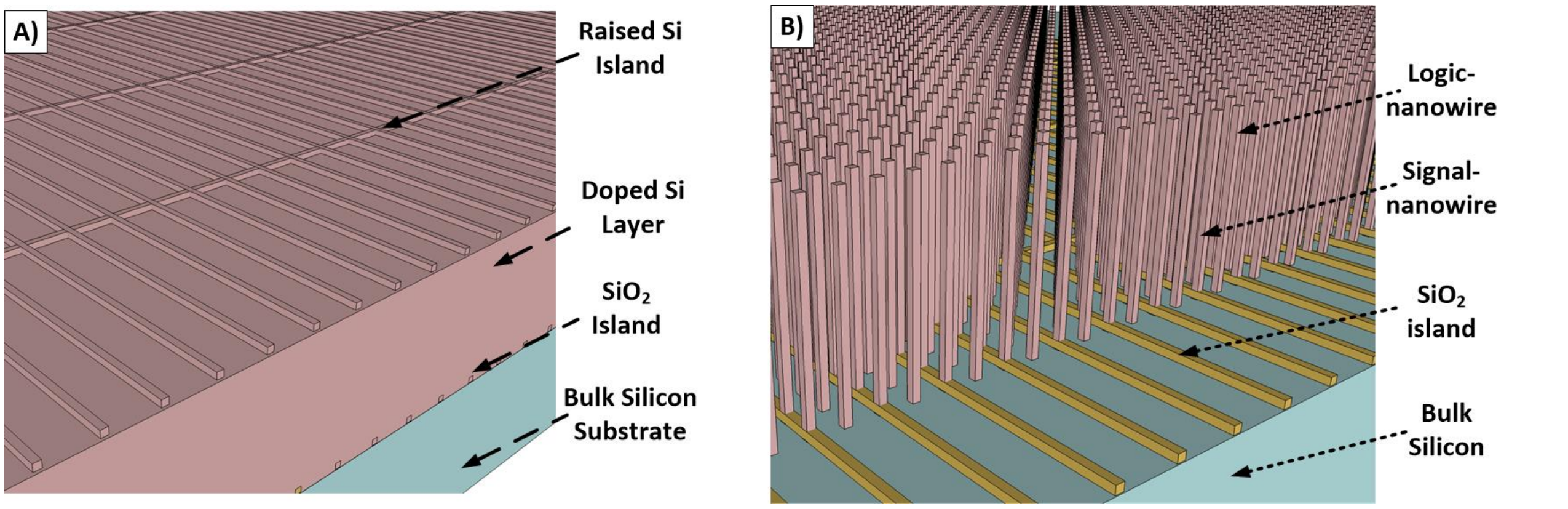

**Fig. S8.4 | Starting wafer and nanowire patterning.** A) Bulk silicon wafer with $SiO_2$ islands and doped silicon layer on top; B) High aspect ratio nanowire patterning with lithography; signal-nanowire pillars are isolated from bulk silicon by $SiO_2$ islands, whereas logic-nanowires connect directly with the bottom bulk.

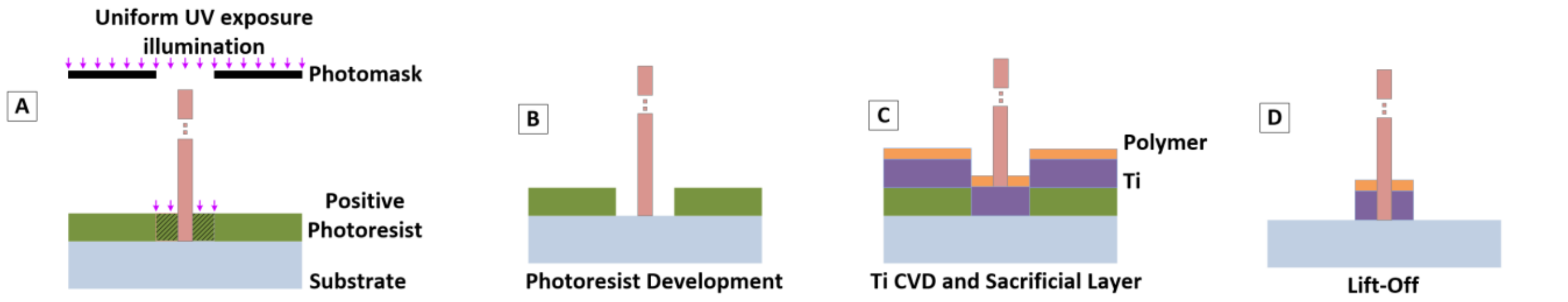

**Fig. S8.5 | Contact formation:** Ohmic contact for power rail (i.e., GND) is made using A) photoresist spinning and UV exposure; B) Photoresist development in developer solution; C) Ti deposition for Ohmic contact, followed by sacrificial polymer deposition; D) Metal lift-off.

*Contact Formation*

Nanowire patterning is followed by a contact formation step for connecting the nanowire with power rails at the bottom. Ohmic contacts at different heights are also formed for input/output and power rail (VDD, GND) connections. In order to make an Ohmic contact, first a region surrounding the nanowire is exposed using UV lithography (Figs. S8.5A, S8.5B); the region of exposure is determined by the minimum material dimension requirements for the Ohmic contact. Ti, a widely used material for Ohmic contacts to heavily-doped n-silicon, is chosen for this purpose. The required Ti thickness and length are derived from 3-D TCAD simulations (see Section 2.1). The UV exposure step is followed by anisotropic Ti deposition (i.e., no step coverage on the side of nanowire, see Fig. S8.5C). Next, a layer of sacrificial polymer [55] is deposited or spun on top of the Ti layer followed by a lift-off process (Fig. S8.5D). During lift-off, the photoresist is removed along with the material deposited on top.

*VDD/GND/Output Signal Carrying Bridges*

In Skybridge, signals are carried from one nanowire to another through *Bridges following* circuit requirements. The manufacturing flow for these Bridges differs depending on their placement (e.g., input signal carrying Bridges connect to transistor gates while output/power signal Bridges connect to logic gate output/power rail contacts).

Fig S8.6 shows the manufacturing steps required to form Bridges that connect to Ohmic contacts. After Photoresist spinning, the lithographic pattern for interconnection is created by UV exposure (Fig. S8.6A) and photoresist development (Fig. S8.6B). Noticeably, the exposure is such that it overlaps previously created Ohmic contacts (Fig. S8.6D) by a small portion; this is done to ensure proper metal-metal contact. After exposure and photoresist development, Tungsten (W) is deposited anisotropically (Fig. S8.6C) using CVD [50]. Tungsten has excellent electrical and thermal properties, and is widely used in industry today as Metal1 and Via filling material. This step is followed by a lift-off process (Fig. S8.6D) and polymer removal step (Fig. S8.6E), removing excess material.

The anisotropic metal deposition requirements for Contact and Bridge formation can be met by optimizing PVD or CVD processes. Various research groups have demonstrated anisotropic material depositions; Urakawa et. al. demonstrated anisotropic deposition in high aspect ratio trench using H assisted RF CVD [58], whereas Marathe et. al. showed anisotropic trench filling methods by controlling target height in sputtering system [59]. Similarly, we envision that anisotropic metal deposition can be achieved for Skybridge's manufacturing, since the manufacturing flow uses most commonly used metal and metal alloys such as Ti, TiN, and W, which are usually deposited using CVD and PVD systems.

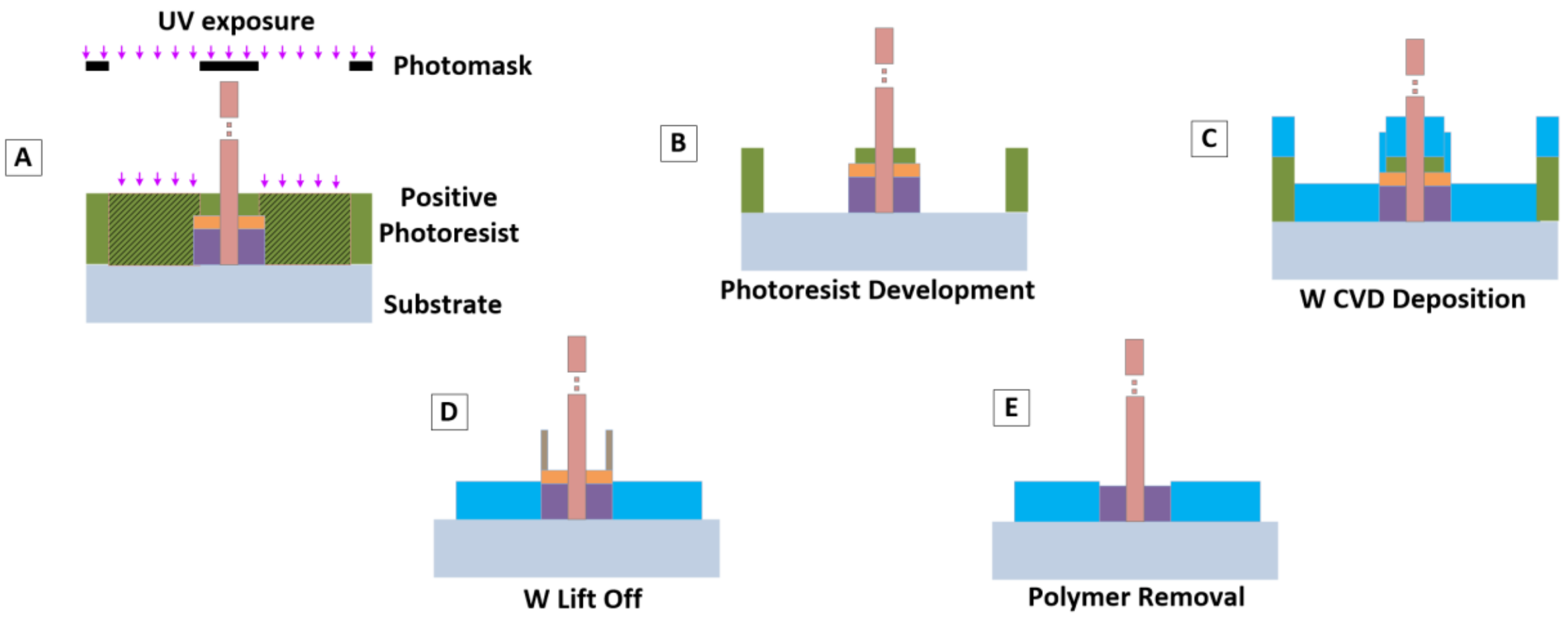

**Fig. S8.6 | Formation of VDD/GND/Output signal carrying Bridges:** A) Photoresist spinning and UV exposure to define regions for Bridges; B) Photoresist development; C) Anisotropic deposition of Tungsten (W) using CVD; D) W lift-off; E) Sacrificial polymer removal to get rid of excess metal.

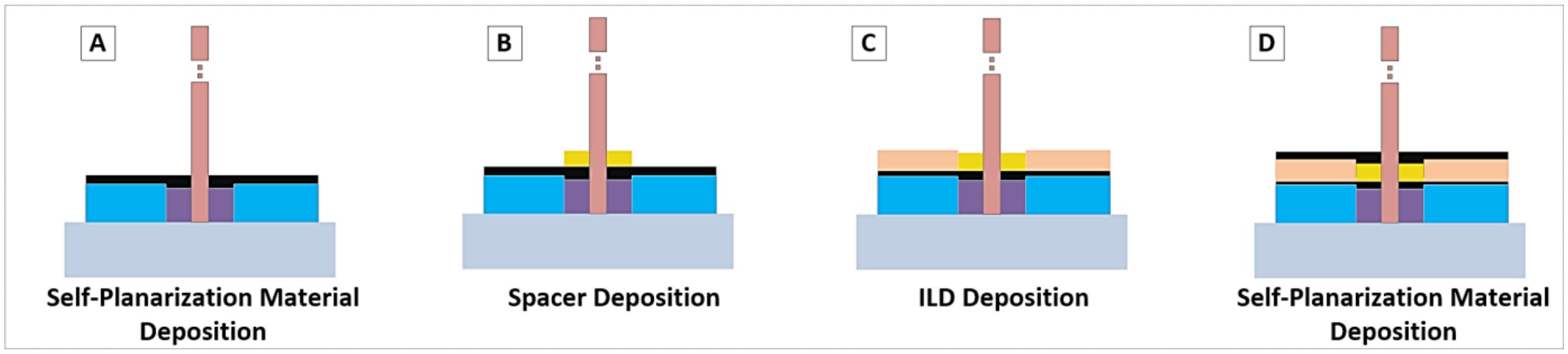

**Fig. S8.7 | Planarization and interlayer dielectric deposition:** A) Self-planarization material deposition to planarize surface; B) Spacer deposition using UV exposure (like Fig. S2); C) ILD (i.e., C-$SiO_2$) deposition (like Fig. S3); D) Self-planarization material deposition.

*Planarization, Interlayer Dielectric Deposition*

Planarization after deposition is an important step since non-planar surfaces cause lithographic focus imbalance, and alignment errors, which can easily result in causing distortion in printed features. Planarization with chemical mechanical polishing is avoided in this Skybridge manufacturing flow to prevent structural damage to standing single crystal vertical nanowires. Alternative planarization techniques such as etch back planarization [53] or self-planarization materials [52] can be used for the same purpose. In etch back planarization techniques [59], generally oxide or multilayer photoresists are deposited first and are gradually etched back. In self-planarization material based technique [52], novel materials are deposited to planarize the surface. Both these techniques are well established and are in commercial use.

In this manufacturing flow we describe the usage of self-planarization materials. These materials planarize themselves regardless of the underlying topology. For example, Fig. S8.7A shows the resultant planarized surface after a self-planarization material is applied; the top surface is plane and smooth even though there is variation in the height of underlying features. This step is followed by spacer (Fig. S8.7B) and interlayer dielectric (C-$SiO_2$, dielectric constant 2.2 [54]) deposition (Fig. S8.7C). After these steps, the surface is expected to be planarized as shown in Fig. S8.7D.

*Gate Stack Deposition*

3-D photoresist structure creation for selective gate material deposition is a key process step in Skybridge assembly. High-*k* gate dielectric and gate metal are deposited in selective places of the nanowires using 3-D resist structures. Several techniques were demonstrated over the years to define 3-D resist structures. In [60] No. et. al., developed diffuser lithography based technique for 3-D structures, and in [55], Yun et. al. used two types of photoresists with different solubility to create 3-D structures. In diffuser lithography, photoresist is exposed in an angular manner to create 3-D shapes. Whereas, in the dual photoresist method, easily soluble resist is spun at the bottom and relatively less soluble resist is put on top; subsequently both photoresists are exposed and development time is controlled such that the bottom resist is washed away, leaving a 3-D structure.

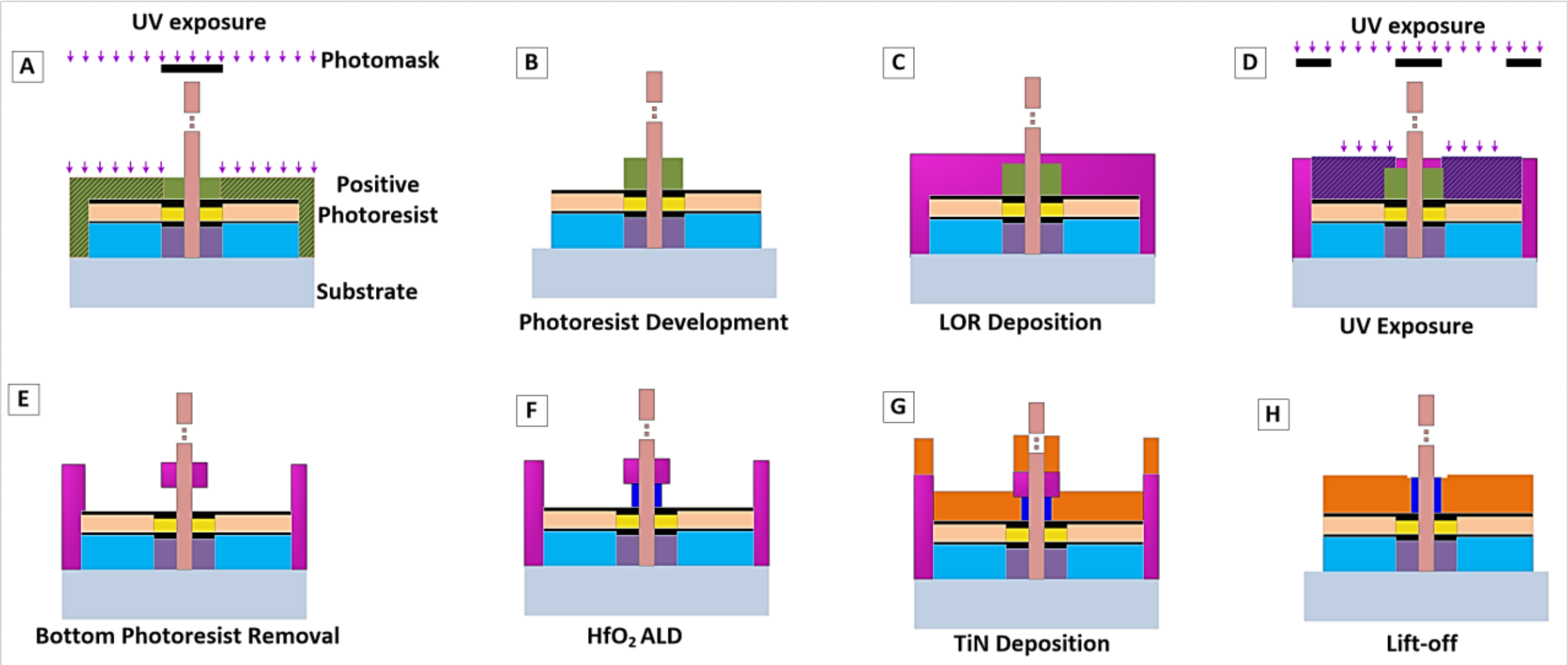

**Fig. S8.8 | Gate stack deposition:** A) Photoresist spinning and UV exposure; B) Resist development; C) Low resolution lift-off Resist deposition; D) Second UV exposure; E) Controlled resist development to remove first Photoresist; F) $HfO_2$ deposition using ALD; G) TiN deposition using CVD; H) Metal lift-off.

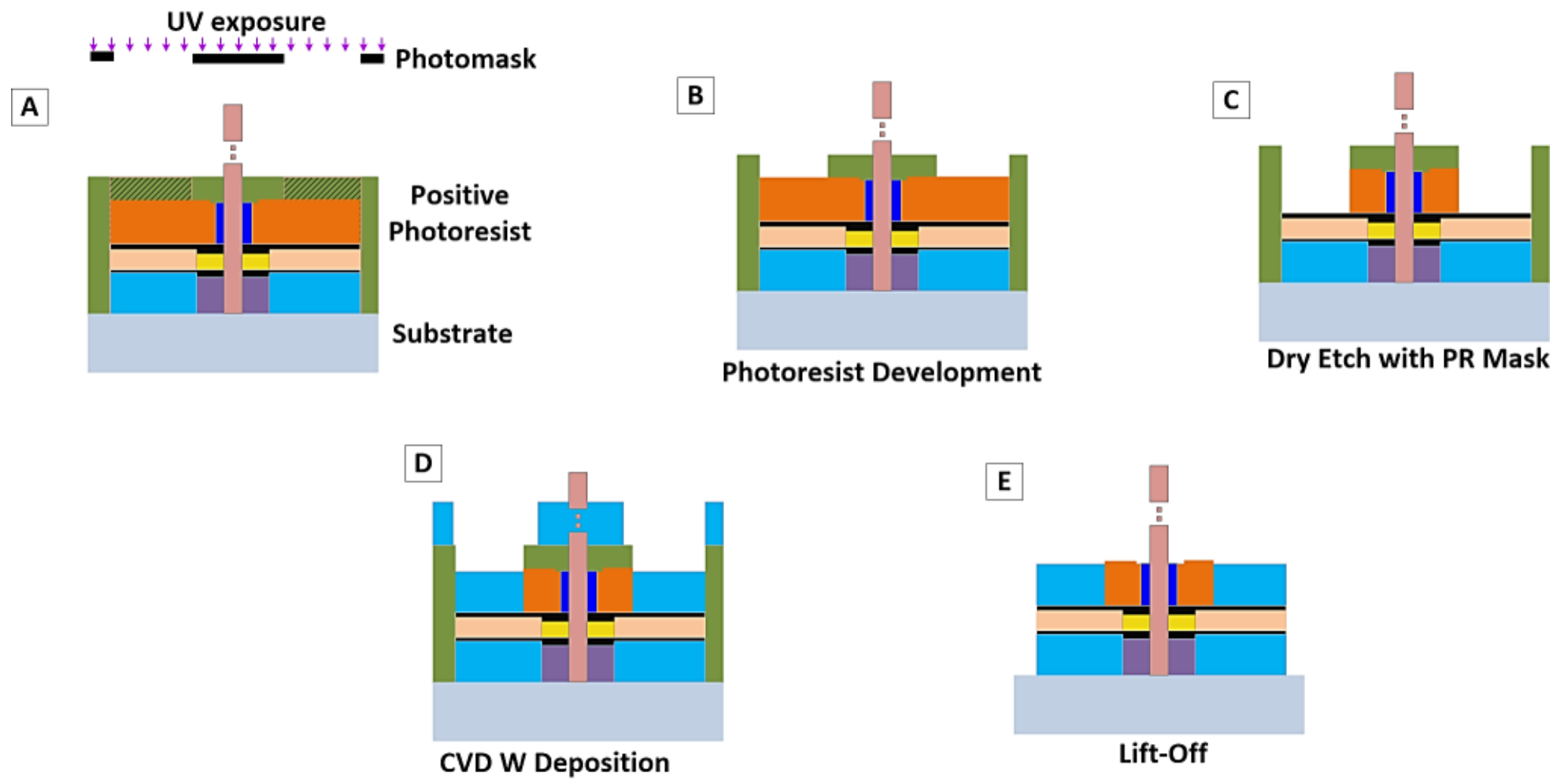

**Fig. S8.9 | Formation of input signal carrying bridges:** A) Photoresist spinning and UV exposure for Bridges; B) Photoresist development; C) TiN dry etch using photoresist as etch-mask; D) anisotropic deposition of W using CVD; E) metal lift-off.

In this manufacturing flow we employ the dual photoresist based technique for 3-D photoresist structures. These structures are used for gate oxide and gate electrode deposition. Both deposition steps use the same lithographically defined pattern. In the beginning, 16 nm thick (requirement per 16-nm V-GAA Junctionless transistor channel length) standard Photoresist is spun and is followed by UV exposure (Figs. S8.8A, S8.8B) to create the desired pattern for selective deposition. Next, a thicker layer of low resolution Photoresist is spun on top (Fig. S8.8C) and UV exposure is done (Fig. S8.8D). During the Photoresist development step (Fig. S8.8E) one standard resist develops faster than the other, and by controlling resist development time 3-D Photoresist shapes can be formed. After creating 3-D structures with Photoresist, the gate stack is deposited. $HfO_2$ is deposited using Atomic Layer Deposition (ALD) (Fig. S8.8F); in this step, $HfO_2$ is deposited only on the exposed Si surface. Next, TiN is deposited anisotropically using CVD [49] (Fig. S8.8G). Gate stack material choices are specific to V-GAA Junctionless devices, and are derived from 3-D TCAD simulations (see Section 2.1). The last step in this process is lift-off (Fig. S8.8H) to remove the excess material on top of the Photoresist.

*Input Signal Carrying Bridges*

Manufacturing steps for input signal carrying Bridges begin with Photoresist spinning and lithographic exposure (Figs. S8.9A, S8.9B). Next, TiN from the exposed region is etched away using dry etch with Photoresist as etch-mask (Fig. S8.9C). Afterwards, Tungsten (W) is deposited anisotropically on the exposed region (Fig. S8.9D). This step is followed by a W lift-off process (Fig. S8.9E).

Other Bridge structures such as Heat Extraction Bridges, routing Bridges follow similar methodology for fabrication.

*Mask Registration and Alignment*

Maintaining alignment precision in multiple layers of processing is a critical requirement, and is different from the CMOS alignment methodology. In CMOS, new alignment markers are created after each layer of processing; these new markers are larger in dimensions compared to previous ones to accommodate Mask Registration offset. In contrast, the same alignment markers can be used in all layers of processing for Skybridge; they are created at the very first step, during nanowire patterning. Different Mask Registration with respect to same alignment markers allow features to be built with same alignment precision across multiple layers. The approach is illustrated in Fig. S8.10 – alignment markers on the periphery of a die are shown to the have same height as the nanowires. This alignment methodology is unique to Skybridge, and is enabled due to aforementioned manufacturing flow, which does not require mechanical planarization processes.

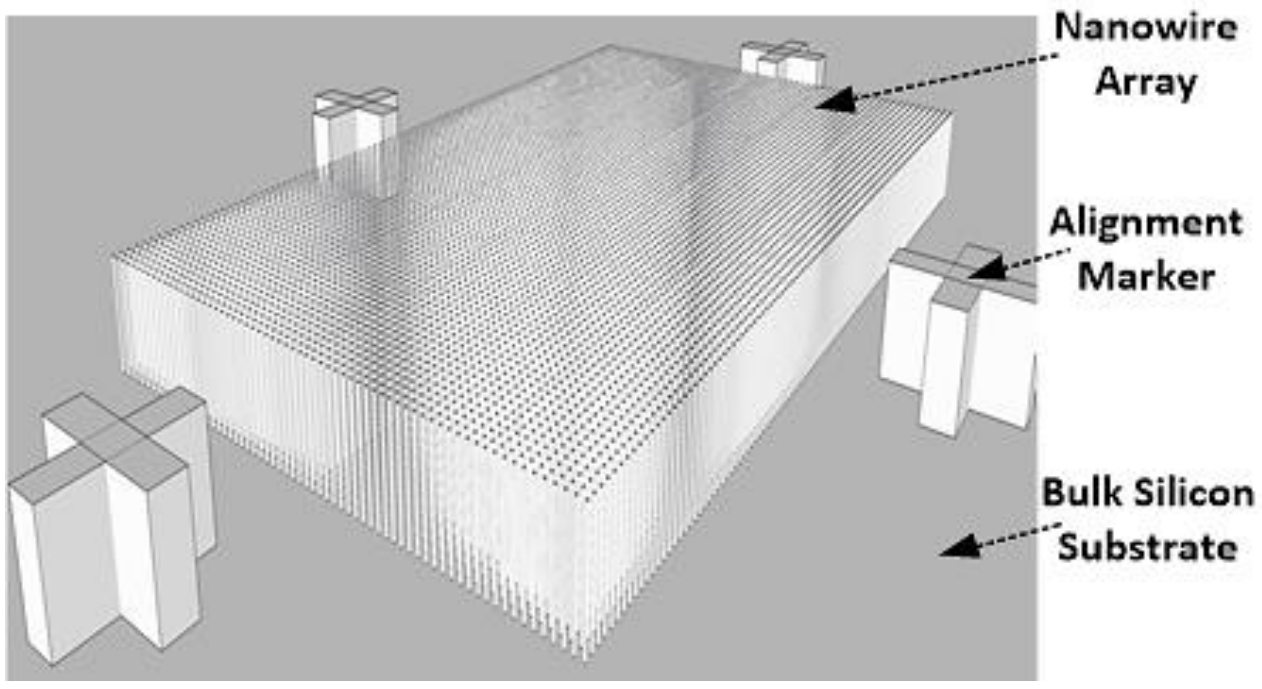

**Fig. S8.10 | Alignment.** Skybridge alignment step using same alignment markers for Mask Registration across all layer of processing.

*8.4 Sensitivity Analysis for Relaxed Wafer Scale Manufacturing Requirements/Parameters*

Manufacturability at nanoscale is a key concern for CMOS. Due to stringent lithography requirements and underlying CMOS architecture, manufacturing imprecision is becoming more prominent. As discussed in the previous section, Skybridge's manufacturing flow is significantly different from CMOS, while staying within established processes. It primarily depends on material deposition on uniform arrays of vertical nanowires as opposed to lithography. Deposition processes are known for their precision to a few Angstroms. Hence, we anticipate lithography related variations to be significantly less for Skybridge.

In our prior work [20][21][62][63], we have done detailed variability analysis for nanowire based integrated circuits considering lithographic aberrations and doping fluctuations, and proposed several redundancy based fault-tolerance schemes. In this sub-section, we focus on specific sources of manufacturing imperfections related to vertical nanowire pattern definition and material deposition, and show their implications on benchmarking metrics: area, power and performance. Our analysis captures both lithographic as well as deposition related aspects and addresses the question: how are the benefits of Skybridge affected when the manufacturing requirements are substantially relaxed.

**Supplementary Table S8.2 | Implications of nanowire aspect ratio**

| Region | Power (µW) | Throughput (Ops/sec) | Area ($\mu m^2$) |
|---|---|---|---|
| CMOS | 235 | 9.9e09 | 18.7 |
| SB (AR: 1:54) | 19.4 | 10.4e09 | 0.76 |
| SB (AR:1:27) | 21.9 | 12.8e09 | 1.11 |
| SB (AR:1:13) | 32.4 | 6.7e09 | 2.27 |

*8.4.1 Sensitivity Analysis to Nanowire Aspect Ratio*

Nanowire pitch and aspect ratio are key factors that determine the vertical integration benefits. In Skybridge's circuit design rules and guidelines (Section 9), experimentally demonstrated nanowire aspect ratios were used, and worst-case lithographic tolerance along with minimum material geometries for reliable circuit operation were considered for nanowire pitch calculations. For large-scale circuits (Section 6 and 7), the initial nanowire aspect ratio was 1:54 (16nm width, 868nm height) – accommodating two 8 fan-in logic gates on each nanowire. This is not a requirement and other aspect ratios can be supported. For example, either reducing the number of gates per vertical nanowire or by reducing the fan-in per gate, we can reduce aspect ratio requirements.

To understand the implications of nanowire aspect ratio, we have carried out detailed analysis using 4-bit CLA as an example circuit. The 4-bit CLA was implemented in 1:54 (16nm width, 868nm height), 1:27 (16nm width, 434nm height), and 1:13.5 (32nm width, 434nm height) aspect ratio nanowires, and was compared with CMOS equivalent at 16nm. As shown in Table S8.2, reducing nanowire aspect ratio lowers the density benefit, but it is still very significant compared to CMOS. The 1:27 aspect ratio based design has a higher throughput vs. the 1:54 version. This improvement is mainly due to less number of coaxial routing structures for routing in vertical direction, since only one logic gate is accommodated in each nanowire. The throughput is lower for 1:13.5 aspect ratio primarily due to slower devices at 32 nm (see Section 8.4.4 for device discussions). Active power consumption is significantly lower in all cases; it is 12x, 10x and 7.25x less compared to projected 16nm CMOS for 1:54, 1:27 and 1:13.5 aspect ratio based Skybridge designs, respectively.

*8.4.2 Sensitivity Analysis Considering Nanowire Spacing for Material Deposition*

Anisotropic material deposition is a key requirement for Skybridge's bottom-up assembly. Various groups have already demonstrated anisotropic [57][58] material deposition techniques for trench filling applications with similar requirements. In this sub-section, we evaluate the implications of nanowire spacing enabling relaxed anisotropic material deposition.

Two scenarios were considered: when the spacing is 2x (32nm) and 3x (48nm) larger than ITRS defined minimum feature limits. The design rules for both cases are illustrated in Fig. S8.11A (for detailed Design Rules Derivation please refer to Section 9). The area of 4, 8 and 16-bit CLA designs was evaluated and compared with respect to the nominal case for Skybridge and CMOS at 16nm.

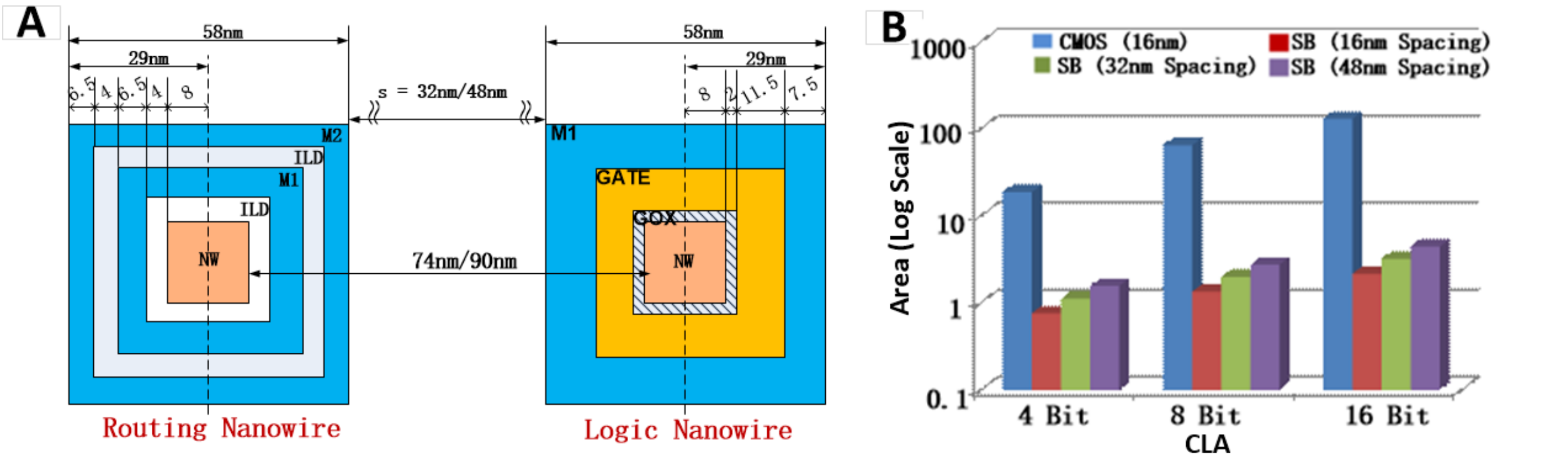

**Fig. S8.11 | Sensitivity to nanowire spacing for material deposition.** A) Design rules accounting for increased nanowire spacing (32nm and 48nm); B) Implications of increased nanowire spacing on Skybridge designs; comparison is shown with respect to nominal case for Skybridge (16nm spacing) and CMOS at 16nm.

As shown in Fig. S8.11B, for all designs the nanowire spacing has a linear effect on area. In the worst case, when the spacing between nanowires is 3x larger (48nm) than nominal (16nm), the area increases by 2x. For 16-bit CLA, the area for 32nm and 48nm spacing are 3.18um$^2$ and 4.41 um$^2$ respectively, which are still 40.9x and 29.5x better compared to equivalent CMOS design at 16nm. These results indicate that due to Skybridge's unique vertical integration, significant benefits can be maintained even when these design rules are relaxed. The relative impact on performance and power is negligible as discussed in sub-section 8.4.3 and 8.4.4.

### *8.4.3 Sensitivity Analysis Considering Nanowire Profile Variation Supporting Structural Rigidity*

In order to achieve high-density arrays of high aspect ratio vertical nanowires, often nanowire features are patterned such that the bottom region is wider than the top. Taper shapes in nanowire profile allow higher mechanical rigidity to withstand processing conditions. In [47], Mirza et. al. reported arrays of high aspect ratio (1:50) nanowires with taper shapes. We have evaluated the effects of such nanowire geometry on Skybridge circuits. The nanowire configuration considered for this sensitivity analysis is shown in Fig. S8.12.

As shown in Fig. S8.12, the bottom of the nanowire is wider than the top. The widths at bottom, middle and top were considered to be 32nm, 22nm and 16nm respectively, which are similar to the experimental demonstration in [47]. An 8 input NAND gate was implemented. The taper shapes resulted in variation in widths for V-GAA Junctionless transistors at different regions; the bottom two transistors had 32nm width, followed by two 22nm and four 16nm wide transistors. Doping for 22nm and 32nm wide transistors were assumed to be $10^{18}$ dopants/cm$^3$ to ensure proper carrier-depletion in channel during OFF state. Fig. S8.13 shows HSPICE simulation results for the NAND gate in the worst-case scenario, when the inputs are all one and the output is switching from "1" to "0". Comparison with equivalent NAND gate with uniform 16nm wide transistors is also shown.

As expected, the NAND gate with varying device geometry is slower than the NAND gate with uniform devices as shown in Fig. S8.13; the delay for the former case is 45ps compared to 34ps with uniform transistors. While with increasing width, transistor performance increases, the OFF current also increases; to ensure $I_{on}/I_{off} \sim 10^5$, the doping concentration at the bottom was lowered than at the top (this does not introduce much additional complexity since it would be done at the wafer level). The performance degradation is mainly due to higher threshold voltage of 22nm devices in the middle, which had the same doping as the bottom 32nm devices to minimize doping complexity in taper geometry. The power consumption is 38% lower for the combined case, primarily due to higher delay. The impact of nanowire profile variation on density is discussed in the next sub-section. In the worst case, for 16-bit CLA design, the area increases by 2x compared to nominal 16nm Skybridge case, which is still 29.5x compared to 16nm CMOS. These results indicate that the benefits of Skybridge design are expected to be maintained even with variations in nanowire profiles.

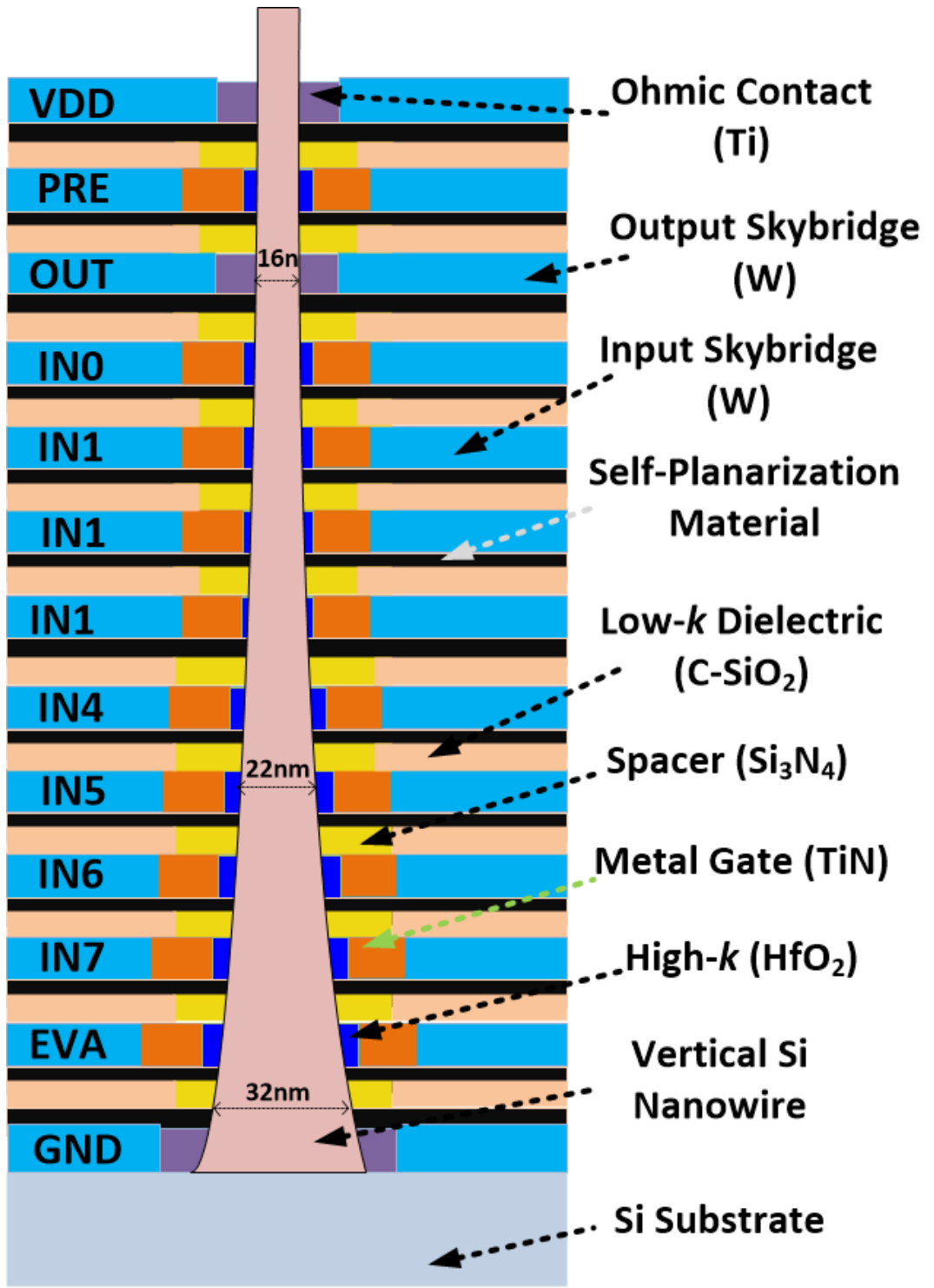

**Fig. S8.12 | Sensitivity to tapered nanowire profile.** Taper geometry to maintain mechanical rigidity, the width is highest at the bottom (32nm) and gradually decreases (middle: 22nm, top: 16nm) to top.

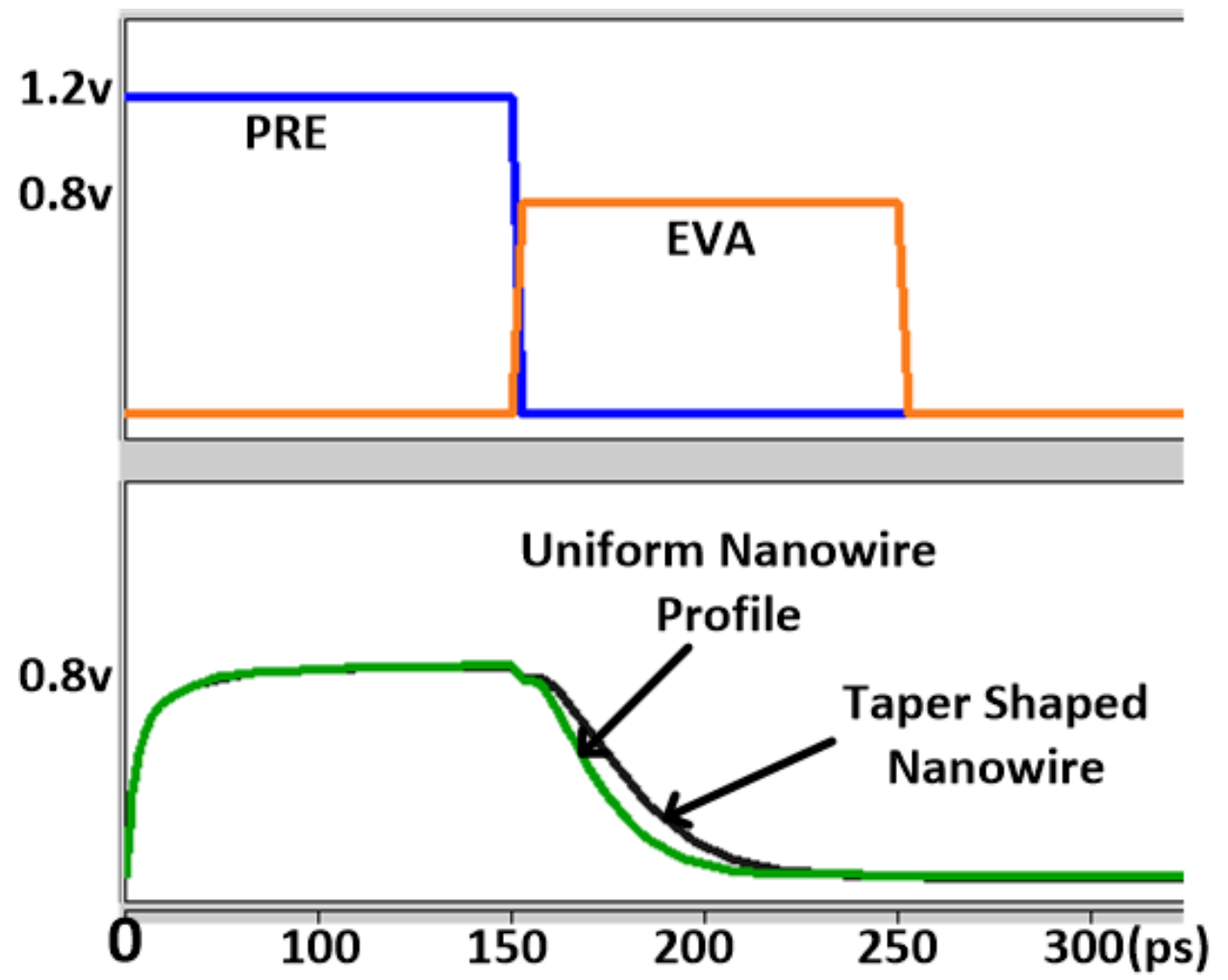

**Fig. S8.13 | Impact of nanowire profile variation.** HSPICE simulation results are shown for 8-input NAND gate, taper shaped nanowire contained two 32nm, two 22nm and four 16nm transistors for inputs at the bottom, middle and top respectively. Comparison is shown with uniform transistor based NAND gate design.

*8.4.4 Sensitivity Analysis – Relaxing Lithographic Precision vs. CMOS; Older Technology Nodes*

Lithography precision in Skybridge is required only to define nanowire patterns. For our current designs, this lithographic precision is set by ITRS defined minimum feature size at 16nm; the manufacturing steps for which is known to date.

The potential of Skybridge is not however limited by minimum feature definition using advanced lithography; significant benefits can be attained even if lithographic precision is relaxed. To quantify these benefits, we have benchmarked the density of Skybridge circuits at older technology nodes using 4, 8 and 16-bit CLAs; 22nm and 32nm technologies were considered. Skybridge specific design rules for each technology, which were derived following the methodology described in Section 9, is shown in Figs. S8.14A and S8.14B. Minimum feature size and spacing for 22nm and 32nm design rules are 22nm and 32nm respectively.

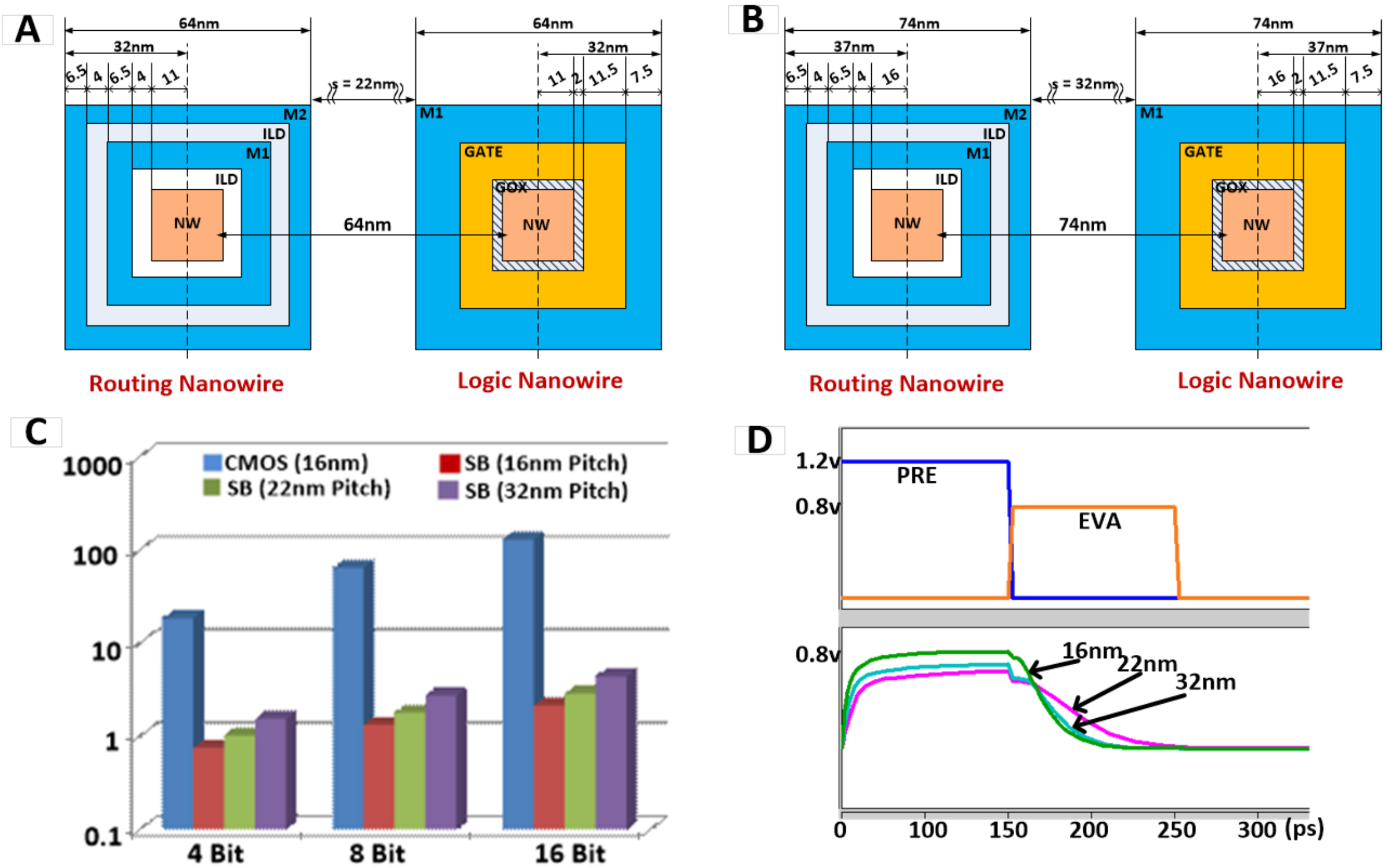

**Fig. S8.14 | Sensitivity to lithography precision.** A-B) Design rules for relaxed lithography at 22nm(A) and 32nm(B). C) Implications of lithography relaxation on Skybridge circuits: 4, 8 and 16-bit CLA designs. Comparisons are shown with Skybridge and CMOS designs at 16nm. D) HSPICE simulation results showing delays due to lithographic relaxations for a 8 input NAND gate.

The density calculations are shown in Fig. 8.14C. With relaxation of lithographic limits, the nanowire pitch increases; this increase has a linear impact on overall area. For 22nm nanowire pitch, the area increases by 1.47x compared to 16nm nanowire pitch designs; whereas for the 32nm nanowire pitch design the area increases by 2.05x. However, even for the pessimistic assumption of 32nm nanowire pitch, the density benefit compared to CMOS is significant, 29.5x for the 16-bit CLA design.

The impact of increase in the nanowire pitch on delay is shown in Fig. 8.14D; delay results are shown for 16nm, 22nm and 32nm pitch based design of an 8 input NAND gate. The delay is heavily dependent on the transistor design. For 16nm design, doping density in V-GAA Junctionless transistors was assumed to be 1e19 dopants/cm$^3$, and for 22nm and 32nm designs the doping density was 1e18 dopants/cm$^3$. Reduced doping concentration resulted in somewhat lower ON currents, 22nm device being the lowest. This is reflected in simulation results in Fig. 8.14D. By optimizing the device, and operating voltage, we expect the delay can be further improved.

## 9. 3-D CIRCUIT DESIGN RULES AND LAYOUT GUIDELINES

Section Summary:

This section presents Skybridge circuit design rules and layout guidelines. Numerical design rules derived from TCAD simulations and manufacturing assumptions are shown. Guidelines for circuit mapping into physical fabric are shown that take into account manufacturability, connectivity, noise mitigation and thermal management.

Overview

All Circuit design and layout in Skybridge adhere to 3-D specific design rules and guidelines. The design rules ensure conformity to necessary material structure requirements and manufacturing assumptions, as presented earlier. The guidelines allow efficient mapping of circuits in this 3-D fabric without routing congestion, helps in mitigating coupling noise, ensures thermal management and manufacturability.

The design rules are a set of numerical rules for circuit layout derived from TCAD simulations and envisioned manufacturing pathway. These design rules set the standard for minimum length, width, thickness, and spacing of nanowires, transistors, and metal layers. The guidelines for 3-D circuit mapping and layout are based on Skybridge's circuit style, global and intermediate signal routing, heat extraction, and manufacturability. Ease of implementations of dynamic circuits in 3-D is emphasized in these guidelines; careful considerations are taken to enable high fan-in logic implementations and to prevent long intra-logic interconnections that are detrimental to performance. Basic guidelines are discussed for routing signals using intrinsic features in Skybridge (signal carrying nanowires, Bridges and Coaxial routing structures) and considerations are taken to mitigate coupling noise through incorporating GND shielding layers on signal routing paths. Circuit design guidelines also take into account 3-D heat extraction requirements. Heat extraction features are used synergistically with other active components to prevent hotspot development in 3-D. Ensuring fabric manufacturability is precursor to all these guidelines.

### *9.1 Design Rules*

Design rules used for behavioral and thermal simulations of Skybridge circuits were derived from material requirements and the manufacturing pathway presented in Section 8. Materials required and their dimensions are specific to design choices, and are validated by simulations; for example: choice of 2nm thick $HfO_2$ as gate-dielectric for vertical J-GAA device was validated by detailed 3-D TCAD Sentaurus based modeling and simulations (see Section 2.1). Similarly material dimensions were selected for spacer, contact formation, inter-layer dielectric, and interconnect and heat junctions. Fig. S9.1 shows cross-section of routing-nanowire and logic-nanowire, and illustrates dimensions and spacing of different material regions. These dimensions are based on their core requirements and manufacturability. For example, as shown in Fig. S9.1, the 11.5nm thickness of TiN layer (gate electrode for vertical J-GAA devices) is determined both by minimum gate electrode thickness requirement for device functionality and lithographic alignment precision (± 3.3nm at 16nm node [46]) required for UV exposure (Section 7.6).

**Supplementary Table S9.1 | Design rules**

| | Width (nm) X | Length (nm) Z | Thickness (nm) Y | Spacing (nm) |
|---|---|---|---|---|
| Bridge **(X,Y,Z)** | 16-58 | 16 | 16-58 | 16-37 |
| Transistor Channel **(X,Y,X)** | 16 | 16 | 16 | 58 |
| Transistor Spacing (Z) | - | - | - | 16 |
| Gate Electrode **(Z)** | 29 | 16 | 11.5 | - |
| Contact **(X,Y,Z)** | 26 | 16 | 16 | 39 |
| Heat Junction **(X,Y,Z)** | 22 | 16 | 6 | - |
| Coaxial (Si-M1) **(X,Y)** | 37 | - | 37 | 4 (Si-M1) |
| Coaxial (M1-M2) **(X,Y)** | 58 | - | 58 | 4 (M1-M2) |

Table S9.1 lists design rules that are specific to each fabric component. Since Skybridge is a 3-D fabric, design rules are required in all X, Y and Z directions as presented in Table S9.1. Some choices are customizable to individual circuit designs, such as Coaxial routing layer length, heat junction spacing etc.; these are not listed in Table S9.1.

*9.2 Additional Guidelines*

An abstract view of the Skybridge fabric with key aspects is shown in Fig. S9.2. As illustrated, local interconnections for input, output and power rails are through Bridges and Coaxial structures. Intermittent Heat dissipating power pillars are also shown on the periphery of logic blocks.

Circuit mapping into the Skybridge fabric involves placement of device, contacts and power rails, and local, semi-global and global interconnections. This 3-D circuit mapping is made compatible with heat extraction and manufacturing requirements.

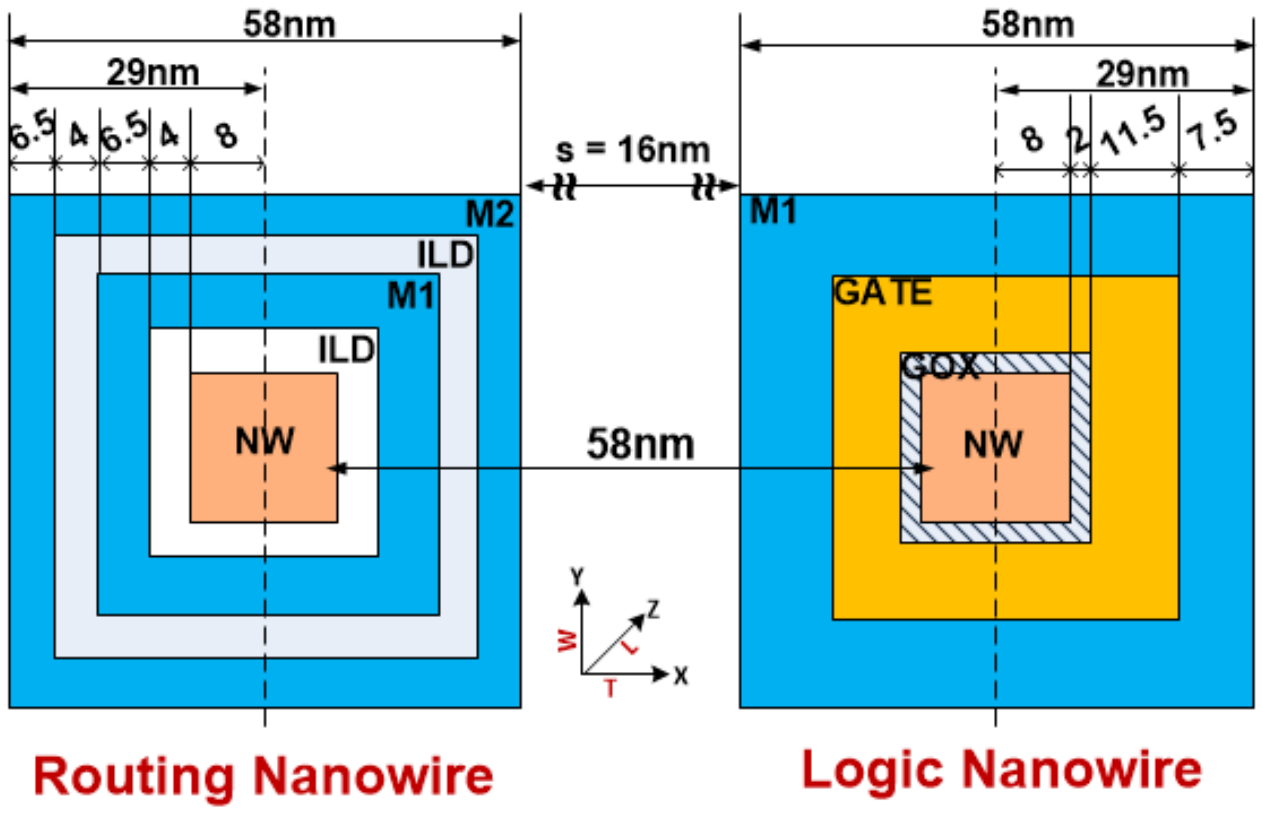

**Fig. S9.1 | Illustration of design rules.** A pair of nanowires are shown: one logic nanowire and another signal nanowire. Transistors are stacked in logic nanowires, whereas signal nanowires are primarily used for signal routing. The figure depicts different materials and dimensions. Logic nanowires outer dimensions are determined by transistor gate electrode thickness, gate contact requirements; signal nanowires outer dimensions are specified by ILD and different metal layer thicknesses.

For circuit mapping, arrays of regular vertical nanowires are partitioned into logic and signal routing nanowires. Logic nanowires are dedicated for containing transistors stacks, and signal nanowires are primarily used for signal routing.

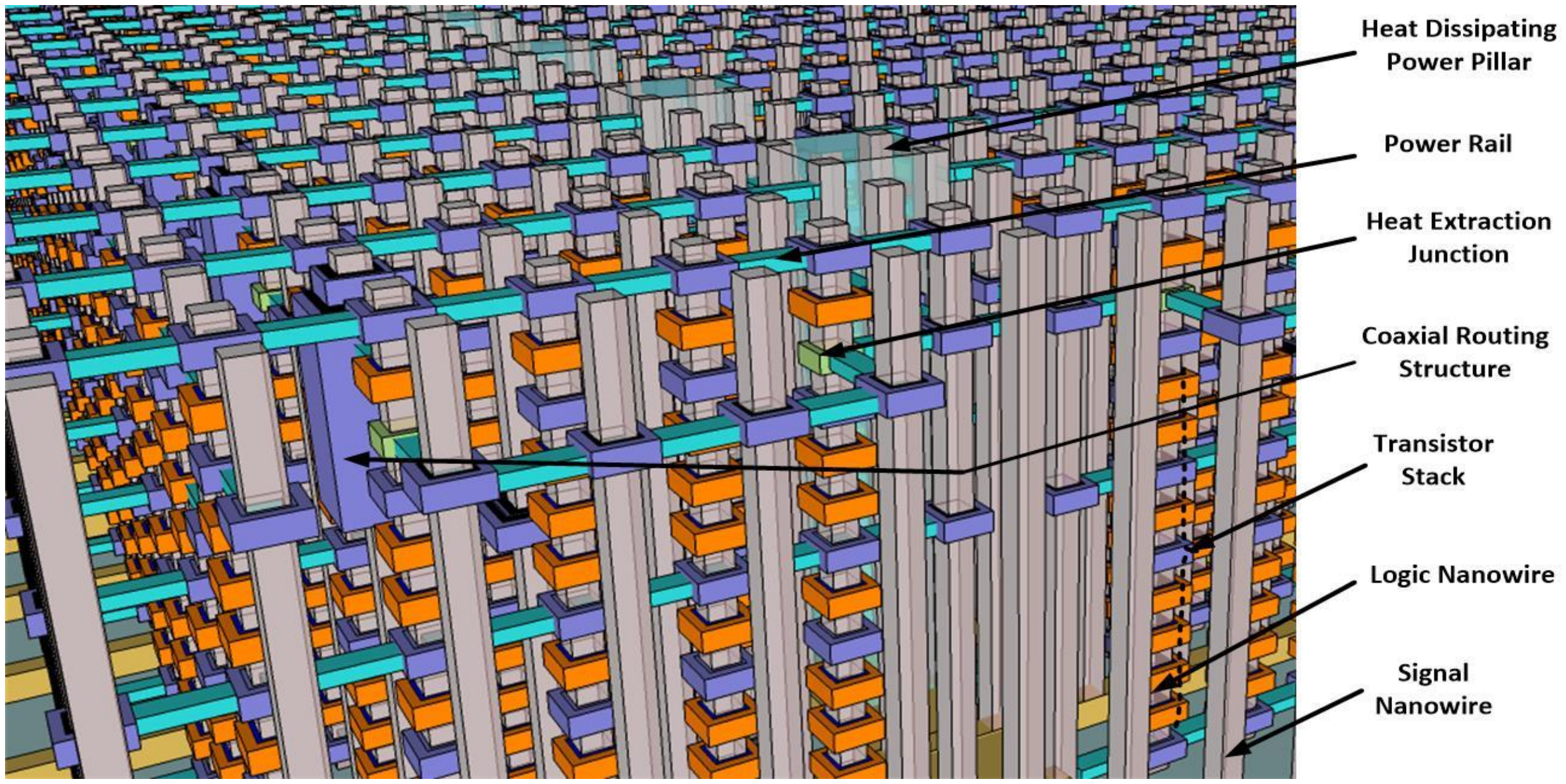

**Fig. S9.2 | Skybridge fabric representation.** The figure shows abstract layout of Skybridge fabric incorporating all fabric components. Logic and signal nanowires are separated, and are interleaved with each other. Logic nanowires contain transistor stacks, and have power rail contacts at top, middle and bottom. Signal nanowires carry signals themselves and also facilitate routing through Coaxial routing structures and Bridges. Coaxial routing structures have dedicated GND signal layer for noise shielding. Heat Extraction features ensure thermal management. As illustrated, Heat Extraction Junctions are placed on selective places on logic nanowires; extracted heat is dissipated through Heat Extraction Bridges and Heat Dissipating Power Pillars.

Placements of logic and signal nanowires are periodic, and are interleaved with each other. All nanowires were assumed to have a fixed height of 886nm. The logic nanowires were partitioned to have at most two logic stages, each having maximum fan-in of 9, and occupying half of maximum nanowire height. Interconnection in-between logic stages is through Bridges and Coaxial routing structures, and utilizes signal nanowires. Bridges form links between nanowires, and Coaxial routing structures that are placed on signal nanowires allow signal hopping and provide noise shielding. Three signals can be routed with one signal nanowire and surrounding metal shells in current designs; one of the three signals is dedicated for GND signal to provide noise shielding. In addition to these routing requirements, logic stages that are used in same logic block are placed in close proximity to reduce long intra-block connections, and thus to reduce delay.

Global signals in Skybridge are primarily clock and power signals. Power signal contacts (VDD, GND) are made at the top, middle, and bottom of the logic nanowires. GND contacts are made at the top and bottom, and VDD contacts are made in the middle; this configuration allows heat flow from the top of the nanowire towards the bottom bulk (details on thermal management on Section 5). Clock signals are routed in parallel to power signals.

Heat Extraction Junctions are placed at the output of every logic stages or one per logic nanowire, depending on the requirements. One input out of a fan-in of 9 is reserved in every logic stage for the Heat Junction. Extracted heat is dissipated through Bridges and Heat Dissipating Power Pillars. The large area Heat Pillars are placed on the periphery of logic blocks, and are separated by an average distance of 10 nanowire pitches from each other. Circuit mapping in the fabric takes into consideration the placement of these pillars.